\documentclass[journal]{IEEEtran}

\usepackage{amsmath}
\usepackage{amssymb}
\usepackage{times}
\usepackage{graphics}
\usepackage{graphicx}
\usepackage{epsfig}
\usepackage{enumerate}
\usepackage{algorithm}
\usepackage{algorithmic}
\usepackage{array}
\usepackage{verbatim}
\usepackage{filecontents}
\usepackage[T1]{fontenc}
\usepackage{tikz,pgfplots}
\usepackage{color,xcolor, soul}
\usepackage{mathrsfs}
\usepackage{url}
\usepackage{caption}
\usepackage{accents}
\usepackage{subcaption}
\usepackage{dsfont}
\usepackage[noadjust]{cite}
\usepackage{physics}
\usetikzlibrary{arrows,shapes,positioning}
\usepackage[T1]{fontenc}

\newcolumntype{M}{>{\centering\arraybackslash}m{\dimexpr.25\linewidth-2\tabcolsep}}
\newlength{\dhatheight}

\newcommand{\mmat}[1]{\ensuremath{\mathrm{\mathbf{\MakeUppercase{#1}}}}}
\newcommand{\mvec}[1]{\ensuremath{\mathrm{{\mathbf{\MakeLowercase{#1}}}}}}

\newcommand{\nth}{^{\textrm{\tiny{th}}}}
\renewcommand{\tr}{{\mathsf T}}				
\newcommand{\hr}{{\mathsf H}}

\newenvironment{MyColorPar}[1]{%
    \leavevmode\color{#1}\ignorespaces%
}{%
}%

\newcommand\scalemath[2]{\scalebox{#1}{\mbox{\ensuremath{\displaystyle #2}}}}

\usepackage[labelformat=simple]{subcaption}

\newtheorem{lemma}{Lemma}
\newtheorem{theorem}{Theorem}
\newtheorem{assumption}{Assumption}
\newtheorem{definition}{Definition}
\newtheorem{remark}{Remark}

\newtheorem{example}{Example}

\graphicspath{{Figures/}}

\begin{document}
\title{Generalized Power Iteration with Application to Distributed Connectivity Estimation of Asymmetric Networks}

\author{M.~Mehdi~Asadi$^{1}$ (\textit{Member, IEEE}),
        Mohammad~Khosravi$^{2}$ (\textit{Member, IEEE}),
        Hesam~Mosalli$^{3}$ (\textit{Member, IEEE}),
        St\'{e}phane~Blouin$^{4}$ (\textit{Senior Member, IEEE}) and
        Amir~G.~Aghdam$^{3}$ (\textit{Fellow, IEEE}) 
\thanks{$^{1}$M. M. Asadi is with the Department of Electrical Engineering, Polytechnique Montr\'{e}al, 2500 chemin de Polytechnique, Montr\'{e}al, QC, Canada H3T 1J4 (e-mail: \tt \small mehdiasadi2007@gmail.com).}
\thanks{$^{2}$M. Khosravi is with the Delft Center for Systems and  Control, Delft University of Technology, Delft, Netherlands (e-mail: \tt \small mohammad.khosravi@tudelft.nl).}
\thanks{$^{3}$H. Mosalli and A. G. Aghdam are with the Department of Electrical and Computer Engineering, Concordia University, Montr\'{e}al, QC, Canada H3G 1M8 (e-mail:\tt \small hesam.mosalli@mail.concordia.ca,  aghdam@ece.concordia.ca).}
\thanks{$^{4}$S. Blouin is with the Defence Research and Development Canada, Atlantic Research Centre, Dartmouth, NS, Canada B3A 3C5 (e-mail: \tt \small stephane.blouin2@forces.gc.ca).}
\thanks{This work has been supported by PWGSC contract no. W7707-145674/001/HAL, funded by Defence Research and Development Canada.}}

\markboth{}
{Shell \MakeLowercase{\textit{et al.}}: Bare Demo of IEEEtran.cls for IEEE Journals}

\maketitle



\title{Generalized Power Iteration - Application to Distributed Connectivity of Asymmetric Networks}

\author{M.~Mehdi~Asadi$^{1}$,
        Mohammad~Khosravi$^{2}$,
        Amir~G.~Aghdam$^{3}$,  
        and St\'{e}phane~Blouin$^{4}$ 
\thanks{$^{1}$M. M. Asadi is with the Department of Electrical Engineering, Polytechnique Montr\'{e}al, 2500 chemin de Polytechnique, Montr\'{e}al, QC, Canada H3T 1J4 (e-mail: \tt \small mehdiasadi2007@gmail.com).}
\thanks{$^{2}$M. Khosravi is with the Delft Center for Systems and  Control, Delft University of Technology, Delft, Netherlands (e-mail: \tt \small mohammad.khosravi@tudelft.nl).}
\thanks{$^{3}$A. G. Aghdam is with the Department of Electrical and Computer Engineering, Concordia University, Montr\'{e}al, QC, Canada H3G 1M8 (e-mail: \tt \small aghdam@ece.concordia.ca).}
\thanks{$^{4}$S. Blouin is with the Defence Research and Development Canada, Atlantic Research Centre, Dartmouth, NS, Canada B3A 3C5 (e-mail: \tt \small stephane.blouin2@forces.gc.ca).}
\thanks{This work has been supported by PWGSC contract no. W7707-145674/001/HAL, funded by Defence Research and Development Canada.}}
\maketitle

\begin{abstract}
The problem of connectivity assessment in an asymmetric network represented by a weighted directed graph is investigated in this article. 
A power iteration algorithm in a centralized implementation is developed first to compute the generalized algebraic connectivity of asymmetric networks. After properly transforming the Laplacian matrix of the network, two sequences of one-dimensional and two-dimensional subspaces are generated iteratively, one of which converges to the desired subspace spanned by the eigenvector(s) associated with the eigenvalue(s) representing the network's generalized algebraic connectivity. A distributed implementation of the proposed power iteration algorithm is then developed to compute the generalized algebraic connectivity from the viewpoint of each node, which is scalable to any asymmetric network of any size with a fixed message length per node. The convergence analysis of these algorithms is subsequently provided under some weak assumptions. The efficiency of the developed algorithms in computing the network connectivity is then demonstrated by simulations.
\end{abstract}

\IEEEpeerreviewmaketitle

\section{Introduction}
{A}{d-hoc} wireless networks are composed of a group of fixed and/or mobile nodes, capable of exchanging data without the support of a pre-existing infrastructure \cite{Akyildiz_02,Santi_05,Martinez_10}.
Cooperative algorithms have been developed for various applications over ad-hoc networks, including consensus, parameter estimation, target localization and data diffusion \cite{Moreau_05,Schizas_08,Notarstefano_11,Cattivelli_10}.
In parameter estimation algorithms, particularly, a set of unknown parameters are estimated using sensor data corrupted by measurement noise \cite{Schizas_08}, \cite{Xiao_05}. Estimation algorithms performing distributed computations have significant advantages compared to methods based on a centralized fusion scheme in terms of scalability and resilience to node failure \cite{Xiao_06}.
The connectivity degree of the network can significantly affect the convergence speed of cooperative algorithms used for various objectives over ad-hoc networks \cite{Moreau_05}, \cite{Cattivelli_10}. Thus, network connectivity is an important factor in the design and implementation of cooperative algorithms. In general, a network with a higher degree of connectivity is able to diffuse information faster and more effectively throughout the network \cite{Dilorenz_13}. 
%
%
%


A network is symmetric if all of its node-to-node communication links are bi-directional, i.e., a communication link $A\!\rightarrow\!B$ implies the existence of its reciprocal link $B\!\rightarrow\!A$ (with the same weight, for weighted graphs). This type of network is represented by an undirected graph (which may be weighted too). 
The algebraic connectivity of a symmetric network is defined as the smallest nonzero eigenvalue of the Laplacian matrix of the network graph \cite{Fiedler_73}. 
This connectivity measure, known as Fiedler's algebraic connectivity, is known to reflect the asymptotic convergence rate of consensus algorithms running on such networks \cite{Yang_10}.
A decentralized orthogonal iteration algorithm is introduced in \cite{Kempe_04} for computing the eigenvectors corresponding to the $k$ dominant eigenvalues of a symmetric network. A distributed method based on the singular value decomposition theory is proposed in \cite{Chen_20} to estimate the graph eigenspectra of mobile communication networks. 
A distributed procedure is developed in \cite{Dilorenzo_14} to estimate and control the algebraic connectivity of symmetric ad-hoc networks with random topology. The authors in \cite{Hayhoe_22} present a centralized method to estimate the eigenvalues of any weighted directed network with interacting dynamical agents using dynamical observations. 
A generalization of Fiedler's algebraic connectivity to directed graphs (digraphs) is introduced in \cite{Wu_05}, where the relationship between the algebraic connectivity and different properties of the graph is investigated. A simple extension of algebraic connectivity to digraphs is proposed in \cite{Li_13}, where the magnitude of the smallest nonzero eigenvalue of the Laplacian matrix is introduced as a measure of network connectivity. A decentralized algorithm is also developed in \cite{Li_13}, requiring the solution of a set of nonlinear equations with relatively high computational complexity, which limits the applicability of the approach to many real-world networks. 
Due to the importance of network connectivity on the performance of distributed algorithms running on such networks, the problem of maximizing algebraic connectivity under different constraints for undirected graphs is investigated in~\cite{shahbaz23,ortega2023maximal}.

A so-called \textit{generalized power iteration} (GPI) approach has been developed in this work to compute the connectivity of directed networks. In addition to computing the connectivity of networked systems, the power iteration method has many applications in ranking algorithms and principal component analysis \cite{Pan_21}, \cite{Huang_22}. A distributed scheme based on the power iteration method is proposed in \cite{Aragues_14} to derive upper and lower bounds for the algebraic connectivity of undirected graphs. In \cite{Gusrialdi_21}, a data-driven power iteration approach is proposed to estimate the dominant eigenvalues of an unknown linear time-invariant (LTI) system in a distributed manner. A distributed procedure to estimate the upper and lower bounds on algebraic connectivity of a directed graph is investigated in \cite{LI_19}.
%
%
The \emph{generalized algebraic connectivity} (GAC) was proposed in~\cite{Asadi_16} as a measure of connectivity for asymmetric networks represented by weighted digraphs. A similar concept is also introduced in \cite{AsadiTSMC_20} which characterizes GAC as the expected rate of convergence to consensus in asymmetric networks. A distributed algorithm based on the Krylov subspace method is proposed in \cite{AsadiTSMC_20} to compute the GAC of asymmetric networks with no convergence proof. The \textit{subspace consensus} approach with a convergence analysis is also proposed in \cite{AsadiTechRep_17} to develop a distributed procedure in order to compute the GAC of asymmetric networks. However, the algorithms presented in \cite{AsadiTSMC_20} and \cite{AsadiTechRep_17} are not scalable to networks of large size due to their message length of order $O(n)$ bits per node for a network of size $n$.

A novel algorithm based on the GPI approach is introduced in the present work to compute the GAC of asymmetric networks in both centralized and distributed manners. To this end, the Laplacian matrix of the network is transformed such that the original problem is reduced to finding the farthest eigenvalue of a matrix from the origin, referred to as the \emph{dominant} eigenvalue. Since a real-valued asymmetric matrix can have either a real or a pair of complex conjugate dominant eigenvalues in general, two sequences of one-dimensional and two-dimensional subspaces corresponding to cases where the dominant eigenvalue is real and complex, respectively, are generated in both centralized and distributed fashions. It is shown that one of the subspace sequences converges to the desired subspace associated with the GAC of the network in both centralized and distributed implementations. 
To circumvent the shortcomings of the approach in \cite{AsadiTSMC_20}, the convergence analysis of the proposed algorithm for estimating the GAC of the network for both centralized and distributed implementations is elaborated as one of the main contributions of this article. In addition, the proposed distributed GPI approach is scalable with a fixed message size of order $O(1)$ bits per node in each iteration, providing a significant improvement in terms of message complexity compared to the distributed approaches in \cite{AsadiTSMC_20,AsadiTechRep_17}. The efficacy of the developed algorithm in estimating the connectivity of asymmetric networks is subsequently demonstrated by simulations.

The remainder of the paper is organized as follows. Some background information and definitions are given in Section~\ref{Sec:II}. In Section \ref{Sec:III}, the problem statement and required assumptions are spelled out. The centralized GPI algorithm to compute the GAC of asymmetric networks is introduced in Section~\ref{Sec:IV} and its convergence proof is elaborated in Section~\ref{Sec:V}.
The distributed GPI algorithm and its convergence analysis are then explained in Sections~\ref{Sec:VI} and \ref{Sec:VII}, respectively. Simulation results are presented in Section~\ref{Sec:VIII}, and finally, conclusions are drawn in Section~\ref{Sec:IX}.



\begin{MyColorPar}{blue}
\end{MyColorPar}


\section{Preliminaries}\label{Sec:II}

This section reviews preliminary concepts prior to introducing the main results. 

\subsection{Symbols and Notation}

Let the sets of natural numbers, non-negative integers, real numbers, and complex numbers be denoted by $\mathbb{N}$, $\mathbb{Z}$, $\mathbb{R}$ and $\mathbb{C}$, respectively. Moreover, let $\mathbb{N}_{n}:=\{1,2,\ldots,n\}$. 
The complex conjugate, magnitude, real part and imaginary part of a complex scalar $c\!\in\!\mathbb{C}$ are, respectively, shown by $c^{\hr}$, $|c|$, $\Re(c)$ and $\Im(c)$. The transpose and conjugate transpose of a vector/matrix are represented by superscripts ${}^{\tr}$ and ${}^{\hr}$, respectively. Bold letters are utilized to distinguish matrices (uppercase letters) and arrays (lowercase letters) from scalar values. 
When applied to a set, the operator $|\cdot|$ denotes its cardinality. The symbol $\|\cdot\|$ denotes the Euclidean norm of a vector/matrix, and the symbol $\not\perp$ means not perpendicular.
%
%
The diacritic marks $\check{}$ and $\hat{}$ above $\mmat{Q}_{k}$, $\mmat{P}_{k}$, $\mmat{R}_{k}$, $d_{k}$ and $\lambda_{k}$, $k\!\in\!\mathbb{N}$, indicate the quantities derived from one-dimensional and two-dimensional subspaces, respectively, whereas a subscript $0$ represents the initial condition of a scalar/vector/matrix. A superscript $i$ denotes a quantity that is derived from the perspective of node $i$ and a subscript $k$ refers to the $k\nth$ iteration step of an iterative procedure. The inner product of two complex vectors $\mvec{v},\mvec{w}\in \mathbb{C}^{n}$ is denoted by $\langle\mvec{v},\mvec{w}\rangle = \mvec{w}^{\hr} \mvec{v}$. The all-zero and all-one column vectors of size $n\!\in\!\mathbb{N}$ are shown by $\mvec{0}_{n}$ and $\mvec{1}_{n}$, respectively, and $\mmat{I}_{n}$ represents the square identity matrix of size $n$. Furthermore, $e$ denotes the Euler number and $\log(\cdot)$, $\log_{2}(\cdot)$ and $\log_{10}(\cdot)$ are, respectively, the natural, binary and decimal logarithms. The trace and determinant of matrix $\mmat{A}$ are denoted by $\mathrm{tr}(\mmat{A})$ and $\mathrm{det}(\mmat{A})$, respectively. Given a matrix $\mmat{A} \in \mathbb{R}^{n \times n}$, define $\Psi(\mmat{A})$ as a set of triplets $\Psi(\mmat{A})=\{ (\lambda_{i}(\mmat{A}),\mvec{v}_{i}(\mmat{A}),\allowbreak\mvec{w}_{i}(\mmat{A})) \, | \, i\!\in\!\mathbb{N}_{n}\}$, where $\lambda_{i}(\mmat{A})\!\in\! \mathbb{C}$, $\mvec{v}_{i}(\mmat{A})\!\in\!\mathbb{C}^{n}$ and $\mvec{w}_{i}(\mmat{A})\!\in\!\mathbb{C}^{n}$ denote the $i\nth$ eigenvalue of $\mmat{A}$, and the right and left eigenvectors associated with it, respectively. These  triplets are indexed in nondecreasing order in terms of the magnitude of the eigenvalues, i.e., $|\lambda_{1}(\mmat{A})| \leq |\lambda_{2}(\mmat{A})| \leq \cdots \leq |\lambda_{n}(\mmat{A})|$. The spectrum of matrix $\mmat{A}$ is defined as $\Lambda(\mmat{A})\!=\!\{ \lambda_{i}(\mmat{A}) \, | \, i\!\in\!\mathbb{N}_{n}\}$.

Some useful concepts from linear algebra and graph theory  are reviewed next.

\subsection{Orthogonal Projection}

A linear transformation $\mathcal{P}:\mathbb{C}^{n}\rightarrow \mathbb{C}^{n}$, represented by a square matrix $\mmat{P}\!\in\!\mathbb{C}^{n\times n}$, is called an \emph{orthogonal projector} if $\mmat{P}^{2}=\mmat{P}=\mmat{P}^{\hr}$ \cite{Varga_00}. 
Let $\mmat{Q}\!\in\!\mathbb{C}^{n \times k}$ be a matrix whose $k$ columns ($k\leq n$) represent a basis for the $k$-dimensional complex subspace $\mathbb{S}\!\subseteq\! \mathbb{C}^{n}$. 
Then, the orthogonal projection from the subspace $\mathbb{S}$ is defined by a \emph{projection matrix} $\mmat{P}=\mathfrak{f}(\mmat{Q})$, where function $\mathfrak{f}(\cdot):\mathbb{C}^{n \times k} \rightarrow \mathbb{C}^{n \times n}$ is given by
\begin{equation}\label{f(Q)1}
\mathfrak{f}(\mmat{Q}) = \mmat{Q}(\mmat{Q}^{\hr}\mmat{Q})^{-1} \mmat{Q}^{\hr}.
\end{equation}
Moreover, the projection of a matrix $\mmat{A}\in \mathbb{C}^{n \times n}$ onto the $k$-dimensional subspace $\mathbb{S}$ (spanned by the column vectors of matrix $\mmat{Q}$ and represented by the projection matrix $\mmat{P}=\mathfrak{f}(\mmat{Q})$) is denoted by matrix $\mmat{R}=\mathfrak{g}(\mmat{A},\mmat{Q})$, where function $\mathfrak{g}(\cdot,\cdot):\mathbb{C}^{n\times n}\times\mathbb{C}^{n\times k}\rightarrow\mathbb{C}^{k\times k}$ is defined as
\begin{equation}\label{g(M,Q)1}
\mathfrak{g}(\mmat{A},\mmat{Q}) = (\mmat{Q}^{\hr}\mmat{Q})^{-1} \mmat{Q}^{\hr} \mmat{A} \mmat{Q}.
\end{equation}
Consider two distinct subspaces $\mathbb{S}_{1},\mathbb{S}_{2}\!\subseteq\!\mathbb{C}^{n}$.
The distance between $\mathbb{S}_{1}$ and $\mathbb{S}_{2}$ is defined as $\|\mmat{P}_{1} - \mmat{P}_{2}\|$, where $\mmat{P}_{1}$ and $\mmat{P}_{2}$ are the projection matrices associated with the orthogonal projectors $\mathcal{P}_{1}$ and $\mathcal{P}_{2}$ onto the subspaces $\mathbb{S}_{1}$ and $\mathbb{S}_{2}$, respectively \cite{Gohberg_86}. 
We now define functions $\Theta(\cdot)$ and $O(\cdot)$, which will prove useful in the development of the main results. The equality $f(x)=\Theta(g(x))$ is said to hold for two functions $f(\cdot)$ and $g(\cdot)$ if there exist finite positive constants $m$, $M$ and $x_{0}$, such that $m\, g(x)\!\leq\!f(x)\!\leq\!M\, g(x)$ for all $x\geq x_{0}$. In addition, $f(x)=O(g(x))$ is said to hold for two functions $f(\cdot)$ and $g(\cdot)$ if there exist finite positive constants $M$ and $x_{0}$, such that $f(x)\!\leq\!M\, g(x)$ for all $x\geq x_{0}$.

\subsection{Graph Theory}

Let $\mathcal{G}=(\mathcal{V},\mathcal{E},\mmat{W})$ represent a weighted directed graph (digraph) composed of $n$ vertices with vertex set $\mathcal{V}$, edge set $\mathcal{E}$, and weight matrix $\mmat{W}=[w_{ij}]\in \mathbb{R}^{n\times n}$, such that
\begin{subequations}
\begin{align}
\mathcal{V} &= \{1,2,\ldots,n \}, \\
\mathcal{E} &= \{ (i,j)\!\in\!\mathcal{V} \times \mathcal{V} \;|\; i\neq j,\; w_{ji}\neq 0 \}.
\end{align}
\end{subequations}
Note that $w_{ji}$ is a finite positive constant for each ordered pair $(i,j)\in \mathcal{E}$ of two distinct vertices $i,j\in \mathcal{V}$. The vertex $j$ is said to belong to the \emph{in-neighbor} set of vertex $i$, denoted by $\mathcal{N}_{i}$, if there is a directed edge pointing from $j$ to $i$, i.e., $(j,i)\in \mathcal{E}$.
A digraph $\mathcal{G}$ is said to be \emph{strongly connected} if there exists a directed path between every ordered pair of distinct vertices $i,j\in \mathcal{V}$. 
The \emph{Laplacian} of the weighted digraph $\mathcal{G}$ is an $n \times n$ real matrix $\mmat{L}=[l_{ij}]$ whose $(i,j)\nth$ entry is given by \cite{Godsil_01}
\begin{equation}\label{eq:Laplacian1}
l_{ij}=
\begin{cases}
-w_{ij},              & \mathrm{if}\; (j,i)\!\in\! \mathcal{E}, \\
\sum\limits_{k=1,k\neq i}^{n}w_{ik}, & \mathrm{if}\; j\!=\!i,\\
0,                    & \mathrm{otherwise}.
\end{cases}
\end{equation}
The notion of \textit{generalized algebraic connectivity} (GAC) is introduced in~\cite{Asadi_16} as a measure of connectivity for asymmetric networks. The definition is repeated here for completeness.
\begin{definition}\label{def:11}
Given a strongly connected weighted digraph $\mathcal{G}$ with Laplacian matrix $\mmat{L}$, the GAC of $\mathcal{G}$ is defined as
\begin{equation}\label{eqn:GACdef1}
\tilde{\lambda}(\mmat{L})= \min_{\lambda_{i}(\mmat{L}) \in \Lambda(\mmat{L}),\; \lambda_{i}(\mmat{L})\neq 0} \Re(\lambda_{i}(\mmat{L})).
\end{equation}
\end{definition}
In other words, the GAC is the smallest nonzero real part of the eigenvalues of the Laplacian matrix.


\section{Problem Statement}\label{Sec:III} 

The primary objective of this publication is to derive an algorithmic scheme for computing the GAC of an asymmetric network in both centralized and distributed manners. To this end, the authors aim at exploiting the \emph{power iteration} method which has been used extensively in the literature to compute the algebraic connectivity of symmetric networks in both centralized and distributed fashions \cite{Dilorenzo_14,Li_13}. 
The power iteration algorithm provides a recursive approach to compute the dominant eigenvalue (defined in the previous section) of a real-valued symmetric matrix. It is assumed that the matrix has a dominant eigenvalue with an algebraic multiplicity of one whose magnitude is strictly greater than the magnitude of the other eigenvalues \cite{Saad_92}. 
There are two main challenges concerning the power iteration method for computing the GAC of asymmetric networks, as follows
\begin{enumerate}[i)]
\item the power iteration method computes the eigenvalue of a symmetric matrix with maximum (not minimum) magnitude (not real part), and
\item the convergence of the power iteration procedure is not guaranteed for asymmetric real matrices with a pair of complex conjugate dominant eigenvalues~\cite{Asadi_16}.
\end{enumerate}
The GPI algorithm is developed in this work as a new generalized version of the power iteration algorithm in such a way that
\begin{enumerate}[1)]
\item it can be used to identify the dominant eigenvalue(s) of an asymmetric matrix, representing the Laplacian of an asymmetric network, unlike the existing power iteration procedures which can only be applied to symmetric matrices \cite{Dilorenz_13}, \cite{Dilorenzo_14}, and
\item it can be used in a distributed setting with a fixed message size of order $O(1)$ bits per node, making the algorithm scalable, unlike the approaches presented in  \cite{Gusrialdi_21,Spong_15,AsadiTSMC_20,AsadiTechRep_17}.
\end{enumerate}
Prior to addressing challenges (i) and (ii) described earlier, a matrix transformation borrowed from \cite{Asadi_16} is presented in the next lemma, which will be employed to develop a procedure to compute the GAC.
\begin{lemma}\label{lemm:MLM}
Let $\mmat{L}\!\in\!\mathbb{R}^{n\times n}$ denote the Laplacian matrix of a weighted digraph $\mathcal{G}$ composed of $n$ nodes. Assume that the zero eigenvalue of $\mmat{L}$ has an algebraic multiplicity of one, and define the \emph{modified Laplacian matrix} of $\mathcal{G}$ as
\begin{equation}\label{eqn:MLM}
\tilde{\mmat{L}} =e^{\mmat{I}_{n}-\delta \mmat{L}}-e\, \mvec{w}_{1}\!(\mmat{L})\, \mvec{w}_{1}^{\tr}\!(\mmat{L}),
\end{equation}
where 
\begin{equation}
0\!<\!\delta \!<\! \Delta^{-1},  \label{eq:delta}
\end{equation}
and 
\begin{equation}
\Delta \!=\! \max_{i\in \mathcal{V}} \sum_{j\in \mathcal{N}_{i}}\!w_{ij}.
\end{equation}
It then follows that
\begin{equation}\label{def:GAC}
\tilde{\lambda}(\mmat{L}) = \frac{1}{\delta}\Big{[}1-\log\big{(}\max_{\lambda_{i}(\tilde{\mmat{L}})\in \Lambda(\tilde{\mmat{L}})}|\lambda_{i}(\tilde{\mmat{L}})|\big{)}\Big{]}.
\end{equation}
\end{lemma}

The following assumptions are made throughout the paper.
\begin{assumption}\label{Assump1}
The weighted digraph $\mathcal{G}$, representing the information flow structure of the network, is strongly connected, and the left eigenvector $\mvec{w}_{1}(\mmat{L})$ associated with the zero eigenvalue of $\mmat{L}$ is assumed to be known \emph{a priori}.
\end{assumption}
%
%
\begin{assumption}\label{Assump2}
The dominant eigenvalue of the modified Laplacian matrix $\tilde{\mmat{L}}$ of the network digraph has an algebraic multiplicity of one.
\end{assumption}
\begin{assumption}\label{Assump3}
The initial state vector $\mvec{x}_{0}$ of the network satisfies the property $\mvec{x}_{0} \not\perp \mathrm{span}\{\mvec{v}_{n-1}\!(\tilde{\mmat{L}}),\mvec{v}_{n}\!(\tilde{\mmat{L}})\}$.
\end{assumption}



\section{Generalized Power Iteration Algorithm} \label{Sec:IV}

In this section, a novel iterative algorithm is presented for computing the GAC of a weighted digraph $\mathcal{G}$ in a centralized setting. 

\subsection{An approach to address the challenges of Section \ref{Sec:III}}

To address the first challenge described in Section~\ref{Sec:III}, the Laplacian matrix $\mmat{L}$ is transformed into the modified Laplacian matrix $\tilde{\mmat{L}}$ according to Lemma~\ref{lemm:MLM}. Using Assumption~\ref{Assump1}, this transformation converts the problem of finding $\tilde{\lambda}(\mmat{L})$ into the problem of finding the dominant eigenvalue of $\tilde{\mmat{L}}$. Since it is assumed that the network is represented by a weighted digraph $\mathcal{G}$ with a real-valued asymmetric weight matrix $\mmat{W}$, the dominant eigenvalue(s) of $\tilde{\mmat{L}}$ will be either a real number or a complex conjugate pair under Assumption~\ref{Assump2}. 
Therefore, $\tilde{\mmat{L}}$ could have one or two dominant eigenvalues, and the conventional power iteration method cannot be used to find them (we will hereafter refer to this as a single eigenvalue, noting that in the case of complex conjugate eigenvalues either of the two can be considered as the dominant one). 

Regarding the second challenge, note that it is not known \textit{a priori} whether the dominant eigenvalue of $\tilde{\mmat{L}}$ is a real number or a complex conjugate pair. 
Thus, the algorithm constructs a one-dimensional subspace $\mathcal{V}_{k}$ and a two-dimensional subspace $\mathcal{W}_{k}$ at the $k\nth$ iteration of the algorithm corresponding to cases when the dominant eigenvalue of $\tilde{\mmat{L}}$ is real and complex, respectively. 
Note that the subspaces $\mathcal{V}_{k}$ and $\mathcal{W}_{k}$ are spanned by the latest state vectors which are obtained at the $k\nth$ iteration of the GPI algorithm. Therefore, a sequence of one-dimensional subspaces $\{ \mathcal{V}_{k}\}_{k\in \mathbb{N}}$ and a sequence of two-dimensional subspaces $\{\mathcal{W}_{k}\}_{k\in \mathbb{N}}$ are generated in this procedure.
Under Assumptions~\ref{Assump1}-\ref{Assump3}, it can be shown that if the dominant eigenvalue of $\tilde{\mmat{L}}$ is a real number, the subspace sequence $\{\mathcal{V}_{k}\}_{k\in \mathbb{N}}$ converges asymptotically to the one-dimensional subspace $\mathcal{V}^{\ast}:=\mathrm{span}\{\mvec{v}_{n}(\tilde{\mmat{L}})\}$ spanned by the right eigenvector of $\tilde{\mmat{L}}$ associated with its real dominant eigenvalue, while the subspace sequence $\{\mathcal{W}_{k}\}_{k\in \mathbb{N}}$ is not convergent. 
However, if $\tilde{\mmat{L}}$ has a pair of complex conjugate dominant eigenvalues, only the subspace sequence $\{\mathcal{W}_{k}\}_{k\in \mathbb{N}}$ converges asymptotically to the two-dimensional subspace $\mathcal{W}^{\ast}:=\mathrm{span}\{\mvec{v}_{n-1}\!(\tilde{\mmat{L}}),\mvec{v}_{n}\!(\tilde{\mmat{L}})\}$ spanned by the two right eigenvectors associated with the pair of complex conjugate dominant eigenvalues of $\tilde{\mmat{L}}$, while the subspace sequence $\{\mathcal{V}_{k}\}_{k\in \mathbb{N}}$ is not convergent.
By defining matrices $\check{\mmat{R}}_{k}$ and $\hat{\mmat{R}}_{k}$ as the projections of $\tilde{\mmat{L}}$ onto the subspaces $\mathcal{V}_{k}$ and $\mathcal{W}_{k}$, respectively, it can be demonstrated that the Euclidean norm of sequence $\{\check{\mmat{R}}_{k}\}_{k\in \mathbb{N}}$ converges to the magnitude of the real dominant eigenvalue of $\tilde{\mmat{L}}$, while that of sequence $\{\hat{\mmat{R}}_{k}\}_{k\in \mathbb{N}}$ converges to the magnitude of the complex conjugate dominant eigenvalues of $\tilde{\mmat{L}}$. 
Using \eqref{def:GAC} and the obtained sequence of the magnitude of the dominant eigenvalue of $\tilde{\mmat{L}}$,
the sequence $\{\tilde{\lambda}_{k}\}_{k\in \mathbb{N}}$ is computed which is proved to asymptotically converge to the network's GAC. 


The convergence of the recursive procedure is measured by defining $\check{d}_{k}$ and $\hat{d}_{k}$ as the distance between the pairs of successive one-dimensional subspaces $\{\mathcal{V}_{k-1},\mathcal{V}_{k}\}$ and two-dimensional subspaces $\{\mathcal{W}_{k-1},\mathcal{W}_{k}\}$, respectively. The iterative procedure is said to converge when the value of $\min\{\check{d}_{k},\hat{d}_{k}\}$ is less than a prespecified sufficiently small threshold $\epsilon$. In addition, in each iteration $k\in \mathbb{N}$, the dominant eigenvalue of $\tilde{\mmat{L}}$ is identified as a real number if $\check{d}_{k}<\hat{d}_{k}$ and a complex number otherwise. 

\subsection{Description of the Centralized GPI Algorithm}\label{Sec:IIIB}

%
%


In the centralized implementation of the GPI algorithm, let $\bar{\mvec{x}}_{k},\mvec{x}_{k}\!\in\!\mathbb{C}^{n}$ be defined as the \textit{intermediate state vector} and \textit{state vector} of the network at the $k\nth$ iteration, respectively. 
As per Assumption \ref{Assump3}, let the initial state vector $\mvec{x}_{0}\!\in\! \mathbb{C}^{n}$ with $\|\mvec{x}_{0}\|\!=\!1$ be generated randomly. Therefore, the condition
\begin{equation}\label{x0ini1}
\mvec{x}_{0} \not\perp \mathrm{span}\big{\{} \mvec{v}_{n-1}(\tilde{\mmat{L}}),\mvec{v}_{n}(\tilde{\mmat{L}})\big{\}}
\end{equation}
holds almost surely. Consider a sufficiently small positive scalar $\epsilon$ for termination condition, and let the iteration index $k$, the distance $d_{k}$ between successive subspaces, and the projection matrices $\check{\mmat{P}}_{k}$ and $\hat{\mmat{P}}_{k}$ be initialized as follows
\begin{equation}\label{kd0P1}
k\gets 1,\;\; d_{1}\gets \epsilon,\;\; \check{\mmat{P}}_{0} \gets \mmat{0}_{n\times n},\;\;\hat{\mmat{P}}_{0} \gets \mmat{0}_{n\times n}. 
\end{equation}
As long as $d_{k}\geq \epsilon$, let the vectors $\bar{\mvec{x}}_{k}$ and $\mvec{x}_{k}$ be updated recursively as 
\begin{subequations}\label{xk+11}
\begin{align}
\bar{\mvec{x}}_{k} =&\, \tilde{\mmat{L}}\,\mvec{x}_{k-1}, \\
\mvec{x}_{k} =& \, \frac{\bar{\mvec{x}}_{k}}{\left\|\bar{\mvec{x}}_{k}\right\|},    
\end{align}
\end{subequations}
at the $k\nth$ iteration of the algorithm, for any $k\!\in\!\mathbb{N}$. Moreover, let the updated values of $\check{\mmat{Q}}_{k}$, $\hat{\mmat{Q}}_{k}$, $\check{\mmat{P}}_{k}$, $\hat{\mmat{P}}_{k}$, $\check{\mmat{R}}_{k}$, $\hat{\mmat{R}}_{k}$, $\check{d}_{k}$, $\hat{d}_{k}$, $\check{\lambda}_{k}$ and $\hat{\lambda}_{k}$ be obtained as
\begin{subequations}\label{updatebunch1}
\begin{align}
\check{\mmat{Q}}_{k}&= \mvec{x}_{k}, &&\hat{\mmat{Q}}_{k}= [\mvec{x}_{k-1}\;\;\mvec{x}_{k}], \label{miandef1}\\
\check{\mmat{P}}_{k}&=\mathfrak{f}(\check{\mmat{Q}}_{k}), &&\hat{\mmat{P}}_{k}=\mathfrak{f}(\hat{\mmat{Q}}_{k}), \label{miandef2}\\
\check{\mmat{R}}_{k}&=\mathfrak{g}(\tilde{\mmat{L}},\check{\mmat{Q}}_{k}), &&\hat{\mmat{R}}_{k}=\mathfrak{g}(\tilde{\mmat{L}},\hat{\mmat{Q}}_{k}),\label{miandef3}\\
\check{d}_{k}&=\|\check{\mmat{P}}_{k}-\check{\mmat{P}}_{k-1}\|, &&\hat{d}_{k}=\|\hat{\mmat{P}}_{k}-\hat{\mmat{P}}_{k-1}\|, \label{dcdh1} \\
\check{\lambda}_{k}&=|\check{\mmat{R}}_{k}|, &&\hat{\lambda}_{k}=\Big{|}\tfrac{1}{2} \mathrm{tr}(\hat{\mmat{R}}_{k})\label{lamb1} \\
& && \;\;\;\;\;\;\;\;\;\;\!+\!\sqrt{(\tfrac{1}{2} \mathrm{tr}(\hat{\mmat{R}}_{k}))^{2}-\mathrm{det}(\hat{\mmat{R}}_{k})}\Big{|}, \nonumber
\end{align}
\end{subequations}
for any $k\!\in\!\mathbb{N}$. Using the above updated parameters, the values of $d_{k+1}$ and $\tilde{\lambda}_{k+1}$ are then computed as follows
\begin{subequations}\label{dk+11}
\begin{align}
d_{k+1} &= \min\{\check{d}_{k},\hat{d}_{k}\}, \label{dcdh2} \\
\tilde{\lambda}_{k+1} &=
\begin{cases}
\tfrac{1}{\delta}\big{[}1-\log\small{(}\check{\lambda}_{k}\small{)}\big{]},            & \mathrm{if}\;\; d_{k+1}=\check{d}_{k}, \\
\tfrac{1}{\delta}\big{[}1- \log\small{(} \hat{\lambda}_{k}\small{)} \big{]},  & \mathrm{if}\;\; d_{k+1}=\hat{d}_{k}.
\end{cases} \label{lambdak1}
\end{align}
\end{subequations}
At the end of the $k\nth$ iteration, the iteration index is updated as $k\!\gets\! k+1$, and the procedure continues until the stopping condition $d_{k}\!<\!\epsilon$ is met. Finally, $\tilde{\lambda}_{k}$ is returned as the GAC of the network.  
A pseudo-code of the GPI algorithm for the proposed centralized implementation is provided in Algorithm~\ref{AlgGPI1}. 

\begin{algorithm}[]
\begin{algorithmic}[1]
\STATE {Inputs: $\mvec{w}_{1}(\mmat{L})$, $\delta$, $\epsilon$.}
\STATE {Initialize $\mvec{x}_{0}$ as a random unit vector (Assumption \ref{Assump3}).}
\STATE {Initialize $k$, $d_{1}$, $\check{\mmat{P}}_{0}$ and $\hat{\mmat{P}}_{0}$ using \eqref{kd0P1}.}
\WHILE {$d_{k} \geq \epsilon$}\label{Alg1_Termination1}
\vspace{2.5pt}
\STATE {Update the vectors $\bar{\mvec{x}}_{k}$ and  $\mvec{x}_{k}$ using \eqref{xk+11}.}
\STATE {Update $\check{\mmat{Q}}_{k}$, $\hat{\mmat{Q}}_{k}$, $\check{\mmat{P}}_{k}$, $\hat{\mmat{P}}_{k}$, $\check{\mmat{R}}_{k}$, $\hat{\mmat{R}}_{k}$, $\check{d}_{k}$, $\hat{d}_{k}$, $\check{\lambda}_{k}$ and $\hat{\lambda}_{k}$ using \eqref{updatebunch1}.}
\STATE {Compute $d_{k+1}$ and $\tilde{\lambda}_{k+1}$ using \eqref{dk+11}.}
\STATE $k\gets k+1$.
\ENDWHILE
\vspace{2.5pt}
\STATE {Output: Return $\tilde{\lambda}_{k}$.}
\end{algorithmic}\caption{A centralized procedure to compute the GAC of an asymmetric network using the GPI algorithm.}
\label{AlgGPI1}
\end{algorithm}

\section{Convergence Proof of Centralized GPI Algorithm}\label{Sec:V}

A convergence analysis of the centralized GPI algorithm is elaborated in this section. As the first step, the following lemma is presented.

\begin{lemma}\label{lemm:dcdh1}
Consider $\mathbf{x}_{0}$, $\mathbf{x}_{1}$, $\mathbf{x}_{2}\!\in\!\mathbb{C}^{n}$ as three complex vectors with unit Euclidean norms. 
Define three complex scalars $z_{1},z_{2},z_{3}\!\in\! \mathbb{C}$ as 
\begin{subequations}
\begin{align}
& z_{1}= \langle \mvec{x}_{2},\mvec{x}_{1}\rangle, \label{z1def0}\\ 
& z_{2}= \langle \mvec{x}_{1},\mvec{x}_{0}\rangle, \label{z2def0}\\ 
& z_{3}= \langle \mvec{x}_{2},\mvec{x}_{0}\rangle.  \label{z3def0} 
\end{align}    
\end{subequations}    
By defining $\check{\mmat{Q}}_{j}$, $\hat{\mmat{Q}}_{j}$, $\check{\mmat{P}}_{j}$,  $\hat{\mmat{P}}_{j}$ for $j\!\in\!\{1,2\}$ using \eqref{miandef1} and \eqref{miandef2}, 
it follows that 
\begin{subequations}
\begin{align}
 \|\check{\mmat{P}}_{2} - \check{\mmat{P}}_{1}\| & = \sqrt{1-|z_{1}|^{2}}, \label{lemm15-10}\\
\|\hat{\mmat{P}}_{2} - \hat{\mmat{P}}_{1}\| & = \sqrt{1-\tfrac{|z_{1}z_{2}-z_{3}|^{2}}{(1-|z_{1}|^{2})(1-|z_{2}|^{2})}}. \label{lemm15-20}
\end{align}    
\end{subequations}
\end{lemma}

\begin{proof}
Using the definition of function $\mathfrak{f}(\cdot)$ in \eqref{f(Q)1} and according to the definitions in \eqref{miandef1} and \eqref{miandef2},  $\check{\mmat{P}}_{1},\hat{\mmat{P}}_{1},\check{\mmat{P}}_{2},\hat{\mmat{P}}_{2}$ are obtained as follows
\begin{subequations}
\begin{align}
& \check{\mmat{P}}_{1}= \mvec{x}_{1}\mvec{x}_{1}^{\hr}, \label{Pc10}\\
& \check{\mmat{P}}_{2}= \mvec{x}_{2}\mvec{x}_{2}^{\hr}, \label{Pc20}\\
& \hat{\mmat{P}}_{1}= \tfrac{1}{1-|z_{2}|^{2}}\big{[} \mvec{x}_{0}\mvec{x}_{0}^{\hr}+\mvec{x}_{1}\mvec{x}_{1}^{\hr}-z_{2}\mvec{x}_{0}\mvec{x}_{1}^{\hr}-z_{2}^{\hr}\mvec{x}_{1}\mvec{x}_{0}^{\hr}\big{]}, \label{Ph10}\\
& \hat{\mmat{P}}_{2}= \tfrac{1}{1-|z_{1}|^{2}}\big{[}
\mvec{x}_{1}\mvec{x}_{1}^{\hr}+\mvec{x}_{2}\mvec{x}_{2}^{\hr}-z_{1}\mvec{x}_{1}\mvec{x}_{2}^{\hr}-z_{1}^{\hr}\mvec{x}_{2}\mvec{x}_{1}^{\hr} \big{]}. \label{Ph20}
\end{align}    
\end{subequations}
Define also $\check{\mmat{P}}$ and $\hat{\mmat{P}}$ as \begin{subequations}
\begin{align}
\check{\mmat{P}} &= \check{\mmat{P}}_{2} -\check{\mmat{P}}_{1}, \label{Pchdef0} \\
\hat{\mmat{P}} &= \hat{\mmat{P}}_{2} -\hat{\mmat{P}}_{1}.\label{Phdef0}
\end{align}   
\end{subequations}
To prove the validity of \eqref{lemm15-10} in the first part of the proof, $\check{\mmat{P}}$ is computed from \eqref{Pc10}, \eqref{Pc20} and \eqref{Pchdef0} as
\begin{equation}\label{Pch0}
\check{\mmat{P}}= \mvec{x}_{2}\mvec{x}_{2}^{\hr}-\mvec{x}_{1}\mvec{x}_{1}^{\hr}.
\end{equation}
Using \eqref{Pch0} and the definition of $z_{1}$ in \eqref{z1def0}, $\check{\mmat{P}}^{3}$ is computed as
\begin{align}
\check{\mmat{P}}^{3}=&\; \check{\mmat{P}} - \mvec{x}_{2}\mvec{x}_{2}^{\hr}\mvec{x}_{1}\mvec{x}_{1}^{\hr}\mvec{x}_{2}\mvec{x}_{2}^{\hr} \nonumber \\
&+\mvec{x}_{1}\mvec{x}_{1}^{\hr}\mvec{x}_{2}\mvec{x}_{2}^{\hr}\mvec{x}_{1}\mvec{x}_{1}^{\hr}= \big{[} 1-|z_{1}|^{2}\big{]}\check{\mmat{P}}.\label{Pch30}
\end{align}
After multiplying both sides of \eqref{Pch30} by $\check{\mmat{P}}^{n-3}$, one arrives at
\begin{equation}\label{RES1}
\check{\mmat{P}}^{n}= \big{[} 1-|z_{1}|^{2}\big{]} \check{\mmat{P}}^{n-2}.   
\end{equation}
This results in
\begin{equation}\label{RES2}
\Lambda(\check{\mmat{P}})\!=\!\Big{\{}{\underbrace{0,0,\ldots,0}_{n-2}},\pm \sqrt{1-|z_{1}|^{2}}\Big{\}},    
\end{equation}
and 
\begin{equation}
\|\check{\mmat{P}}\|=\sqrt{1-|z_{1}|^{2}},    
\end{equation}
which completes the proof of \eqref{lemm15-10}. To prove the validity of \eqref{lemm15-20} in the second part of the proof and using \eqref{Ph10}, \eqref{Ph20} and \eqref{Phdef0}, $\hat{\mmat{P}}$ is obtained as 
\begin{align}
\hat{\mmat{P}}&\!=\!\tfrac{1}{1-|z_{1}|^{2}}\big{[}
\mvec{x}_{1}\mvec{x}_{1}^{\hr}\!+\!\mvec{x}_{2}\mvec{x}_{2}^{\hr}\!-\!z_{1}\mvec{x}_{1}\mvec{x}_{2}^{\hr}\!-\!z_{1}^{\hr}\mvec{x}_{2}\mvec{x}_{1}^{\hr} \big{]}\nonumber\\
& -\tfrac{1}{1-|z_{2}|^{2}}\big{[} \mvec{x}_{0}\mvec{x}_{0}^{\hr}+\mvec{x}_{1}\mvec{x}_{1}^{\hr}-z_{2}\mvec{x}_{0}\mvec{x}_{1}^{\hr}-z_{2}^{\hr}\mvec{x}_{1}\mvec{x}_{0}^{\hr}\big{]} \nonumber \\ 
& = \tfrac{1}{(1-|z_{1}|^{2})(1-|z_{2}|^{2})}\Big{[} \mvec{x}_{2}\mvec{x}_{2}^{\hr}(1\!-\!|z_{2}|^{2})+\mvec{x}_{1}\mvec{x}_{1}^{\hr}(|z_{1}|^{2}\!-\!|z_{2}|^{2})\nonumber \\
& -\mvec{x}_{0}\mvec{x}_{0}^{\hr}(1\!-\!|z_{1}|^{2}) - 
(1-|z_{2}|^{2})[z_{1}\mvec{x}_{1}\mvec{x}_{2}^{\hr}+z_{1}^{\hr}\mvec{x}_{2}\mvec{x}_{1}^{\hr}]\nonumber\\
&+ (1-|z_{1}|^{2})[z_{2}\mvec{x}_{0}\mvec{x}_{1}^{\hr}+z_{2}^{\hr}\mvec{x}_{1}\mvec{x}_{0}^{\hr}] \Big{]}.  
\end{align}
Thus, $\hat{\mmat{P}}^{3}$ can be obtained as 
\begin{equation}\label{P30}
\hat{\mmat{P}}^{3}=\hat{\mmat{P}}+\hat{\mmat{P}}_{1} \hat{\mmat{P}}_{2}\hat{\mmat{P}}_{1} - \hat{\mmat{P}}_{2} \hat{\mmat{P}}_{1}\hat{\mmat{P}}_{2}.  
\end{equation}
The second term in the right-hand side of the above equation is given by 
\begin{align} 
& \hat{\mmat{P}}_{1} \hat{\mmat{P}}_{2}\hat{\mmat{P}}_{1} = \tfrac{1}{(1-|z_{1}|^{2})(1-|z_{2}|^{2})^{2}} \Big{[} \mvec{x}_{1}\mvec{x}_{1}^{\hr}\big{[} (z_{1}z_{2}-|z_{2}|^{2}z_{3})(z_{1}^{\hr}z_{2}^{\hr}-z_{3}^{\hr})\nonumber\\
& + (1-|z_{2}|^{2})(1-|z_{1}|^{2}-|z_{2}|^{2}+z_{1}z_{2}z_{3}^{\hr})\big{]}\label{1210} \\ 
& +|z_{1}z_{2}-z_{3}|^{2}(\mvec{x}_{0}\mvec{x}_{0}^{\hr}-\mvec{x}_{0}\mvec{x}_{1}^{\hr}z_{2}-\mvec{x}_{1}\mvec{x}_{0}^{\hr}z_{2}^{\hr}) \Big{]}. \nonumber
\end{align}
Similarly, 
\begin{align}
& \hat{\mmat{P}}_{2} \hat{\mmat{P}}_{1}\hat{\mmat{P}}_{2} = \tfrac{1}{(1-|z_{1}|^{2})^{2}(1-|z_{2}|^{2})}\Big{[} \mvec{x}_{1}\mvec{x}_{1}^{\hr}\big{[} (z_{1}z_{2}-|z_{1}|^{2}z_{3})(z_{1}^{\hr}z_{2}^{\hr}-z_{3}^{\hr}) \nonumber\\
& +(1-|z_{1}|^{2})(1-|z_{1}|^{2}-|z_{2}|^{2}+z_{1}z_{2}z_{3}^{\hr})\big{]} \label{2120} \\ 
& +|z_{1}z_{2}-z_{3}|^{2}(\mvec{x}_{2}\mvec{x}_{2}^{\hr}-\mvec{x}_{1}\mvec{x}_{2}^{\hr}z_{1}-\mvec{x}_{2}\mvec{x}_{1}^{\hr}z_{1}^{\hr}) \Big{]}. \nonumber
\end{align}
By substituting the values of $\hat{\mmat{P}}_{1} \hat{\mmat{P}}_{2}\hat{\mmat{P}}_{1}$ and $\hat{\mmat{P}}_{2} \hat{\mmat{P}}_{1}\hat{\mmat{P}}_{2}$ from \eqref{1210} and \eqref{2120} in \eqref{P30} and simplifying the resultant equation, it follows that
\begin{equation}\label{Ph30}
\hat{\mmat{P}}^{3}=\Big{[} 1- \tfrac{|z_{1}z_{2}-z_{3}|^{2}}{(1-|z_{1}|^{2})(1-|z_{2}|^{2})}\Big{]}\hat{\mmat{P}}.    
\end{equation}
A reasoning similar to that used in equations~\eqref{RES1} and \eqref{RES2} results in
\begin{equation}
\|\hat{\mmat{P}}\|=\sqrt{1- \tfrac{|z_{1}z_{2}-z_{3}|^{2}}{(1-|z_{1}|^{2})(1-|z_{2}|^{2})}},    
\end{equation}
which completes the proof.
\end{proof}


The main result of this section is elaborated in the next theorem. 
\begin{theorem}\label{Th1}
Consider an asymmetric network with $n$ nodes, represented by a weighted digraph $\mathcal{G}$ with Laplacian matrix $\mmat{L}$ for which Assumptions~\ref{Assump1}-\ref{Assump3} hold. Let the centralized GPI algorithm, described in Algorithm~\ref{AlgGPI1}, be applied to this network. It then follows that Algorithm~\ref{AlgGPI1} is convergent such that    
\begin{equation}\label{CenFin}
\lim\limits_{k\rightarrow \infty}\tilde{\lambda}_{k} = \tilde{\lambda}(\mmat{L}).   
\end{equation}
\end{theorem}


\begin{proof}
Consider three consecutive state vectors $\mvec{x}_{k}$, $\mvec{x}_{k-1}$ and $\mvec{x}_{k-2}$ with unit Euclidean norms, generated by the centralized GPI algorithm at the $k\nth$ iteration of the algorithm. More specifically, these three vectors belong to $\mathbb{C}^{n}$ such that $\|\mvec{x}_{k}\|\!=\!\|\mvec{x}_{k-1}\|\!=\!\|\mvec{x}_{k-2}\|\!=\!1$ for any $k\in \mathbb{N}\!\setminus\!\{1\}$. Let also three scalar values $z_{k}^{1}$, $z_{k}^{2}$ and $z_{k}^{3}$ be defined as
\begin{subequations}\label{zk123}
\begin{align}
z_{k}^{1}& =\langle \mvec{x}_{k},\mvec{x}_{k-1} \rangle,\\ 
z_{k}^{2}& = \begin{cases}
\langle \mvec{x}_{k-1},\mvec{x}_{k-2} \rangle, & \mathrm{if}\;\; k\neq 1, \\
0,  & \mathrm{if}\;\; k=1,
\end{cases} \\
z_{k}^{3}& =
\begin{cases}
\langle \mvec{x}_{k},\mvec{x}_{k-2} \rangle, & \mathrm{if}\;\; k\neq 1, \\
0,  & \mathrm{if}\;\; k=1,
\end{cases} 
\end{align}    
\end{subequations}
for any $k\!\in\!\mathbb{N}$. Consider  $\check{\mmat{Q}}_{j}$, $\hat{\mmat{Q}}_{j}$, $\check{\mmat{P}}_{j}$, $\hat{\mmat{P}}_{j}$, $\check{\mmat{R}}_{j}$, $\hat{\mmat{R}}_{j}$, $\check{d}_{j}$, $\hat{d}_{j}$, $\check{\lambda}_{j}$ and $\hat{\lambda}_{j}$, $j\!\in\!\{k\!-\!1,k\}$, defined in \eqref{miandef1}-\eqref{lamb1}. The distance $d_{k+1}$ between successive subspaces and the estimated network's GAC, $\tilde{\lambda}_{k+1}$, are computed using \eqref{dk+11} in the $k\nth$ iteration of the algorithm. Define scenarios $\mathcal{R}$ and $\mathcal{I}$ for the two different possibilities that could occur in the process of implementing the GPI algorithm, corresponding to the cases where the GAC is either associated with a real dominant eigenvalue of $\tilde{\mmat{L}}$ or a pair of complex conjugate dominant eigenvalues of $\tilde{\mmat{L}}$, respectively. 
Let the initial state vector $\mvec{x}_{0}$ be decomposed as
\begin{equation}\label{x0centproof}
\mvec{x}_{0} = \sum_{j=1}^{n}c_{j,0}\,\mvec{v}_{j}(\tilde{\mmat{L}}),  
\end{equation}
where $c_{j,0}\!\in\!\mathbb{C}$ for any $j\!\in\!\mathcal{V}$. 
%
Consider scenario $\mathcal{I}$ first, and define real scalars $r$, $\theta$, $\rho$, $\phi$, $c_{0}$ and $\eta$ as follows 
\begin{subequations}\label{noldef}    
\begin{align}
& r e^{\mathrm{j}\theta}\!=\! \lambda_{n}(\tilde{\mmat{L}}) =\lambda_{n-1}^{\hr}(\tilde{\mmat{L}}), \\
& \rho e^{\mathrm{j} \phi} \!=\!\langle \mvec{v}_{n}(\tilde{\mmat{L}}),\mvec{v}_{n-1}(\tilde{\mmat{L}}) \rangle, \\
& c_{0} e^{\mathrm{j} \eta}\!=\! c_{n,0}=c_{n-1,0}^{\hr}.  
\end{align}
\end{subequations}
As a result, $\mvec{x}_{0}$ in \eqref{x0centproof} can be rewritten as
\begin{equation}
\mvec{x}_{0} = c_{0}e^{\mathrm{j}\eta}\,\mvec{v}_{n}(\tilde{\mmat{L}}) \!+\! c_{0}e^{-\mathrm{j}\eta}\, \mvec{v}_{n-1}(\tilde{\mmat{L}})\!+\! \sum_{j=1}^{n-2}c_{j,0}\,\mvec{v}_{j}(\tilde{\mmat{L}}),   
\end{equation}
where Assumption~\ref{Assump3} implies that $c_{0}\neq 0$. Using the GPI update procedure in \eqref{xk+11} to generate the intermediate state vector $\bar{\mvec{x}}_{k}\!\in\!\mathbb{C}^{n}$ and the state vector $\mvec{x}_{k}\!\in\!\mathbb{C}^{n}$ recursively at the $k\nth$ iteration, one arrives at
\begin{subequations}
\begin{align}
\bar{\mvec{x}}_{k} =&\, c_{k-1}\,r\,e^{\mathrm{j} (\eta+k\theta)}\,\mvec{v}_{n}(\tilde{\mmat{L}}) \!+\! c_{k-1}\,r\,e^{-\mathrm{j}(\eta+k\theta)}\,\mvec{v}_{n-1}(\tilde{\mmat{L}}) \nonumber \\
& + \tfrac{1}{r^{k-1}}\sum_{j=1}^{n-2} c_{j,k-1}\,\lambda_{j}^{k}(\tilde{\mmat{L}})\,\mvec{v}_{j}(\tilde{\mmat{L}}), \\
\mvec{x}_{k} =&\, c_{k}\,e^{\mathrm{j}(\eta+k\theta)}\, \mvec{v}_{n}(\tilde{\mmat{L}}) + c_{k}\,e^{-\mathrm{j} (\eta+k\theta)}\,\mvec{v}_{n-1}(\tilde{\mmat{L}}) \nonumber \\
& + \tfrac{1}{r^{k}}\sum_{j=1}^{n-2} c_{j,k} \,\lambda_{j}^{k}(\tilde{\mmat{L}})\,\mvec{v}_{j}(\tilde{\mmat{L}}),\label{xkmaindef}
\end{align}    
\end{subequations}
where
\begin{subequations}
\begin{align}
c_{k}=& \tfrac{1}{\sqrt{2+2\rho \cos(2\eta +2k\theta +\phi)+h_{\mathcal{I},k}}},\\ 
c_{j,k}=& \tfrac{c_{j,0}}{c_{0}\sqrt{2+2\rho \cos(2\eta +2k\theta +\phi)+ h_{\mathcal{I},k}}},\\
h_{\mathcal{I},k}=& 2\sum_{j=1}^{n-2}\Re\!\!\Big{[} \tfrac{c_{j,k-1}}{c_{k-1}}\tfrac{\lambda_{j}^{k}(\tilde{\mmat{L}})}{r^{k}} e^{-\mathrm{j} (\eta+k\theta)}\mvec{v}_{n}^{\hr}(\tilde{\mmat{L}}) \mvec{v}_{j}(\tilde{\mmat{L}})\nonumber\\
& + \tfrac{c_{j,k-1}}{c_{k-1}}\tfrac{\lambda_{j}^{k}(\tilde{\mmat{L}})}{r^{k}} e^{\mathrm{j} (\eta+k\theta)}\mvec{v}_{n-1}^{\hr}(\tilde{\mmat{L}}) \mvec{v}_{j}(\tilde{\mmat{L}})\Big{]} \label{hikdef} \\
& +\sum_{j=1}^{n-2}\sum_{p=1}^{n-2}\tfrac{c_{j,k-1}^{\hr}\,c_{p,k-1}}{(c_{k-1})^{2}} \tfrac{\lambda_{j}^{k,\hr}(\tilde{\mmat{L}}) \lambda_{p}^{k}(\tilde{\mmat{L}})}{r^{2k}}\mvec{v}_{j}^{\hr}(\tilde{\mmat{L}}) \mvec{v}_{p}(\tilde{\mmat{L}}), \nonumber
\end{align}    
\end{subequations}
for any $j\!\in\!\mathbb{N}_{n-2}$ and $k\!\in\!\mathbb{N}$. As a result, $\mvec{x}_{k}$ in \eqref{xkmaindef} can be rewritten as
\begin{equation}\label{skthisone0}
\mvec{x}_{k} = \tfrac{e^{\mathrm{j} (\eta+k\theta)}\mvec{v}_{n}(\tilde{\mmat{L}}) + e^{-\mathrm{j} (\eta+k\theta)}\mvec{v}_{n-1}(\tilde{\mmat{L}})+\sum_{j=1}^{n-2} 
\tfrac{c_{j,0}}{c_{0}} \tfrac{\lambda_{j}^{k}(\tilde{\mmat{L}})}{r^k}\,\mvec{v}_{j}(\tilde{\mmat{L}})}{\sqrt{2+2\rho \cos(2\eta+2k\theta +\phi)+h_{\mathcal{I},k}}},   
\end{equation}
for any $k\in \mathbb{N}$. 
By defining $\zeta_{\mathcal{I}}$ as
\begin{equation}\label{zetaIdef}
\zeta_{\mathcal{I}}=\Big{|}\tfrac{\lambda_{n-2}(\tilde{\mmat{L}})}{\lambda_{n}(\tilde{\mmat{L}})}\Big{|}=\tfrac{|\lambda_{n-2}(\tilde{\mmat{L}})|}{r},    
\end{equation}
it follows from Assumptions~\ref{Assump1} and \ref{Assump2} that $0\!<\!\zeta_{\mathcal{I}}\!<\!1$. It then follows from \eqref{hikdef} and \eqref{zetaIdef} that $h_{\mathcal{I},k}$ is a real value satisfying $h_{\mathcal{I},k}=\Theta(\zeta_{\mathcal{I}}^{k})$ for any $k\!\in\!\mathbb{N}$, given that $\tfrac{|\lambda_{j}(\tilde{\mmat{L}})|}{r}\!\leq\!\zeta_{\mathcal{I}}\!<\!1$ 
for all $j\!\in\! \mathbb{N}_{n-2}$. As a result, $\mvec{x}_{k}$ in \eqref{skthisone0} can be simplified as 
\begin{equation}\label{skthisone}
\mvec{x}_{k} = \tfrac{e^{\mathrm{j} (\eta+k\theta)}\mvec{v}_{n}(\tilde{\mmat{L}}) + e^{-\mathrm{j} (\eta+k\theta)}\mvec{v}_{n-1}(\tilde{\mmat{L}})+a_{k}\mvec{s}_{k}}{\sqrt{2+2\rho \cos(2\eta+2k\theta +\phi)+h_{\mathcal{I},k}}},   
\end{equation}
where 
$a_{k}=\Theta(\zeta_{\mathcal{I}}^{k})$
and $\mvec{s}_{k}\!\in\!\mathbb{C}^{n}$ is a unit vector whose direction is the same as that of
$\sum_{j=1}^{n-2} 
c_{j,0}\lambda_{j}^{k}(\tilde{\mmat{L}})\,\mvec{v}_{j}(\tilde{\mmat{L}})$
for any $k\!\in\!\mathbb{N}$. Since $\zeta^{k}_{\mathcal{I}}\rightarrow 0$ as $k\rightarrow \infty$, after linear approximation of the function in the right-hand side of \eqref{skthisone} around $a_{k}=0$ and $h_{\mathcal{I},k}=0$, one obtains
\begin{equation}\label{x1234a}
\mvec{x}_{k}= \tfrac{e^{\mathrm{j}(\eta + k \theta)}\mvec{v}_{n}(\tilde{\mmat{L}}) + e^{-\mathrm{j}(\eta + k \theta)}\mvec{v}_{n-1}(\tilde{\mmat{L}})}{\sqrt{2+2\rho \cos(2\eta +2k \theta + \phi)}} + b_{k}\,\hat{\mvec{s}}_{k},
\end{equation}
where the unit vector $\hat{\mvec{s}}_{k}\!\in\! \mathbb{C}^{n}$ has the same direction as $2a_{k}\mvec{s}_{k}-\tfrac{h_{\mathcal{I},k}(e^{\mathrm{j}(\eta+k \theta)}\mvec{v}_{n}(\tilde{\mmat{L}})+e^{-\mathrm{j}(\eta + k \theta)}\mvec{v}_{n-1}(\tilde{\mmat{L}}))}{2+2\rho \cos(2\eta +2k \theta + \phi)}$ and $b_{k}\!=\!\Theta(\zeta_{\mathcal{I}}^{k})$. Following a similar procedure, the state vectors $\mvec{x}_{k-1}$ and $\mvec{x}_{k-2}$ are also derived as 
\begin{subequations}\label{x1234}
\begin{align}
\mvec{x}_{k-1} \!=\!& \, \tfrac{e^{\mathrm{j}(\eta + (k-1) \theta)}\mvec{v}_{n}(\tilde{\mmat{L}}) + e^{-\mathrm{j}(\eta + (k-1) \theta)}\mvec{v}_{n-1}(\tilde{\mmat{L}})}{\sqrt{2+2\rho \cos(2\eta + 2(k-1)\theta +\phi)}} \nonumber \\
& +b_{k-1}\, \hat{\mvec{s}}_{k-1}, \label{x1234b}\\
\mvec{x}_{k-2} \!=\!& \, \tfrac{e^{\mathrm{j}(\eta + (k-2) \theta)}\mvec{v}_{n}(\tilde{\mmat{L}}) + e^{-\mathrm{j}(\eta + (k-2) \theta)}\mvec{v}_{n-1}(\tilde{\mmat{L}})}{\sqrt{2+2\rho \cos(2\eta +2(k-2)\theta +\phi)}} \nonumber\\ 
&+ b_{k-2}\, \hat{\mvec{s}}_{k-2}, \label{x1234c}
\end{align}   
\end{subequations}
for two unit vectors $\hat{\mvec{s}}_{k-1}, \hat{\mvec{s}}_{k-2}\!\in\!\mathbb{C}^{n}$ and two scalars
$b_{k-1}=\Theta(\zeta^{k-1}_{\mathcal{I}})$,
$b_{k-2}=\Theta(\zeta^{k-2}_{\mathcal{I}})$. 
By substituting from \eqref{x1234a} and \eqref{x1234} in \eqref{zk123} while keeping the asymptotically dominant terms only, $z_{k}^{1}$, $z_{k}^{2}$ and $z_{k}^{3}$ are obtained as
\begin{subequations}\label{z123}
\begin{align}
z_{k}^{1}\!&=\!\!\tfrac{\cos \theta + \rho \cos(\gamma +(2k-1)\theta)}{\sqrt{(1+\rho\cos(\gamma+2k\theta))(1+\rho\cos(\gamma+2(k-1)\theta ))}}\!+\!\mathfrak{z}_k^1, \\
z_{k}^{2}\!&=\!\! \tfrac{\cos \theta + \rho \cos(\gamma +(2k-3)\theta )}{\sqrt{(1+\rho\cos(\gamma+2(k-1)\theta ))(1+\rho\cos(\gamma+2(k-2)\theta))}}\!+\!\mathfrak{z}_k^2, \\
z_{k}^{3}\!&=\!\!\tfrac{\cos \theta + \rho \cos(\gamma +(2k-2)\theta )}{\sqrt{(1\!+\!\rho\cos(\gamma+2k\theta ))(1+\rho\cos(\gamma+2(k-2)\theta ))}}\!+\!\mathfrak{z}_k^3,
\end{align}
\end{subequations}
where $\gamma:=2\eta+\phi$,
$|\mathfrak{z}_k^1|=\Theta(\zeta^{k-1}_{\mathcal{I}})$,
$|\mathfrak{z}_k^2|=\Theta(\zeta^{k-2}_{\mathcal{I}})$ and
$|\mathfrak{z}_k^3|=\Theta(\zeta^{k-2}_{\mathcal{I}})$ for any $k\in\mathbb{N}$.
By considering the results of Lemma~\ref{lemm:dcdh1} and since $\check{d}_{k}=\|\check{\mmat{P}}_{k}-\check{\mmat{P}}_{k-1}\|$ and $\hat{d}_{k}=\|\hat{\mmat{P}}_{k}-\hat{\mmat{P}}_{k-1}\|$ according to \eqref{dcdh1}, the following values for $\check{d}_{k}$ and $\hat{d}_{k}$ are found
\begin{subequations}\label{dckdhk}
\begin{align}
\check{d}_{k} & \!=\! \sqrt{1-|\langle \mvec{x}_{k},\mvec{x}_{k-1} \rangle|^{2} },\label{dhk219}\\
\hat{d}_{k} & \!=\! \sqrt{1-\tfrac{\big{|}\langle \mvec{x}_{k},\mvec{x}_{k-1} \rangle \, \langle \mvec{x}_{k-1},\mvec{x}_{k-2} \rangle - \langle \mvec{x}_{k},\mvec{x}_{k-2} \rangle \big{|}^{2}}{\big{(}1-|\langle \mvec{x}_{k},\mvec{x}_{k-1} \rangle|^{2}\big{)}\big{(}1-|\langle \mvec{x}_{k-1},\mvec{x}_{k-2}\rangle|^{2}\big{)}}}.\label{dhk219b}
\end{align}   
\end{subequations}
Using the definitions of $z_{k}^{1}$, $z_{k}^{2}$ and $z_{k}^{3}$ in \eqref{zk123} results in
\begin{subequations}\label{eqdesired}
\begin{align}
\check{d}_{k} & \!=\! \sqrt{1-|z_{k}^{1}|^{2} },\label{dhk220}\\
\hat{d}_{k} & \!=\! \sqrt{1-\tfrac{\big{|}z_{k}^{1}z_{k}^{2}-z_{k}^{3}\big{|}^{2}}{\big{(}1-|z_{k}^{1}|^{2}\big{)}\big{(}1-|z_{k}^{2}|^{2}\big{)}}}.\label{dhk220b}
\end{align}   
\end{subequations}
Substituting $z_{k}^{1}$, $z_{k}^{2}$ and $z_{k}^{3}$ from \eqref{z123} into \eqref{eqdesired} and only keeping the asymptotically dominant terms yields
\begin{subequations}\label{dkdkI}    
\begin{align}
(\check{d}_{k})^{2} & \!=\! \tfrac{\sin^{2}\theta +\rho^{2}[\cos(\gamma+2k\theta)\cos(\gamma+(2k-2)\theta)-\cos^{2}(\gamma+(2k-1)\theta)]}{(1+\rho \cos(\gamma+2k\theta))(1+\rho \cos(\gamma+2(k-1)\theta))} \nonumber \\
&\;\; + \Theta(\zeta_{\mathcal{I}}^{k-1}), \\
(\hat{d}_{k})^{2} & \!=\! \Theta(\zeta_{\mathcal{I}}^{k-2}).
\end{align}
\end{subequations}
Since $\zeta_{\mathcal{I}}^{k-1}\!\rightarrow\!0$ and $\zeta_{\mathcal{I}}^{k-2}\!\rightarrow\!0$ as $k\!\rightarrow\!\infty$, it follows from \eqref{dkdkI} that
\begin{align}\label{dkdkI2}
\lim\limits_{k\rightarrow \infty}\!\!\check{d}_{k}\!\! =& \sqrt{\tfrac{\sin^{2}\theta +\rho^{2}[\cos(\gamma+2k\theta)\cos(\gamma+(2k-2)\theta)-\cos^{2}(\gamma+(2k-1)\theta)]}{(1+\rho \cos(\gamma+2k\theta))(1+\rho \cos(\gamma+2(k-1)\theta))}}, \nonumber \\
\lim\limits_{k\rightarrow \infty}\!\!\hat{d}_{k}\!\! =& 0,
\end{align}
in an asymptotic manner. Given that $\check{d}_{k}$ converges to a nonzero positive value while $\hat{d}_{k}$ converges to zero in scenario $\mathcal{I}$ according to \eqref{dkdkI2}, it follows from the definition of $d_{k+1}$ in \eqref{dcdh2} that 
\begin{equation}\label{dkdkI3}
\lim\limits_{k\rightarrow \infty}d_{k+1} \!=\! 0,   
\end{equation}
guaranteeing the termination of Algorithm~\ref{AlgGPI1} in finite number of iterations in scenario $\mathcal{I}$, noting that the positive threshold $\epsilon$ is used in the termination condition of the algorithm. It is then implied from \eqref{dkdkI2} and \eqref{dkdkI3} that 
%
\begin{equation}
\lim\limits_{k\rightarrow \infty}d_{k+1}\!=\! \lim\limits_{k\rightarrow \infty}\hat{d}_{k},   
\end{equation}
which means that the GPI algorithm successfully estimates the magnitude of the complex conjugate eigenvalues $\lambda_{n}(\tilde{\mmat{L}})$ and $\lambda_{n-1}(\tilde{\mmat{L}})$ as the pair of dominant eigenvalues of $\tilde{\mmat{L}}$, corresponding to the two-dimensional subspace $\mathcal{W}^{\ast}:=\mathrm{span}\{\mvec{v}_{n-1}(\tilde{\mmat{L}}),\mvec{v}_{n}(\tilde{\mmat{L}})\}$, 
as $k\rightarrow \infty$ in scenario $\mathcal{I}$. It then follows from the definition of $\hat{\lambda}_{k}$ in \eqref{lamb1} that 
\begin{equation}
\lim\limits_{k\rightarrow \infty}\hat{\lambda}_{k}\!=\!|\lambda_{n}(\tilde{\mmat{L}})|.
\end{equation}
In other words, the two-dimensional subspace sequence 
$\{\mathcal{W}_{k}\}_{k\in \mathbb{N}}$, $\mathcal{W}_{k}\!=\!\mathrm{span}\{\mvec{x}_{k-1},\mvec{x}_{k}\}$, converges to $\mathcal{W}^{\ast}$ while the one-dimensional subspace sequence $\{\mathcal{V}_{k}\}_{k\in \mathbb{N}}$, $\mathcal{V}_{k}=\mathrm{span}\{\mvec{x}_{k}\}$, is not convergent.
Using the definition of $\tilde{\lambda}_{k+1}$ in \eqref{lambdak1} results in a successful estimation of the GAC of the network which proves the validity of \eqref{CenFin} in scenario $\mathcal{I}$.

Now consider scenario $\mathcal{R}$ for which the dominant eigenvalue of $\tilde{\mmat{L}}$ is a real scalar. To this end, let the real scalars $r$ and $c_{0}$ be defined as 
\begin{subequations}
\begin{align}
r=&\, \lambda_{n}(\tilde{\mmat{L}}), \\
c_{0}=&\,|c_{n,0}|.
\end{align}    
\end{subequations}
As a result, the initial state vector $\mvec{x}_{0}$ in \eqref{x0centproof} is rewritten as 
\begin{equation}
\mvec{x}_{0} = c_{0}\,\mvec{v}_{n}(\tilde{\mmat{L}}) +\sum_{j=1}^{n-1}c_{j,0}\,\mvec{v}_{j}(\tilde{\mmat{L}}),  
\end{equation}
where $c_{0}\neq 0$ holds due to Assumption~\ref{Assump3}. 
After successive implementation of the GPI update procedure \eqref{xk+11} in scenario $\mathcal{R}$, the intermediate state vector $\bar{\mvec{x}}_{k}\!\in\!\mathbb{C}^{n}$ and the state vector $\mvec{x}_{k}\!\in\!\mathbb{C}^{n}$ at the $k\nth$ iteration are generated in a recursive manner as follows
\begin{subequations}
\begin{align}
\bar{\mvec{x}}_{k} =&\, c_{k-1}\,r\,\mvec{v}_{n}(\tilde{\mmat{L}}) + \tfrac{1}{r^{k-1}}\sum_{j=1}^{n-1} c_{j,k-1}\,\lambda_{j}^{k}(\tilde{\mmat{L}})\,\mvec{v}_{j}(\tilde{\mmat{L}}), \\
\mvec{x}_{k} =&\, c_{k}\,\mvec{v}_{n}(\tilde{\mmat{L}}) + \tfrac{1}{r^{k}}\sum_{j=1}^{n-1} c_{j,k} \,\lambda_{j}^{k}(\tilde{\mmat{L}})\,\mvec{v}_{j}(\tilde{\mmat{L}}),\label{xkmaindefrR}
\end{align}    
\end{subequations}
where
\begin{subequations}\label{TotalR}
\begin{align}
c_{k}=& \tfrac{1}{\sqrt{1+h_{\mathcal{R},k}}},\\ 
c_{j,k}=& \tfrac{c_{j,0}}{c_{0}\sqrt{1+ h_{\mathcal{R},k}}},\\
h_{\mathcal{R},k}=& 2\sum_{j=1}^{n-1}\Re\!\!\Big{[} \tfrac{c_{j,k-1}}{c_{k-1}}\tfrac{\lambda_{j}^{k}(\tilde{\mmat{L}})}{r^{k}} \mvec{v}_{n}^{\hr}(\tilde{\mmat{L}}) \mvec{v}_{j}(\tilde{\mmat{L}})\Big{]}  \label{hikdefR}\\
& +\sum_{j=1}^{n-1}\sum_{p=1}^{n-1}\tfrac{c_{j,k-1}^{\hr}\,c_{p,k-1}}{(c_{k-1})^{2}} \tfrac{\lambda_{j}^{k,\hr}(\tilde{\mmat{L}}) \lambda_{p}^{k}(\tilde{\mmat{L}})}{r^{2k}}\mvec{v}_{j}^{\hr}(\tilde{\mmat{L}}) \mvec{v}_{p}(\tilde{\mmat{L}}), \nonumber
\end{align}    
\end{subequations}
for any $j\!\in\!\mathbb{N}_{n-1}$ and $k\!\in\!\mathbb{N}$. Using \eqref{xkmaindefrR} and \eqref{TotalR}, the vector $\mvec{x}_{k}$ is then obtained as 
\begin{equation}\label{skthisoneR0}
\mvec{x}_{k} = \tfrac{1}{\sqrt{1+h_{\mathcal{R},k}}}
\Big{[}\mvec{v}_{n}(\tilde{\mmat{L}})\!+\!\sum_{j=1}^{n-1}\tfrac{c_{j,0}}{c_{0}}\tfrac{\lambda_{j}^{k}(\tilde{\mmat{L}})}{r^{k}}\mvec{v}_{j}(\tilde{\mmat{L}})\Big{]},    
\end{equation}
for any $k\!\in\!\mathbb{N}$. 
By defining $\zeta_{\mathcal{R}}$ as 
\begin{equation}\label{zetaIdefR}
\zeta_{\mathcal{R}}=\Big{|}\tfrac{\lambda_{n-1}(\tilde{\mmat{L}})}{\lambda_{n}(\tilde{\mmat{L}})}\Big{|}=\tfrac{|\lambda_{n-1}(\tilde{\mmat{L}})|}{r},    
\end{equation}
it follows from Assumptions~\ref{Assump1} and \ref{Assump2} that $0\!<\! \zeta_{\mathcal{R}}\!<\!1$. 
It then follows from \eqref{hikdefR} and \eqref{zetaIdefR} that $h_{\mathcal{R},k}$ is a real value satisfying
$h_{\mathcal{R},k}=\Theta(\zeta_{\mathcal{R}}^{k})$ 
for any $k\!\in\!\mathbb{N}$, given that $\tfrac{|\lambda_{j}(\tilde{\mmat{L}})|}{r}\!\leq\!\zeta_{\mathcal{R}}\!<\!1$ for all $j\!\in\! \mathbb{N}_{n-1}$.  
As a result, $\mvec{x}_{k}$ in \eqref{skthisoneR0} can be simplified as
\begin{equation}\label{skthisoneR}
\mvec{x}_{k} = \tfrac{\mvec{v}_{n}(\tilde{\mmat{L}})+ a_{k}\mvec{t}_{k}}{\sqrt{1+h_{\mathcal{R},k}}},   
\end{equation}
where $a_{k}=\Theta(\zeta^{k}_{\mathcal{R}})$ and $\mvec{t}_{k}\!\in\! \mathbb{C}^{n}$ is defined as a unit vector whose direction is the same as that of $\sum_{j=1}^{n-1} c_{j,0}\,\lambda_{j}^{k}(\tilde{\mmat{L}})\,\mvec{v}_{j}(\tilde{\mmat{L}})$ for any $k\!\in\!\mathbb{N}$. 
Since $\zeta^{k}_{\mathcal{R}}\rightarrow 0$ 
as $k\rightarrow \infty$, after linear approximation of the function in the right-hand side of \eqref{skthisoneR} around $a_{k}=0$ and $h_{\mathcal{R},k}=0$, one arrives at
\begin{equation}\label{x1234aR}
\mvec{x}_{k}=  \mvec{v}_{n}(\tilde{\mmat{L}}) + b_{k}\,\hat{\mvec{t}}_{k},
\end{equation}
where
$b_{k}=\Theta(\zeta^{k}_{\mathcal{R}})$
and the unit vector $\hat{\mvec{t}}_{k}\!\in\! \mathbb{C}^{n}$ has the same direction as $2a_{k}\mvec{t}_{k}-h_{\mathcal{R},k}\mvec{v}_{n}(\tilde{\mmat{L}})$ for any $k\!\in\!\mathbb{N}$. Following a similar procedure, the state vectors $\mvec{x}_{k-1}$ and $\mvec{x}_{k-2}$ are also obtained as 
\begin{subequations}\label{x1234R}
\begin{align}
\mvec{x}_{k-1} =& \, \mvec{v}_{n}(\tilde{\mmat{L}}) + b_{k-1}\, \hat{\mvec{t}}_{k-1}, \label{x1234bR} \\
\mvec{x}_{k-2} =& \, \mvec{v}_{n}(\tilde{\mmat{L}}) + b_{k-2}\, \hat{\mvec{t}}_{k-2}, \label{x1234cR}
\end{align}   
\end{subequations}
for the two unit vectors $\hat{\mvec{t}}_{k-1}, \hat{\mvec{t}}_{k-2}\!\in\!\mathbb{C}^{n}$ and two scalars
$b_{k-1}=\Theta(\zeta^{k-1}_{\mathcal{R}})$, $b_{k-2}=\Theta(\zeta^{k-2}_{\mathcal{R}})$. 
Substituting the values of state vectors from \eqref{x1234aR} and \eqref{x1234R} in \eqref{zk123} while keeping the asymptotically dominant terms only yields
\begin{equation}\label{z123rn}
z_{k}^{j} = 1- \mathfrak{z}_{k}^{j}, 
\end{equation}
where
$|\mathfrak{z}_k^1|=\Theta(\zeta^{k-1}_{\mathcal{R}})$,
$|\mathfrak{z}_k^2|=\Theta(\zeta^{k-2}_{\mathcal{R}})$ and
$|\mathfrak{z}_k^3|=\Theta(\zeta^{k-2}_{\mathcal{R}})$ for any $j\!\in\!\mathbb{N}_{3}$ and $k\in\mathbb{N}$, and the values of the above-mentioned inner products belong to the interval $[-1,1]$. After substituting the above equalities into \eqref{eqdesired}, $\check{d}_{k}$ and $\hat{d}_{k}$ are derived as
\begin{subequations}\label{dckn}
\begin{align}
\check{d}_{k} =&\,
\sqrt{1-\big{|}1- \mathfrak{z}_{k}^{1}\big{|}^2}, \\
\hat{d}_{k} =&\, \sqrt{1-\tfrac{\big{|} (1-\mathfrak{z}_{k}^{1})(1-\mathfrak{z}_{k}^{2})-(1-\mathfrak{z}_{k}^{3})\big{|}^2}{\big{(} 1-|1-\mathfrak{z}_{k}^{1}|^2\big{)} \big{(} 1-|1-\mathfrak{z}_{k}^{2}|^2\big{)}}}.  \label{dcknb} 
\end{align}    
\end{subequations}
After ignoring all but the asymptotically dominant terms and 
using the definition of function $\Theta(\cdot)$, it can be concluded from \eqref{dckn} that
\begin{subequations}\label{dckn2}
\begin{align}
(\check{d}_{k})^{2} =&\,\Theta(\zeta_{\mathcal{R}}^{k-1}),\\
(\hat{d}_{k})^{2} =&\, 1-\tfrac{\Theta(\zeta_{\mathcal{R}}^{2k-4})}{\Theta(\zeta_{\mathcal{R}}^{2k-3})}=1-\Theta(1).    
\end{align}    
\end{subequations}
Since $0\!<\!\zeta_{\mathcal{R}}\!<\!1$ and $\zeta_{\mathcal{R}}^{k-1}\!\rightarrow\!0$ as $k\!\rightarrow\!\infty$, it can be concluded from \eqref{dckn2} and the definition of function $\Theta(\cdot)$ that there exists a positive constant $\kappa$, such that  
\begin{subequations}\label{Thisone5}
\begin{align}
&\lim_{k\rightarrow\infty}\check{d}_{k} = 0, \\
&\lim_{k\rightarrow\infty}\hat{d}_{k}>\kappa.
\end{align}   
\end{subequations}
This implies that $\check{d}_{k}$ converges to zero while $\hat{d}_{k}$ is lower-bounded by a nonzero positive scalar $\kappa$ as $k\!\rightarrow\!\infty$.
From the definition of $d_{k+1}$ in \eqref{dcdh2}, it follows that $\lim_{k\rightarrow \infty}d_{k+1}\!=\!0$, which means that the GPI algorithm terminates after a finite number of iterations in scenario $\mathcal{R}$. Since 
\begin{equation}\label{thisonecen}
\lim\limits_{k\rightarrow \infty}d_{k+1}\!=\! \lim\limits_{k\rightarrow \infty}\check{d}_{k},
\end{equation}
the GPI algorithm successfully estimates the magnitude of $\lambda_{n}(\tilde{\mmat{L}})$ as the single dominant eigenvalue of $\tilde{\mmat{L}}$, corresponding to the one-dimensional subspace $\mathcal{V}^{\ast}\!:=\!\mathrm{span}\{\mvec{v}_{n}(\tilde{\mmat{L}})\}$ as $k\!\rightarrow\!\infty$ in scenario $\mathcal{R}$. It then follows from the definition of $\check{\lambda}_{k}$ in \eqref{lamb1} that 
\begin{equation}\label{thisonecenF}
\lim\limits_{k\rightarrow \infty}\check{\lambda}_{k}\!=\!|\lambda_{n}(\tilde{\mmat{L}})|.
\end{equation}
In other words, the one-dimensional subspace sequence $\{\mathcal{V}_{k}\}_{k\in \mathbb{N}}$, $\mathcal{V}_{k}\!=\!\mathrm{span}\{\mvec{x}_{k}\}$, converges to $\mathcal{V}^{\ast}$ whereas the two-dimensional subspace sequence $\{\mathcal{W}_{k}\}_{k\in \mathbb{N}}$, $\mathcal{W}_{k}\!=\!\mathrm{span}\{\mvec{x}_{k-1} ,\mvec{x}_{k}\}$, is not convergent. 
From \eqref{thisonecenF} and the definition of $\tilde{\lambda}_{k+1}$ in \eqref{lambdak1}, a proper estimation of the GAC as well as the validity of \eqref{CenFin} in scenario $\mathcal{R}$ follows, which completes the convergence proof of the centralized GPI algorithm in scenario $\mathcal{R}$.
\end{proof}
\section{Distributed Implementation of GPI Algorithm}\label{Sec:VI}

The main challenges concerning the distributed implementation of the proposed GPI algorithm stem from the lack of knowledge about the global parameters, including
\begin{enumerate}[1)]
\item node $i$ only has direct access to the $i\nth$ element of the state vector $\mvec{x}_{k}$, denoted by $x_{k}^{i}$, for any $i\!\in\!\mathcal{V}$ and $k\!\in\! \mathbb{N}$, 
\item the exponential term of the modified Laplacian matrix $\tilde{\mmat{L}}$ has to be approximated in a distributed manner, and
\item only the $i\nth$ row of the weight matrix $\mmat{W}$, representing the weights associated with the in-neighbors of the $i\nth$ node, is known to node $i$.
\end{enumerate}

\begin{remark}
To implement the GPI algorithm in a distributed manner, every node $i$ of the network requires the following information about the unit left eigenvector $\mvec{w}_{1}(\mathbf{L})=[w_{1}^{i}(\mathbf{L})]$ associated with the zero eigenvalue of $\mmat{L}$ 
\begin{enumerate}[i)]
\item $\langle \mvec{w}_{1}(\mathbf{L}),\mvec{1}_{n} \rangle$, which is the summation of all elements in $\mvec{w}_{1}(\mathbf{L})$, and 
\item $w_{1}^{i}(\mathbf{L})$, which is the $i\nth$ element of $\mvec{w}_{1}(\mathbf{L})$. 
\end{enumerate}
These two scalar values can be acquired by each node using the existing methods in the literature, prior to running the GPI algorithm \cite{Qu_14, Rabbat_15}.
\end{remark}


In the distributed implementation of the GPI algorithm, the iteration index $k\!\in\!\mathbb{N}$ is associated with the main sequence of the GPI algorithm, inspired by the power iteration procedure. Also, $l$ and $m$ are defined as two new iteration indices, running up to the prespecified integers $l^{\ast}$ and $m^{\ast}$, respectively, in each iteration of the main sequence. 
%
%
In the distributed version of Algorithm~\ref{AlgGPI1}, a new set of local variables
\begin{equation}\label{local}
\bar{x}^{i}_{k},\, x^{i}_{k},\, \check{\mmat{R}}^i_{k-1},\, \hat{\mmat{R}}^i_{k-1},\, \check{d}^i_{k},\, \hat{d}^i_{k},\, d^i_{k+1},\, \check{\lambda}^i_{k},\, \hat{\lambda}^i_{k},\, \tilde{\lambda}^i_{k+1}    
\end{equation}
are introduced, which provide a local estimation of the following set of variables (as their global counterparts), previously used in the centralized GPI algorithm 
\begin{equation}\label{global}
\bar{\mvec{x}}_{k},\, \mvec{x}_{k},\, \check{\mmat{R}}_{k-1},\, \hat{\mmat{R}}_{k-1},\, \check{d}_{k},\, \hat{d}_{k},\, d_{k+1},\, \check{\lambda}_{k},\, \hat{\lambda}_{k},\, \tilde{\lambda}_{k+1}    
\end{equation}
from the perspective of each node $i\!\in\!\mathcal{V}$ at each iteration $k\!\in\!\mathbb{N}$. In the sequel, the values of the variables in \eqref{local} are locally computed based only upon information locally-available to any node $i\!\in\!\mathcal{V}$ at each iteration $k\!\in\!\mathbb{N}$. 
The goal of generating the local values in equation~\eqref{local} (as estimates of their global counterparts in \eqref{global}) in a distributed setting is achieved upon discovering that all the needed global information in \eqref{global} can be evaluated as functions of the following set of inner product scalars
\begin{equation}\label{GVdistgpi}
\begin{split}
& \langle\bar{\mvec{x}}_{k-2},\bar{\mvec{x}}_{k-2}\rangle,\;\; \langle\bar{\mvec{x}}_{k-1},\bar{\mvec{x}}_{k-1}\rangle,\;\; \langle\bar{\mvec{x}}_{k-1},\bar{\mvec{x}}_{k-2}\rangle,\\
& \langle\bar{\mvec{x}}_{k-1},\mvec{w}_{1}(\mmat{L})\rangle,\;\; \langle\bar{\mvec{x}}_{k},\bar{\mvec{x}}_{k}\rangle,\;\; \langle\bar{\mvec{x}}_{k},\bar{\mvec{x}}_{k-1}\rangle, \;\; \langle\bar{\mvec{x}}_{k},\bar{\mvec{x}}_{k-2}\rangle,
\end{split}
\end{equation}
for any $k\!\in\!\mathbb{N}$. In other words, all global quantities in \eqref{global} are substituted by functions of proper scalars in \eqref{GVdistgpi} as their inputs at the $k\nth$ iteration, for any $k\!\in\! \mathbb{N}$. As a result, the local estimates in \eqref{local} can be evaluated in a distributed manner by locally estimating the scalar quantities in \eqref{GVdistgpi}, and then using the perceived relationships between \eqref{global} and \eqref{GVdistgpi}.

Let $\tilde{\mmat{L}}_{l^{\ast}}$ be defined as the \textit{$l^{\ast}$-approximate modified Laplacian} matrix which provides an approximation of the matrix $\tilde{\mmat{L}}$ in \eqref{eqn:MLM} for a given positive integer $l^{\ast}$. More specifically, the first term in the right-hand side of equation~\eqref{eqn:MLM} is replaced by the sum of the first $l^{\ast}$ terms in the Taylor series expansion of the matrix exponential $e^{\mmat{I}_{n}-\delta \mmat{L}}\!-\!\mmat{I}_{n}$. In other words, 
\begin{equation}\label{eqn:Ll11}
\tilde{\mmat{L}}_{l^{\ast}} = \sum_{j=0}^{l^{\ast}}\tfrac{1}{j!}(\mmat{I}_{n}- \delta \mmat{L})^{j} - e\,\mvec{w}_{1}\!(\mmat{L})\, \mvec{w}_{1}^{\tr}\!(\mmat{L}).
\end{equation}
The iteration index $l\!\in\!\mathbb{N}_{l^{\ast}}$ is then used to count the number of terms included in the Taylor series expansion of $e^{\mmat{I}_{n}-\delta \mmat{L}}\!-\!\mmat{I}_{n}$ in order to approximate $\tilde{\mmat{L}}$ with $\tilde{\mmat{L}}_{l^{\ast}}$. The iteration index $m\!\in\!\mathbb{N}_{m^{\ast}}$, on the other hand, is used to indicate the number of steps that the \textit{consensus observer} procedure \cite{AsadiTSMC_20} is run to disseminate the aforementioned inner product scalars in \eqref{GVdistgpi} throughout the network in order to evaluate the local estimates of \eqref{local}, which are required to realize a distributed implementation of the GPI algorithm.

Prior to the distributed implementation of the GPI algorithm, a number of scalar parameters are required to be known \textit{a priori} to every node. This includes $w_{1}^{i}(\mmat{L})$, $\langle \mvec{w}_{1}(\mmat{L}),\mvec{1}_{n} \rangle$, positive scalar $\delta$ defined in \eqref{eq:delta} (which is used in the update procedures of the nested loops with iteration indices $l$ and $m$ as well as the definitions of $\tilde{\mmat{L}}$ and $\tilde{\mmat{L}}_{l^{\ast}}$), a sufficiently small positive scalar $\epsilon$ (which is used in the termination condition of the GPI algorithm), and two positive integers $l^{\ast}$ and $m^{\ast}$ (which determine the acceptable accuracies for the estimation objectives in the two nested loops, and are set as design parameters). 
The distributed implementation of the GPI algorithm from the viewpoint of the $i\nth$ node is subsequently initialized as follows
\begin{equation}\label{kd0P1dist}
k\!\gets\!1,\, d_{1}^{i}\!\gets\!\epsilon,\, \bar{z}_{-1}^{1,i}\!\gets\!1,\, \bar{z}_{0}^{1,i}\!\gets\!1,\, \bar{z}_{0}^{2,i}\!\gets\!0,\, \bar{z}_{0}^{4,i}\!\gets\!0,\, \forall\,i\!\in\!\mathcal{V}. 
\end{equation}

Let $\bar{\mvec{x}}_{k}\!=\![\bar{x}_{k}^{i}]\!\in\!\mathbb{C}^{n}$ and $\mvec{x}_{k}\!=\![x_{k}^{i}]\!\in\!\mathbb{C}^{n}$ denote the intermediate state vector and the state vector of the network at the $k\nth$ iteration of the distributed GPI algorithm, respectively, where $\tilde{\mmat{L}}$ is approximated by $\tilde{\mmat{L}}_{l^{\ast}}$ and the update procedure of the consensus observer is repeated $m^{\ast}$ times to disseminate the global information in \eqref{GVdistgpi} throughout the network. Let $\bar{x}_{k}^{i}$ and $x_{k}^{i}$ represent the $i\nth$ element of the vectors $\bar{\mvec{x}}_{k}$ and $\mvec{x}_{k}$, respectively, which are computed by the $i\nth$ node of the network for any $i\!\in\! \mathcal{V}$ and $k\!\in\!\mathbb{N}$. The steps required to recursively update the values of $\bar{x}_{k}^{i}$, $x_{k}^{i}$, $\check{d}_{k}^{i}$, $\hat{d}_{k}^{i}$, $\check{\mmat{R}}_{k-1}^{i}$, $\hat{\mmat{R}}_{k-1}^{i}$, $\check{\lambda}_{k}^{i}$, $\hat{\lambda}_{k}^{i}$, $d_{k+1}^{i}$ and $\tilde{\lambda}_{k+1}^{i}$ by the $i\nth$ node in the $k\nth$ iteration of the GPI algorithm via local information exchange with its neighbors in a distributed setting is elaborated in the sequel, for any $i\!\in\!\mathcal{V}$ and $k\!\in\!\mathbb{N}$. 
%
%
%
%
Using the definition of $\tilde{\mmat{L}}_{l^{\ast}}$ in \eqref{eqn:Ll11}, an approximate of the intermediate state vector $\bar{\mvec{x}}_{k}$ is computed as
\begin{align}\label{Imedsv}
\bar{\mvec{x}}_{k} & = \tilde{\mmat{L}}_{l^{\ast}}\,\mvec{x}_{k-1}\\ 
&= \sum_{j=0}^{l^{\ast}}\tfrac{1}{j!}(\mmat{I}_{n}- \delta \mmat{L})^{j} \,\mvec{x}_{k-1} - e\,\mvec{w}_{1}\!(\mmat{L})\tfrac{\langle \bar{\mvec{x}}_{k-1},\mvec{w}_{1}(\mmat{L}) \rangle}{\sqrt{\langle \bar{\mvec{x}}_{k-1},\bar{\mvec{x}}_{k-1}\rangle}}. \nonumber
\end{align}
Normalizing the vector $\bar{\mvec{x}}_{k}$ in \eqref{Imedsv} results in the updated state vector $\mvec{x}_{k}$ as follows
\begin{equation}\label{eqn:x_k1}
\mvec{x}_{k} = \tfrac{\bar{\mvec{x}}_{k}}{\|\bar{\mvec{x}}_{k}\|}=\tfrac{\bar{\mvec{x}}_{k}}{\sqrt{\langle\bar{\mvec{x}}_{k},\bar{\mvec{x}}_{k}\rangle}}, 
\end{equation}
for any $k\!\in\!\mathbb{N}$. 
Define $\bar{\mvec{y}}_{k}\!=\![\bar{y}_{k}^{i}]\!\in\!\mathbb{C}^{n}$ as 
\begin{equation}\label{ybar}
\bar{\mvec{y}}_{k} =\sum_{j=0}^{l^{\ast}}\tfrac{1}{j!}(\mmat{I}_{n}-\delta \mmat{L})^{j}\mvec{x}_{k-1}.
\end{equation}
It follows from \eqref{Imedsv} that
\begin{equation}\label{xbari2}
\bar{x}_{k}^{i}= \bar{y}_{k}^{i} - e\,w_{1}^{i}(\mmat{L})\,\tfrac{\langle \bar{\mvec{x}}_{k-1},\mvec{w}_{1}(\mmat{L}) \rangle}{\sqrt{\langle \bar{\mvec{x}}_{k-1},\bar{\mvec{x}}_{k-1}\rangle}},
\end{equation}
for all $i\!\in \!\mathcal{V}$ and $k\!\in\!\mathbb{N}$. 
Note that $\bar{y}_{k}^{i}$ as the $i\nth$ element of the vector $\bar{\mvec{y}}_{k}$ and the first term in the right-hand side of \eqref{xbari2} can be generated in a distributed manner as 
\begin{equation}\label{ybarDIS}
\bar{y}_{k}^{i} = \sum_{l=0}^{l^{\ast}}\tfrac{1}{l!}\,y^{i}(l),    
\end{equation}
for any $i\!\in\! \mathcal{V}$ and $k\!\in\! \mathbb{N}$. In addition, the $i\nth$ element of the vector $\mvec{y}(l)\!=\![y^{i}(l)]$ in the right-hand side of \eqref{ybarDIS} is initialized and recursively updated by employing
\begin{subequations}\label{Eq212}
\begin{align}
& y^{i}(l=0)=x_{k-1}^{i},\label{Eq212a} \\
& y^{i}(l) = y^{i}(l\!-\!1) + \delta\!\sum_{j\in \mathcal{N}_{i}}w_{ij}\big{(}y^{j}(l\!-\!1)-y^{i}(l\!-\!1)\big{)},\label{Eq212b}
\end{align}
\end{subequations}
for any $i\!\in\!\mathcal{V}$, $l\!\in\!\mathbb{N}_{l^{\ast}}$ and $k\!\in\! \mathbb{N}$. The approximation of $\tilde{\mmat{L}}$ can be made as accurate as necessary by choosing a sufficiently large $l^{\ast}$.

Subsequently, the consensus observer procedure is applied in the $k\nth$ iteration of the distributed GPI algorithm to estimate the following terms 
\begin{equation}\label{InnerP}
\begin{array}{llll}
& \hat{z}_{k}^{1}\!=\!\langle \bar{\mvec{x}}_{k},\bar{\mvec{x}}_{k} \rangle, 
&& \;\;\hat{z}_{k}^{2}\!=\!\langle \bar{\mvec{x}}_{k},\bar{\mvec{x}}_{k-1} \rangle, \\
& \hat{z}_{k}^{3}\!=\!\langle \bar{\mvec{x}}_{k},\bar{\mvec{x}}_{k-2} \rangle,
&& \;\;\hat{z}_{k}^{4}\!=\!\langle \bar{\mvec{x}}_{k},\mvec{w}_{1}(\mmat{L}) \rangle,
\end{array}
\end{equation}
in order to update some of the inner products in \eqref{GVdistgpi}, impacted by the new updates in the elements of $\bar{\mvec{x}}_{k}$ in the $k\nth$ iteration. In other words, the estimates of the global values in \eqref{InnerP} are further used at the $k\nth$ iteration of the distributed GPI algorithm in the absence of global information on their exact values. 
To this end, $\mvec{z}_{k}^{s}(m)\!=\![z_{k}^{s,i}(m)]\!\in\!\mathbb{C}^{n}$ is defined for any $s\!\in\!\mathbb{N}_{4}$ and $m\!\in\! \mathbb{N}_{m^{\ast}}$, as the state vector of the consensus observer, initialized as 
\begin{subequations}\label{z1234INI}
\begin{align}
z_{k}^{1,i}(m\!=\!0) &= \tfrac{\langle \mvec{w}_{1}(\mmat{L}),\mvec{1}_{n} \rangle}{w_{1}^{i}(\mmat{L})}\bar{x}_{k}^{i,\hr}\bar{x}_{k}^{i}, \\
z_{k}^{2,i}(m\!=\!0) &= \tfrac{\langle \mvec{w}_{1}(\mmat{L}),\mvec{1}_{n} \rangle}{w_{1}^{i}(\mmat{L})}\bar{x}_{k-1}^{i,\hr}\bar{x}_{k}^{i}, \\
z_{k}^{3,i}(m\!=\!0) & = 
\begin{cases}
\tfrac{\langle \mvec{w}_{1}(\mmat{L}),\mvec{1}_{n} \rangle}{w_{1}^{i}(\mmat{L})}\bar{x}_{k-2}^{i,\hr}\bar{x}_{k}^{i}, & \mathrm{if}\;\; k\neq 1, \\
0,  & \mathrm{if}\;\; k=1.
\end{cases} \\
z_{k}^{4,i}(m\!=\!0) & = \langle \mvec{w}_{1}(\mmat{L}),\mvec{1}_{n} \rangle
\bar{x}_{k}^{i},
\end{align}
\end{subequations}
for any $i\!\in\!\mathcal{V}$ and $k\!\in\!\mathbb{N}$. 
Now, to implement the consensus observer procedure in a distributed fashion from the viewpoint of the $i\nth$ node at the $k\nth$ iteration of the distributed GPI algorithm, the following update rule is used
\begin{align}\label{z1234UPDP}
z^{s,i}_{k}(m)= &\, z^{s,i}_{k}(m-1)\, + \\
& \delta\! \sum_{j\in \mathcal{N}_{i}}w_{ij}(z^{s,j}_{k}(m-1)-z^{s,i}_{k}(m-1)),\nonumber
\end{align}
with the iteration index $m$ for any $i\!\in\! \mathbb{N}_{n}$, $s\!\in\! \mathbb{N}_{4}$, $m\!\in\! \mathbb{N}_{m^{\ast}}$ and $k\!\in\!\mathbb{N}$. 
As $m\!\rightarrow\!\infty$ in the update procedure of the consensus observer procedure in \eqref{z1234UPDP}, the vector $\mvec{z}_{k}^{s}(m)$ converges to $\hat{z}_{k}^{s}\times\mvec{1}_{n}$, where $\hat{z}_{k}^{s}$ is defined in \eqref{InnerP} for any $s\!\in\!\mathbb{N}_{4}$ and $k\!\in\!\mathbb{N}$. Define $\bar{\mvec{z}}_{k}^{s}\!=\![\bar{z}_{k}^{s,i}]\!\in\!\mathbb{C}^{n}$ as the output of the consensus observer at the end of the $m^{\ast}{}\nth$ iteration, i.e., 
\begin{equation}\label{zbarmst}
\bar{\mvec{z}}_{k}^{s}=\mvec{z}_{k}^{s}(m^{\ast}), \end{equation}
for any $s\!\in\!\mathbb{N}_{4}$ and $k\!\in\!\mathbb{N}$. 
Using \eqref{zbarmst}, estimates of the inner product scalars in \eqref{GVdistgpi} can be obtained using the following outputs of the consensus observer procedure from the viewpoint of the $i\nth$ node
\begin{equation}\label{zestimates}
\bar{z}_{k-2}^{1,i},\,\bar{z}_{k-1}^{1,i},\,\bar{z}_{k-1}^{2,i},\,\bar{z}_{k-1}^{4,i},\,\bar{z}_{k}^{1,i},\,\bar{z}_{k}^{2,i},\,\bar{z}_{k}^{3,i}, 
\end{equation}
for any $i\!\in\!\mathcal{V}$ and $k\!\in\!\mathbb{N}$. Note that the above terms are the outputs of the consensus observer at the $(k-2)\nth$, $(k-1)\nth$ and $k\nth$ iterations of the GPI algorithm by the $i\nth$ node, and are used by the $i\nth$ node to compute the values of the variables in \eqref{local} at the $k\nth$ iteration of the distributed GPI algorithm, for any $i\!\in\!\mathcal{V}$ and $k\!\in\!\mathbb{N}$. 
In essence, the distributed GPI algorithm contains a triplet of iteration indices, i.e., $k\!\in\!\mathbb{N}$ for the main sequence of the GPI algorithm, $l\!\in\!\mathbb{N}_{l^{\ast}}$ for obtaining the elements of the vector $\mvec{y}(l)$ and subsequently $\bar{\mvec{y}}_{k}$, 
and lastly, $m\!\in\!\mathbb{N}_{m^{\ast}}$ for finding the elements of the vector $\mvec{z}^{s}_{k}(m)$ as an estimate of the vector $\hat{z}_{k}^{s}\times\mvec{1}_{n}$, for all $s\!\in\!\mathbb{N}_{4}$ and $k\!\in\!\mathbb{N}$, used to disseminate the key global information required for the distributed implementation of the GPI algorithm throughout the network. 

Considering \eqref{xbari2}, the updated scalar $\bar{x}_{k}^{i}$ at the $k\nth$ iteration of the algorithm is computed as
\begin{equation}\label{eqn:x_k_i}
\bar{x}_{k}^{i} = \bar{y}_{k}^{i} - e\;w_{1}^{i}(\mmat{L})\;\tfrac{\bar{z}_{k-1}^{4,i}}{\sqrt{\bar{z}_{k-1}^{1,i}}}, 
\end{equation}
using \eqref{ybarDIS} and the outputs of the consensus observer in \eqref{zestimates}. It then results from \eqref{eqn:x_k1} and \eqref{zbarmst} that 
\begin{equation}\label{UPDsvdist}
x_{k}^{i}= \tfrac{\bar{x}_{k}^{i}}{\sqrt{\bar{z}_{k}^{1,i}}}=\tfrac{1}{\sqrt{\bar{z}_{k}^{1,i}}}\Big{[}\bar{y}_{k}^{i}-e\,w_{1}^{i}(\mmat{L})\,\tfrac{\bar{z}_{k-1}^{4,i}}{\sqrt{\bar{z}_{k-1}^{1,i}}}\Big{]},
\end{equation}
for any $i\!\in\!\mathcal{V}$ and $k\!\in\!\mathbb{N}$. 
It follows from the definitions of $\check{d}_{k}$ and $\hat{d}_{k}$ in \eqref{dckdhk}, and the relation between vectors $\mvec{x}_{k}$ and $\bar{\mvec{x}}_{k}$ in \eqref{eqn:x_k1} that 
\begin{align}
&\check{d}_{k}\!=\!\sqrt{1\!-\!\tfrac{|\langle \bar{\mvec{x}}_{k},\bar{\mvec{x}}_{k-1}\rangle|^{2}}{\| \bar{\mvec{x}}_{k}\|^{2}\,\|\bar{\mvec{x}}_{k-1}\|^{2}}}, \\
&\scalemath{0.90}{\hat{d}_{k}\!=\! \left[{1\!-\! \tfrac{{|} \langle \bar{\mvec{x}}_{k},\bar{\mvec{x}}_{k-1}\rangle \langle \bar{\mvec{x}}_{k-1},\bar{\mvec{x}}_{k-2}\rangle - \langle \bar{\mvec{x}}_{k},\bar{\mvec{x}}_{k-2}\rangle \| \bar{\mvec{x}}_{k-1}\|^{2} {|}^{2}}{{(}\| \bar{\mvec{x}}_{k}\|^2 \| \bar{\mvec{x}}_{k-1}\|^2 - |\langle \bar{\mvec{x}}_{k},\bar{\mvec{x}}_{k-1}\rangle|^{2}{)}{(} \| \bar{\mvec{x}}_{k-1}\|^2 \| \bar{\mvec{x}}_{k-2}\|^2 - |\langle \bar{\mvec{x}}_{k-1},\bar{\mvec{x}}_{k-2}\rangle|^{2} {)}}}\right]^{\frac{1}{2}}}.\nonumber
\end{align}
Using the outputs of the consensus observer in \eqref{zestimates}, the values of $\check{d}_{k}^{i}$ and $\hat{d}_{k}^{i}$ from the viewpoint of node $i$ in a distributed implementation are then computed as 
\begin{subequations}\label{dhdcDISTdist}
\begin{align}
\check{d}_{k}^{i} & = \sqrt{1-\tfrac{|\bar{z}_{k}^{2,i}|^{2}}{\bar{z}_{k}^{1,i} \bar{z}_{k-1}^{1,i}}}, \\
\hat{d}_{k}^{i} & = \sqrt{1-\tfrac{\big{|}\bar{z}_{k}^{2,i}\bar{z}_{k-1}^{2,i}-\bar{z}_{k}^{3,i}\bar{z}_{k-1}^{1,i}\big{|}^{2}}{\big{(}\bar{z}_{k}^{1,i}\bar{z}_{k-1}^{1,i}-|\bar{z}_{k}^{2,i}|^{2}\big{)}\big{(}\bar{z}_{k-1}^{1,i}\bar{z}_{k-2}^{1,i}-|\bar{z}_{k-1}^{2,i}|^{2}\big{)}}},
\end{align}
\end{subequations}
for any $i\!\in\! \mathcal{V}$ and $k\!\in\!\mathbb{N}$. In order to obtain an estimate of $\check{\mmat{R}}$ at the $k\nth$ iteration of the algorithm in a distributed fashion from the viewpoint of the $i\nth$ node, the values of $\check{\mmat{R}}_{k-1}$ and $\hat{\mmat{R}}_{k-1}$ are computed (instead of $\check{\mmat{R}}_{k}$ and $\hat{\mmat{R}}_{k}$, respectively) to ensure that all the required global information, necessary for the distributed implementation of the GPI algorithm, are available to each node. To this end, $\check{\mmat{R}}_{k-1}$ and $\hat{\mmat{R}}_{k-1}$ are described as functions of the state vectors and intermediate state vectors of the GPI algorithm as follows
\begin{subequations}\label{RhRck-1}
\begin{align}
&\check{\mmat{R}}_{k-1} = \mathfrak{g}(\tilde{\mmat{L}},\check{\mmat{Q}}_{k-1}) = \mvec{x}_{k-1}^{\hr} \tilde{\mmat{L}} \mvec{x}_{k-1} = \mvec{x}_{k-1}^{\hr} \bar{\mvec{x}}_{k}, \\
&\hat{\mmat{R}}_{k-1} = \mathfrak{g}(\tilde{\mmat{L}},\hat{\mmat{Q}}_{k-1}) \\
& =\begin{bmatrix}
1 & \mvec{x}_{k-2}^{\hr}\mvec{x}_{k-1} \\
\mvec{x}_{k-1}^{\hr}\mvec{x}_{k-2} & 1
\end{bmatrix}^{-1} \begin{bmatrix}
\mvec{x}_{k-2}^{\hr}\tilde{\mmat{L}}\mvec{x}_{k-2} & \mvec{x}_{k-2}^{\hr}\tilde{\mmat{L}}\mvec{x}_{k-1} \\
\mvec{x}_{k-1}^{\hr}\tilde{\mmat{L}}\mvec{x}_{k-2} & \mvec{x}_{k-1}^{\hr}\tilde{\mmat{L}}\mvec{x}_{k-1}
\end{bmatrix} \nonumber\\
& =\begin{bmatrix}
1 & \mvec{x}_{k-2}^{\hr}\mvec{x}_{k-1} \\
\mvec{x}_{k-1}^{\hr}\mvec{x}_{k-2} & 1
\end{bmatrix}^{-1} \begin{bmatrix}
\mvec{x}_{k-2}^{\hr}\bar{\mvec{x}}_{k-1} & \mvec{x}_{k-2}^{\hr}\bar{\mvec{x}}_{k} \\
\mvec{x}_{k-1}^{\hr}\bar{\mvec{x}}_{k-1} & \mvec{x}_{k-1}^{\hr}\bar{\mvec{x}}_{k}
\end{bmatrix}. \nonumber 
\end{align}
\end{subequations}
Using the definitions of the inner product and the intermediate state vectors, it follows from \eqref{RhRck-1} that
\begin{subequations}\label{RhRck-12}
\begin{align}
&\check{\mmat{R}}_{k-1} \!=\! \tfrac{\langle \bar{\mvec{x}}_{k},\bar{\mvec{x}}_{k-1}\rangle}{\sqrt{\langle\bar{\mvec{x}}_{k-1},\bar{\mvec{x}}_{k-1}\rangle}}= \tfrac{\langle \bar{\mvec{x}}_{k},\bar{\mvec{x}}_{k-1}\rangle}{\|\bar{\mvec{x}}_{k-1}\|}, \\
&\hat{\mmat{R}}_{k-1} \!=\!
\begin{bmatrix}
1 & \tfrac{\langle \bar{\mvec{x}}_{k-1},\bar{\mvec{x}}_{k-2} \rangle}{\|\bar{\mvec{x}}_{k-1}\|\,\|\bar{\mvec{x}}_{k-2}\|} \\
\tfrac{\langle \bar{\mvec{x}}_{k-2},\bar{\mvec{x}}_{k-1} \rangle}{\|\bar{\mvec{x}}_{k-1}\|\,\|\bar{\mvec{x}}_{k-2}\|} & 1 \end{bmatrix}^{\!-1} \!\!\times \\ 
&\;\;\;\;\;\;\;\;\;\;\;\, \begin{bmatrix}
\tfrac{\langle \bar{\mvec{x}}_{k-1},\bar{\mvec{x}}_{k-2} \rangle}{\|\bar{\mvec{x}}_{k-2}\|} &  \tfrac{\langle \bar{\mvec{x}}_{k},\bar{\mvec{x}}_{k-2} \rangle}{\|\bar{\mvec{x}}_{k-2}\|} \\
\tfrac{\langle \bar{\mvec{x}}_{k-1},\bar{\mvec{x}}_{k-1} \rangle}{\|\bar{\mvec{x}}_{k-1}\|} &  \tfrac{\langle \bar{\mvec{x}}_{k},\bar{\mvec{x}}_{k-1} \rangle}{\|\bar{\mvec{x}}_{k-1}\|}
\end{bmatrix}. \nonumber 
\end{align}
\end{subequations}
After substituting the inner product values of equation~\eqref{RhRck-12} with the outputs of the consensus observer procedure 
from the viewpoint of the $i\nth$ node according to \eqref{zbarmst} and \eqref{zestimates}, $\check{\mmat{R}}_{k-1}^{i}$ and $\hat{\mmat{R}}_{k-1}^{i}$ are obtained as the distributed estimations of $\check{\mmat{R}}_{k-1}$ and $\hat{\mmat{R}}_{k-1}$ from the perspective of node $i$, respectively, as 
\begin{align}\label{RhRc3}
&\check{\mmat{R}}_{k-1}^{i} \!=\!\tfrac{\bar{z}_{k}^{2,i}}{\sqrt{\bar{z}_{k-1}^{1,i}}},\\
&\hat{\mmat{R}}_{k-1}^{i} \!=\!\begin{bmatrix}
1 & \tfrac{\bar{z}_{k-1}^{2,i}}{\sqrt{\bar{z}_{k-1}^{1,i} \bar{z}_{k-2}^{1,i}}} \\
\tfrac{\bar{z}_{k-1}^{2,i}}{\sqrt{\bar{z}_{k-1}^{1,i} \bar{z}_{k-2}^{1,i}}} & 1
\end{bmatrix}^{\!\!-1}\!\!\! \begin{bmatrix}
\tfrac{\bar{z}_{k-1}^{2,i}}{\sqrt{\bar{z}_{k-2}^{1,i}}}& \tfrac{\bar{z}_{k}^{3,i}}{\sqrt{\bar{z}_{k-2}^{1,i}}} \\
\tfrac{\bar{z}_{k-1}^{1,i}}{\sqrt{\bar{z}_{k-1}^{1,i}}} &  \tfrac{\bar{z}_{k}^{2,i}}{\sqrt{\bar{z}_{k-1}^{1,i}}}
\end{bmatrix}, \nonumber
\end{align}
for any $i\!\in\!\mathcal{V}$ and $k\!\in\!\mathbb{N}$. Using \eqref{RhRc3}, a simplified version of $\hat{\mmat{R}}_{k-1}^{i}$ at the $k\nth$ iteration of the distributed GPI algorithm from the viewpoint of the $i\nth$ node is obtained as 
\begin{equation*}
\scalemath{0.87}{\hat{\mmat{R}}_{k-1}^{i}\!=\!\tfrac{\bar{z}_{k-1}^{1,i}\bar{z}_{k-2}^{1,i}}{\bar{z}_{k-1}^{1,i}\bar{z}_{k-2}^{1,i}-|\bar{z}_{k-1}^{2,i}|^2}\!\! \begin{bmatrix}
0 & \tfrac{\bar{z}_{k}^{3,i}}{\sqrt{\bar{z}_{k-2}^{1,i}}}\!-\!\tfrac{\bar{z}_{k-1}^{2,i}\bar{z}_{k}^{2,i}}{\bar{z}_{k-1}^{1,i}\sqrt{\bar{z}_{k-2}^{1,i}}} \\
\tfrac{\bar{z}_{k-1}^{1,i}}{\sqrt{\bar{z}_{k-1}^{1,i}}}\!-\! \tfrac{|\bar{z}_{k-1}^{2,i}|^2}{\bar{z}_{k-2}^{1,i}\sqrt{\bar{z}_{k-1}^{1,i}}}
 &  \tfrac{\bar{z}_{k}^{2,i}}{\sqrt{\bar{z}_{k-1}^{1,i}}}\!-\!\tfrac{\bar{z}_{k}^{3,i}\bar{z}_{k-1}^{2,i}}{\bar{z}_{k-2}^{1,i}\sqrt{\bar{z}_{k-1}^{1,i}}}
\end{bmatrix}.}    
\end{equation*}
The values of $\check{\lambda}_{k}^{i}$, $\hat{\lambda}_{k}^{i}$, $d_{k+1}^{i}$ and $\tilde{\lambda}_{k+1}^{i}$ are finally computed based on \eqref{lamb1} and \eqref{dk+11}, but using the estimated values from the perspective of the $i\nth$ node, resulting in
\begin{subequations}\label{lambdaDIST}
\begin{align}
\check{\lambda}_{k}^{i} &= |\check{\mmat{R}}_{k-1}^{i}|, \label{Thislambda1}\\
\hat{\lambda}_{k}^{i} &= \Big{|} \tfrac{1}{2} \mathrm{tr}(\hat{\mmat{R}}_{k-1}^{i})\!+\!\sqrt{(\tfrac{1}{2} \mathrm{tr}(\hat{\mmat{R}}_{k-1}^{i}))^{2}\!-\!\mathrm{det}(\hat{\mmat{R}}_{k-1}^{i})}\Big{|}, \label{Thislambda2}\\
d_{k+1}^{i} &= \min\{\check{d}_{k}^{i},\hat{d}_{k}^{i}\}, \label{dk+1dist} \\
\tilde{\lambda}_{k+1}^{i} &=
\begin{cases}
\tfrac{1}{\delta}\big{[} 1-\log( \check{\lambda}_{k}^{i})\big{]},      & \mathrm{if}\; d_{k+1}^{i}=\check{d}_{k}^{i}, \\
\tfrac{1}{\delta}\big{[} 1-\log\small{(}\hat{\lambda}_{k}^{i})\big{]},         & \mathrm{if}\; d_{k+1}^{i}=\hat{d}_{k}^{i}.
\end{cases} \label{lambdak1dist}
\end{align}
\end{subequations}
This ends the $k\nth$ iteration of the algorithm from the perspective of node $i$, and the iteration index is updated by setting $k\!\gets\! k+1$. In practice, to limit the execution time of the distributed GPI algorithm, the main sequence of the procedure keeps running until the stopping condition $\max_{i\in \mathcal{V}}d_{k}^{i}\!<\!\epsilon$ holds at the beginning of the $k\nth$ iteration for a sufficiently small positive constant $\epsilon$. The stopping condition can be evaluated in a distributed manner by each node using the existing approaches \cite{PeiXu_17}. 
This terminates the algorithm and $\tilde{\lambda}_{k}^{i}$ provides an estimation of the GAC of the network from the viewpoint of node $i$ for any $i\!\in\! \mathcal{V}$. 

A pseudo-code of the distributed GPI algorithm from the viewpoint of the $i\nth$ node is described in Algorithm~\ref{AlgGPIdist}, where the integers $l^{\ast}:=k$ and $m^{\ast}:=k$ are used at the $k\nth$ iteration of the algorithm for any $i\!\in\!\mathcal{V}$ and $k\!\in\!\mathbb{N}$. 

\begin{remark}
The message length of the proposed distributed
GPI procedure in Algorithm~\ref{AlgGPIdist} is of order $O(1)$ bits per node due to the fixed size of the messages exchanged between the nodes in each iteration regardless of the network's size. This is in direct contrast with all similar ``distributed'' algorithms in \cite{Gusrialdi_21,Spong_15,AsadiTSMC_20,AsadiTechRep_17} whose message length is of order $O(n)$ bits per node in each iteration for a network of size $n$. This means that Algorithm~\ref{AlgGPIdist} is scalable, and can be even used for very large network applications. 
\end{remark}

\begin{algorithm}[]
\begin{algorithmic}[1]
\STATE {Inputs to node $i$: $w_{1}^{i}(\mmat{L})$, $\langle\mvec{w}_{1}(\mmat{L}),\mvec{1}_{n}\rangle$, $\delta$, $\epsilon$.}  

\STATE {Initialize $x_{0}^{i}$, $k$, $d_{1}^{i}$, $\bar{z}_{-1}^{1,i}$, $\bar{z}_{0}^{1,i}$, $\bar{z}_{0}^{2,i}$ and $\bar{z}_{0}^{4,i}$ based on \eqref{kd0P1dist}.}

\WHILE {$\max_{i\in \mathcal{V}}d_{k}^{i} \geq \epsilon$}
\vspace{2.5pt}

\STATE{$l^{\ast}\gets k$, $m^{\ast}\gets k$.} \label{Thisline}

\STATE{Initialize $y^{i}(l\!=\!0)$ using \eqref{Eq212a} and compute $y_{k}^{i}(l)$ using the update procedure \eqref{Eq212b} for all $l\!\in\!\mathbb{N}_{l^{\ast}}$.}

\STATE{Compute $\bar{y}_{k}^{i}$ using \eqref{ybarDIS}.}

\STATE{Initialize $z_{k}^{s,i}(m\!=\!0)$ according to \eqref{z1234INI} and compute $z_{k}^{s,i}(m)$ using the update procedure \eqref{z1234UPDP} for all $s\!\in\!\mathbb{N}_{4}$ and $m\!\in\!\mathbb{N}_{m^{\ast}}$.}


\STATE{Evaluate $\bar{z}_{k}^{s,i}$ for all $s\!\in\! \mathbb{N}_{4}$ using \eqref{zbarmst} and compute $x_{k}^{i}$ using \eqref{UPDsvdist}.}

\STATE {Update $\check{d}_{k}^{i}$ and $\hat{d}_{k}^{i}$ using \eqref{dhdcDISTdist}.}

\STATE {Compute $\check{\mmat{R}}_{k-1}^{i}$ and $\check{\mmat{R}}_{k-1}^{i}$ using \eqref{RhRc3}.}

\STATE {Compute $\check{\lambda}_{k}^{i}$, $\hat{\lambda}_{k}^{i}$, $d_{k+1}^{i}$ and $\tilde{\lambda}_{k+1}^{i}$ using \eqref{lambdaDIST}.}

\STATE $k\gets k+1$.
\ENDWHILE
\vspace{2.5pt}
\STATE {Output: Return $\tilde{\lambda}_{k}^{i}$.}
\end{algorithmic}\caption{Distributed implementation of GPI algorithm from the viewpoint of the node $i\!\in\!\mathcal{V}$.}
\label{AlgGPIdist}
\end{algorithm}

\section{Convergence Proof of Distributed GPI Algorithm}\label{Sec:VII}
The main result of this section is presented in the next theorem.
\begin{theorem}\label{Th2}
Consider an asymmetric network of $n$ nodes, represented by a weighted digraph $\mathcal{G}$ with Laplacian matrix $\mmat{L}$, and let Assumptions~\ref{Assump1}-\ref{Assump3} hold. Then, the distributed GPI procedure described in Algorithm~\ref{AlgGPIdist} is convergent when applied to this network such that    
\begin{equation}\label{Prdist}
\lim\limits_{k\rightarrow \infty}\tilde{\lambda}_{k}^{i} =\tilde{\lambda}(\mmat{L}),\;\;\; \forall\, i\!\in\! \mathcal{V}. 
\end{equation}
\end{theorem}

\begin{proof}
Similar to the convergence proof of the centralized GPI algorithm in Section~\ref{Sec:V}, two different scenarios $\mathcal{I}$ and $\mathcal{R}$ are investigated in the sequel. Compared to the convergence proof of the centralized GPI algorithm in Theorem~\ref{Th1}, it is necessary here to investigate the impact of two additional sources of error, due to 
\begin{enumerate}[1)]
\item approximation of $\tilde{\mmat{L}}$ with $\tilde{\mmat{L}}_{l^{\ast}}$ in the update procedure of the state vector, and 
\item dissemination of the global information in \eqref{GVdistgpi} throughout the network after a finite number of iterations ($m^{\ast}$) of the consensus observer procedure.
\end{enumerate}
To investigate the impact of these two sources of error for given values of $l^{\ast}$ and $m^{\ast}$, $\bar{\mvec{x}}_{k,l}\!=\![\bar{x}_{k,l}^{i}]$ and $\mvec{x}_{k,l}\!=\![x_{k,l}^{i}]$ are defined, respectively, as the intermediate state vector and the state vector in $\mathbb{C}^{n}$ generated by the distributed GPI algorithm at the $k\nth$ iteration, after the inclusion of the first source of error only. 
In other words, the subscript $l$ in vectors $\bar{\mvec{x}}_{k,l}$ and $\mvec{x}_{k,l}$ indicates that the matrix $\tilde{\mmat{L}}$, which is used to generate $\bar{\mvec{x}}_{k}$ and $\mvec{x}_{k}$ in \eqref{xk+11}, is approximated by $\tilde{\mmat{L}}_{l^{\ast}}$ in \eqref{Imedsv} and \eqref{eqn:x_k1}. Let also the unit initial state vector $\mvec{x}_{0}\!\in\! \mathbb{C}^{n}$ be described as 
\begin{equation}\label{x0dist}
\mvec{x}_{0} = \sum_{j=1}^{n}c_{j,0,l}\,\mvec{v}_{j}(\tilde{\mmat{L}}_{l^{\ast}}),  
\end{equation} 
where $c_{j,0,l}\!\in\! \mathbb{C}$ for any $j\!\in\!\mathcal{V}$.
Let scenario $\mathcal{I}$ be investigated first, where the set of real scalars $r_{l}$, $\theta_{l}$, $\rho_{l}$, $\phi_{l}$, $c_{0,l}$ and $\eta_{l}$ are defined as 
\begin{subequations}\label{withldefDIS}
\begin{align}
& r_{l} e^{\mathrm{j}\theta_{l}}\!=\! \lambda_{n}(\tilde{\mmat{L}}_{l^{\ast}}) =\lambda_{n-1}^{\hr}(\tilde{\mmat{L}}_{l^{\ast}}), \\ 
& \rho_{l} e^{\mathrm{j} \phi_{l}} \!=\!\langle \mvec{v}_{n}(\tilde{\mmat{L}}_{l^{\ast}}),\mvec{v}_{n-1}(\tilde{\mmat{L}}_{l^{\ast}}) \rangle, \\
& c_{0,l} e^{\mathrm{j} \eta_{l}}\!=\! c_{n,0,l}\!=\! c_{n-1,0,l}^{\hr}. 
\end{align}
\end{subequations} 
As a result, $\mvec{x}_{0}$ in \eqref{x0dist} can be rewritten as
\begin{align}
\mvec{x}_{0} = &\,  c_{0,l}\,e^{\mathrm{j}\eta_{l}}\,\mvec{v}_{n}(\tilde{\mmat{L}}_{l^{\ast}}) + c_{0,l}\,e^{-\mathrm{j}\eta_{l}}\, \mvec{v}_{n-1}(\tilde{\mmat{L}}_{l^{\ast}}) \nonumber \\
& + \sum_{j=1}^{n-2}c_{j,0,l}\,\mvec{v}_{j}(\tilde{\mmat{L}}_{l^{\ast}}),   
\end{align}
where $c_{0,l}\neq 0$ holds due to Assumption~\ref{Assump3}. After recursive implementation of the update procedures in \eqref{Imedsv} and \eqref{eqn:x_k1} for $k$ iterations of the distributed GPI algorithm, $\bar{\mvec{x}}_{k,l}$ and $\mvec{x}_{k,l}$ are obtained as
\begin{subequations}
\begin{align}
\bar{\mvec{x}}_{k,l} =&\, c_{k-1,l}\,r_{l}\, e^{\mathrm{j} (\eta_{l}+k\theta_{l})}\,\mvec{v}_{n}(\tilde{\mmat{L}}_{l^{\ast}}) \nonumber \\
& + c_{k-1,l}\,r_{l}\, e^{-\mathrm{j}(\eta_{l}+k\theta_{l})}\,\mvec{v}_{n-1}(\tilde{\mmat{L}}_{l^{\ast}}) \nonumber \\
& + \tfrac{1}{r_{l}^{k-1}}\sum_{j=1}^{n-2} c_{j,k-1,l}\,\lambda_{j}^{k}(\tilde{\mmat{L}}_{l^{\ast}})\,\mvec{v}_{j}(\tilde{\mmat{L}}_{l^{\ast}}), \\
\mvec{x}_{k,l} =&\, c_{k,l}\,e^{\mathrm{j}(\eta_{l}+k\theta_{l})}\, \mvec{v}_{n}(\tilde{\mmat{L}}_{l^{\ast}}) + c_{k,l}\,e^{-\mathrm{j} (\eta_{l}+k\theta_{l})}\,\mvec{v}_{n-1}(\tilde{\mmat{L}}_{l^{\ast}}) \nonumber \\
& + \tfrac{1}{r_{l}^{k}}\sum_{j=1}^{n-2} c_{j,k,l} \,\lambda_{j}^{k}(\tilde{\mmat{L}}_{l^{\ast}})\,\mvec{v}_{j}(\tilde{\mmat{L}}_{l^{\ast}}),\label{xkmaindefDIS}
\end{align}    
\end{subequations}
where
\begin{subequations}
\begin{align}
&c_{k,l}= \tfrac{1}{\sqrt{2+2\rho_{l} \cos(2\eta_{l} +2k\theta_{l} +\phi_{l})+h_{\mathcal{I},k,l}}},\\ 
&c_{j,k,l}= \tfrac{c_{j,0,l}}{c_{0,l}\sqrt{2+2\rho_{l} \cos(2\eta_{l} +2k\theta_{l} +\phi_{l})+ h_{\mathcal{I},k,l}}},\\
&h_{\mathcal{I},k,l}= 2\sum_{j=1}^{n-2}\Re\!\!\Big{[} \tfrac{c_{j,k-1,l}}{c_{k-1,l}}\tfrac{\lambda_{j}^{k}(\tilde{\mmat{L}}_{l^{\ast}})}{r_{l}^{k}} e^{-\mathrm{j} (\eta_{l}+k\theta_{l})}\mvec{v}_{n}^{\hr}(\tilde{\mmat{L}}_{l^{\ast}}) \mvec{v}_{j}(\tilde{\mmat{L}}_{l^{\ast}})\nonumber\\
& + \tfrac{c_{j,k-1,l}}{c_{k-1,l}}\tfrac{\lambda_{j}^{k}(\tilde{\mmat{L}}_{l^{\ast}})}{r_{l}^{k}} e^{\mathrm{j} (\eta_{l}+k\theta_{l})}\mvec{v}_{n-1}^{\hr}(\tilde{\mmat{L}}_{l^{\ast}}) \mvec{v}_{j}(\tilde{\mmat{L}}_{l^{\ast}})\Big{]} \label{hikdefDIS}\\
& +\sum_{j=1}^{n-2}\sum_{p=1}^{n-2}\tfrac{c_{j,k-1,l}^{\hr}\,c_{p,k-1,l}}{(c_{k-1,l})^{2}} \tfrac{\lambda_{j}^{k,\hr}(\tilde{\mmat{L}}_{l^{\ast}})\lambda_{p}^{k}(\tilde{\mmat{L}}_{l^{\ast}})}{r_{l}^{2k}}\mvec{v}_{j}^{\hr}(\tilde{\mmat{L}}_{l^{\ast}}) \mvec{v}_{p}(\tilde{\mmat{L}}_{l^{\ast}}),  \nonumber
\end{align}    
\end{subequations}
for any $j\!\in\!\mathbb{N}_{n-2}$ and $k\!\in\!\mathbb{N}$. This leads to the following description of vector $\mvec{x}_{k,l}$ from \eqref{xkmaindefDIS}
\begin{align}\label{skthisone0ll}
\mvec{x}_{k,l} =& \tfrac{e^{\mathrm{j} (\eta_{l}+k\theta_{l})}\mvec{v}_{n}(\tilde{\mmat{L}}_{l^{\ast}}) + e^{-\mathrm{j} (\eta_{l}+k\theta_{l})}\mvec{v}_{n-1}(\tilde{\mmat{L}}_{l^{\ast}})}{\sqrt{2+2\rho_{l} \cos(2\eta_{l}+2k\theta_{l} +\phi_{l})+h_{\mathcal{I},k,l}}} \nonumber \\
& + \tfrac{\sum_{j=1}^{n-2} 
\tfrac{c_{j,0,l}}{c_{0,l}} \tfrac{\lambda_{j}^{k}(\tilde{\mmat{L}}_{l^{\ast}})}{r_{l}^{k}}\,\mvec{v}_{j}(\tilde{\mmat{L}}_{l^{\ast}})}{\sqrt{2+2\rho_{l} \cos(2\eta_{l}+2k\theta_{l} +\phi_{l})+h_{\mathcal{I},k,l}}},
\end{align}
for any $k\in \mathbb{N}$. By defining $\zeta_{\mathcal{I},l}$ as 
\begin{equation}\label{zetaIdefDIS}
\zeta_{\mathcal{I},l}\!=\! \Big{|}\tfrac{\lambda_{n-2}(\tilde{\mmat{L}}_{l^{\ast}})}{\lambda_{n}(\tilde{\mmat{L}}_{l^{\ast}})}\Big{|}\!=\!\tfrac{|\lambda_{n-2}(\tilde{\mmat{L}}_{l^{\ast}})|}{r_{l}},     
\end{equation}
it follows from Assumptions~\ref{Assump1} and \ref{Assump2} that $0\!<\!\zeta_{\mathcal{I},l}\!<\!1$. It then follows from \eqref{hikdefDIS} and \eqref{zetaIdefDIS} that $h_{\mathcal{I},k,l}=\Theta(\zeta_{\mathcal{I},l}^{k})$ for any $k\!\in\!\mathbb{N}$, given that $\tfrac{|\lambda_{j}(\tilde{\mmat{L}}_{l^{\ast}})|}{r_{l}}\!\leq\! \zeta_{\mathcal{I},l}\!<\!1\;\forall j\!\in\!\mathbb{N}_{n-2}$ and $l^{\ast}:=k$ (note that $h_{\mathcal{I},k,l}$ is a real value). 
Subsequently, $\mvec{x}_{k,l}$ in \eqref{skthisone0ll} is rewritten as
\begin{equation}\label{skthisoneDIS}
\mvec{x}_{k,l} = \tfrac{e^{\mathrm{j} (\eta_{l}+k\theta_{l})}\mvec{v}_{n}(\tilde{\mmat{L}}_{l^{\ast}}) + e^{-\mathrm{j} (\eta_{l}+k\theta_{l})}\mvec{v}_{n-1}(\tilde{\mmat{L}}_{l^{\ast}})+a_{k,l}\,
\mvec{s}_{k,l}}{\sqrt{2+2\rho_{l} \cos(2\eta_{l}+ 2k\theta_{l} +\phi_{l})+ h_{\mathcal{I},k,l}}}, 
\end{equation}
for any $k\!\in\!\mathbb{N}$, where
$a_{k,l}=\Theta(\zeta_{\mathcal{I},l}^{k})$ and $\mvec{s}_{k,l}\!\in\! \mathbb{C}^{n}$ is defined as a unit vector in the same direction as $
\sum_{j=1}^{n-2} c_{j,0,l}\,\lambda_{j}^{k}(\tilde{\mmat{L}}_{l^{\ast}})\,\mvec{v}_{j}(\tilde{\mmat{L}}_{l^{\ast}})$.  
Since $\zeta^{k}_{\mathcal{I},l}\rightarrow 0$ as $k\rightarrow \infty$, the linear approximation of the function in the right-hand side of \eqref{skthisoneDIS} around $a_{k,l}=0$ and $h_{\mathcal{I},k,l}=0$ yields
\begin{equation}\label{x1234aDIS}
\mvec{x}_{k,l}= \tfrac{e^{\mathrm{j}(\eta_{l} + k \theta_{l})}\mvec{v}_{n}(\tilde{\mmat{L}}_{l^{\ast}}) + e^{-\mathrm{j}(\eta_{l} + k \theta_{l})}\mvec{v}_{n-1}(\tilde{\mmat{L}}_{l^{\ast}})}{\sqrt{2+2\rho_{l} \cos(2\eta_{l} +2k \theta_{l} + \phi_{l})}} + b_{k,l}\,\hat{\mvec{s}}_{k,l},
\end{equation}
where the unit vector $\hat{\mvec{s}}_{k,l}\!\in\! \mathbb{C}^{n}$ has the same direction as $2a_{k,l}\,\mvec{s}_{k,l}-\tfrac{h_{\mathcal{I},k,l}(e^{\mathrm{j}(\eta_{l}+k \theta_{l})}\mvec{v}_{n}(\tilde{\mmat{L}}_{l^{\ast}})+e^{-\mathrm{j}(\eta_{l} + k \theta_{l})}\mvec{v}_{n-1}(\tilde{\mmat{L}}_{l^{\ast}}))}{2+2\rho_{l} \cos(2\eta_{l} +2k \theta_{l} + \phi_{l})}$ and $b_{k,l}=\Theta(\zeta_{\mathcal{I},l}^{k})$. The unit state vectors $\mvec{x}_{k-1,l}$ and $\mvec{x}_{k-2,l}$ can also be obtained by following a similar procedure, i.e.,
\begin{subequations}\label{x1234DIS}
\begin{align}
\mvec{x}_{k-1,l} =& \, \tfrac{e^{\mathrm{j}(\eta_{l} + (k-1) \theta_{l})}\mvec{v}_{n}(\tilde{\mmat{L}}_{l^{\ast}}) + e^{-\mathrm{j}(\eta_{l} + (k-1) \theta_{l})}\mvec{v}_{n-1}(\tilde{\mmat{L}}_{l^{\ast}})}{\sqrt{2+2\rho_{l} \cos(2\eta_{l} + 2(k-1)\theta_{l} +\phi_{l})}} \nonumber \\
& +b_{k-1,l}\, \hat{\mvec{s}}_{k-1,l}, \label{x1234bDIS}\\
\mvec{x}_{k-2,l} =& \, \tfrac{e^{\mathrm{j}(\eta_{l} + (k-2) \theta_{l})}\mvec{v}_{n}(\tilde{\mmat{L}}_{l^{\ast}}) + e^{-\mathrm{j}(\eta_{l} + (k-2) \theta_{l})}\mvec{v}_{n-1}(\tilde{\mmat{L}}_{l^{\ast}})}{\sqrt{2+2\rho_{l} \cos(2\eta_{l} +2(k-2)\theta_{l} +\phi_{l})}} \nonumber\\ 
&+ b_{k-2,l}\, \hat{\mvec{s}}_{k-2,l}, \label{x1234cDIS}
\end{align}   
\end{subequations}
for the two unit vectors $\hat{\mvec{s}}_{k-1,l}, \hat{\mvec{s}}_{k-2,l}\!\in\!\mathbb{C}^{n}$ and two scalars
$b_{k-1,l}\!=\!\Theta(\zeta^{k-1}_{\mathcal{I},l})$
and $b_{k-2,l}\!=\!\Theta(\zeta^{k-2}_{\mathcal{I},l})$. 
The impact of the consensus observer, with the iteration index $m$, on the estimation error of the distributed GPI algorithm, as the second source of error, is investigated next. By defining $\mvec{z}_{k,l}^{s}(m)\!=\![z_{k,l}^{s,i}(m)]\!\in\!\mathbb{C}^{n}$ as the state vector of the consensus observer in its $m\nth$ iteration, for any $m\!\in\! \mathbb{N}_{m^{\ast}}$, after the inclusion of the first source of error, it follows from the initialization step in \eqref{z1234INI} and the update procedure in \eqref{z1234UPDP} that
\begin{equation}\label{updatezdist}
\mvec{z}_{k,l}^{s}(m^{\ast})\!=\!\big{(}\mmat{I}_{n}-\delta \mmat{L}\big{)}\,\mvec{z}_{k,l}^{s}(m^{\ast}-1)\!=\!\big{(}\mmat{I}_{n}-\delta \mmat{L}\big{)}^{m^{\ast}}\!\!\mvec{z}_{k,l}^{s}(0), 
\end{equation}
where 
\begin{subequations}\label{z1234prfdis}
\begin{align}
z_{k,l}^{1,i}(m\!=\!0) &\!=\! \tfrac{\langle \mvec{w}_{1}(\mmat{L}),\mvec{1}_{n} \rangle}{w_{1}^{i}(\mmat{L})}\bar{x}_{k,l}^{i,\hr}\bar{x}_{k,l}^{i}, \\
z_{k,l}^{2,i}(m\!=\!0) &\!=\! \tfrac{\langle \mvec{w}_{1}(\mmat{L}),\mvec{1}_{n} \rangle}{w_{1}^{i}(\mmat{L})}\bar{x}_{k-1,l}^{i,\hr}\bar{x}_{k,l}^{i}, \\
z_{k,l}^{3,i}(m\!=\!0) & \!=\! 
\begin{cases}
\tfrac{\langle \mvec{w}_{1}(\mmat{L}),\mvec{1}_{n} \rangle}{w_{1}^{i}(\mmat{L})}\bar{x}_{k-2,l}^{i,\hr}\bar{x}_{k,l}^{i}, & \mathrm{if}\;\; k\neq 1, \\
 0,  & \mathrm{if}\;\; k=1.
\end{cases} \\
z_{k,l}^{4,i}(m\!=\!0) & \!=\! \langle \mvec{w}_{1}(\mmat{L}),\mvec{1}_{n} \rangle
\bar{x}_{k,l}^{i},
\end{align}
\end{subequations}
for prespecified integers $l^{\ast}$, $m^{\ast}$ and any $i\!\in\! \mathcal{V}$, $s\!\in\!\mathbb{N}_{4}$ and $k\!\in\!\mathbb{N}$. Define 
\begin{equation}\label{zhatdesdist}
\begin{array}{llll}
& \hat{z}_{k,l}^{1}\!=\!\langle \bar{\mvec{x}}_{k,l},\bar{\mvec{x}}_{k,l} \rangle, 
&& \;\;\hat{z}_{k,l}^{2}\!=\!\langle \bar{\mvec{x}}_{k,l},\bar{\mvec{x}}_{k-1,l} \rangle, \\
& \hat{z}_{k,l}^{3}\!=\!\langle \bar{\mvec{x}}_{k,l},\bar{\mvec{x}}_{k-2,l} \rangle,
&& \;\;\hat{z}_{k,l}^{4}\!=\!\langle \bar{\mvec{x}}_{k,l},\mvec{w}_{1}(\mmat{L}) \rangle,
\end{array}
\end{equation}
as the desired values that the consensus observer procedure in \eqref{updatezdist} with the initial values in \eqref{z1234prfdis} is aimed at converging to in the $k\nth$ iteration, according to \eqref{InnerP}. The following conclusion on the estimation error of the consensus observer can then be made
\begin{equation}\label{tetmast}
\max_{i\in\mathcal{V},s\in\mathbb{N}_{4}}\big{|}z_{k,l}^{s,i}(m^{\ast})-\hat{z}_{k,l}^{s}\big{|}= \Theta\big{(}\xi^{m^{\ast}}\big{)},
\end{equation}
where
\begin{equation}\label{xidef}
\xi = 1 - \delta\, \tilde{\lambda}(\mmat{L}),    
\end{equation}
for prespecified integers $l^{\ast}$, $m^{\ast}$ and any $k\!\in\!\mathbb{N}$. Without loss of generality, let node $i^{\ast}$ be defined as
\begin{equation}\label{istrdef}
i^{\ast}=\underset{i\in \mathcal{V},s\in \mathbb{N}_{4}}{\mathrm{argmax}}\big{|}z_{k,l}^{s,i}(m^{\ast})\!-\!\hat{z}_{k,l}^{s}\big{|}. 
\end{equation}
To incorporate the effect of the second source of error, in addition to that of the first one, $\bar{\mvec{x}}_{k,l,m}\!=\![\bar{x}_{k,l,m}^{i}]\!\in\!\mathbb{C}^{n}$ and $\mvec{x}_{k,l,m}\!=\![x_{k,l,m}^{i}]\!\in\!\mathbb{C}^{n}$ are defined, respectively, as the intermediate state vector and the state vector obtained by the distributed GPI algorithm at the $k\nth$ iteration, after the inclusion of both the first and second sources of error, for any $k\!\in\!\mathbb{N}$. Moreover, $\bar{\mvec{x}}_{k,l,m} $ and $\mvec{x}_{k,l,m}$ are related by $\mvec{x}_{k,l,m}\!=\! \tfrac{\bar{\mvec{x}}_{k,l,m}}{\|\bar{\mvec{x}}_{k,l,m}\|}$. 
More specifically, $x^{i}_{k,l,m}$ denotes the $i\nth$ element of $\mvec{x}_{k,l,m}$, providing an estimate of the $i\nth$ element of the state vector $\mvec{x}_{k,l}$ by the $i\nth$ node after employing the consensus observer procedure for $m^{\ast}$ iterations in order to disseminate global information \eqref{GVdistgpi} throughout the network in the distributed implementation of the GPI algorithm. Similar to \eqref{zbarmst}, $\bar{\mvec{z}}_{k,l}^{s}\!=\![\bar{z}_{k,l}^{s,i}]\!\in\!\mathbb{C}^{n}$ is defined as the output of the consensus observer after $m^{\ast}$ iterations such that
\begin{equation}
\bar{\mvec{z}}_{k,l}^{s}=\mvec{z}_{k,l}^{s}(m^{\ast}), 
\end{equation}
for any $s\!\in\!\mathbb{N}_{4}$ and $k\!\in\!\mathbb{N}$. After replacing the state vector $\mvec{x}_{k-1}$ with $\mvec{x}_{k-1,l}$ in \eqref{ybar}, the vector $\bar{\mvec{y}}_{k,l}\!=\![\bar{y}_{k,l}^{i}]\!\in\!\mathbb{C}^{n}$ is defined as 
\begin{equation}\label{ykldef}
\bar{\mvec{y}}_{k,l}\!=\! \sum_{j=0}^{l^{\ast}}\tfrac{1}{j!}(\mmat{I}_{n}-\delta \mmat{L})^{j}\,\mvec{x}_{k-1,l}, 
\end{equation}
for any $k\!\in\!\mathbb{N}$. 
By considering the node $i^{\ast}$ (defined in \eqref{istrdef}) to investigate the worst-case scenario for the second source of error according to \eqref{tetmast} and using the definition of $\bar{\mvec{y}}_{k,l}$ in \eqref{ykldef}, 
the ${i^{\ast}}\nth$ element of vector $\mvec{x}_{k,l,m}$ is derived using \eqref{UPDsvdist} as follows  
\begin{equation}\label{xklmiast}
x^{i^{\ast}}_{k,l,m} \!=\! 
\tfrac{\bar{x}^{i^{\ast}}_{k,l,m}}{\|\bar{\mvec{x}}_{k,l,m}\|}
\!=\!\tfrac{\bar{y}_{k,l}^{i^{\ast}}-ew_{1}^{i^{\ast}}\!(\mmat{L})\,\tfrac{\bar{z}_{k-1,l}^{4,i^{\ast}}+\varphi_{k,l,m}^{1}}{\sqrt{\bar{z}_{k-1,l}^{1,i^{\ast}}+\varphi_{k,l,m}^{2}}}}{\sqrt{\bar{z}_{k,l}^{1,i^{\ast}}+\varphi_{k,l,m}^{3}}},
\end{equation}
where $|\varphi_{k,l,m}^{j}|\!=\!\Theta(\xi^{m^{\ast}})$ for any $j\!\in\!\mathbb{N}_{3}$ and $k\!\in\!\mathbb{N}$. 
From Assumption~\ref{Assump1} and on noting that $m^{\ast}:=k$, it follows that $\xi^{m^{\ast}}\!\rightarrow\!0$ as $k\!\rightarrow\!\infty$. 
Using a linear approximation of the function in the right-hand side of \eqref{xklmiast} around $\varphi_{k,l,m}^{1}\!=\!0$, $\varphi_{k,l,m}^{2}\!=\!0$ and $\varphi_{k,l,m}^{3}\!=\!0$ yields 
\begin{equation}\label{xklmiast2}
x^{i^{\ast}}_{k,l,m} 
\!=\!\tfrac{1}{\sqrt{\bar{z}_{k,l}^{1,i^{\ast}}}}\Big{[}\bar{y}_{k,l}^{i^{\ast}}-ew_{1}^{i^{\ast}}\!(\mmat{L})\,\tfrac{\bar{z}_{k-1,l}^{4,i^{\ast}}}{\sqrt{\bar{z}_{k-1,l}^{1,i^{\ast}}}}\Big{]} + b_{k,l,m},
\end{equation}
where $|b_{k,l,m}|=\Theta(\xi^{m^{\ast}})$ for any $k\!\in\!\mathbb{N}$. Performing the same procedure for all elements of vector $\mvec{x}_{k,l,m}$ results in
\begin{equation}\label{xkintermDISRI}
\mvec{x}_{k,l,m} \!=\! \mvec{x}_{k,l}+b_{k,l,m} \, \hat{\mvec{s}}_{k,l,m},   
\end{equation}
where $\hat{\mvec{s}}_{k,l,m}$ represents a unit vector in $\mathbb{C}^{n}$ for any $k\!\in\!\mathbb{N}$. After substituting $\mvec{x}_{k,l}$ from \eqref{x1234aDIS} into \eqref{xkintermDISRI} and performing a similar procedure for the vectors $\mvec{x}_{k-1,l,m}$ and $\mvec{x}_{k-2,l,m}$ while using \eqref{x1234DIS}, one arrives at
\begin{subequations}\label{xklmrealdis23DISRI}
\begin{align}
\mvec{x}_{k,l,m} =\,&\tfrac{e^{\mathrm{j}(\eta_{l} + k \theta_{l})}\,\mvec{v}_{n}(\tilde{\mmat{L}}_{l^{\ast}}) + e^{-\mathrm{j}(\eta_{l} + k \theta_{l})}\,\mvec{v}_{n-1}(\tilde{\mmat{L}}_{l^{\ast}})}{\sqrt{2+2\rho_{l} \cos(2\eta_{l} +2k \theta_{l} + \phi_{l})}} \nonumber \\
& +b_{k,l}\, \hat{\mvec{s}}_{k,l}
+b_{k,l,m}\, \hat{\mvec{s}}_{k,l,m}, \\
\mvec{x}_{k-1,l,m} =\,&\tfrac{e^{\mathrm{j}(\eta_{l} + (k-1) \theta_{l})}\,\mvec{v}_{n}(\tilde{\mmat{L}}_{l^{\ast}}) + e^{-\mathrm{j}(\eta_{l} + (k-1)\theta_{l})}\,\mvec{v}_{n-1}(\tilde{\mmat{L}}_{l^{\ast}})}{\sqrt{2+2\rho_{l} \cos(2\eta_{l} +2(k-1)\theta_{l} + \phi_{l})}} \nonumber \\
& +b_{k-1,l}\, \hat{\mvec{s}}_{k-1,l}
+b_{k-1,l,m}\, \hat{\mvec{s}}_{k-1,l,m}, \\
\mvec{x}_{k-2,l,m} = \,&\tfrac{e^{\mathrm{j}(\eta_{l} + (k-2) \theta_{l})}\,\mvec{v}_{n}(\tilde{\mmat{L}}_{l^{\ast}}) + e^{-\mathrm{j}(\eta_{l} + (k-2)\theta_{l})}\,\mvec{v}_{n-1}(\tilde{\mmat{L}}_{l^{\ast}})}{\sqrt{2+2\rho_{l} \cos(2\eta_{l} +2(k-2)\theta_{l} + \phi_{l})}}\nonumber \\
& +b_{k-2,l}\, \hat{\mvec{s}}_{k-2,l}
+b_{k-2,l,m}\, \hat{\mvec{s}}_{k-2,l,m},
\end{align}    
\end{subequations}
where $b_{j,l}=\Theta(\zeta^{j}_{\mathcal{I},l})$, $|b_{j,l,m}|=\Theta(\xi^{m^{\ast}})$, and $\hat{\mvec{s}}_{j,l},\hat{\mvec{s}}_{j,l,m}\!\in\! \mathbb{C}^{n}$ represent a set of unit vectors for $j\!\in\!\{k\!-\!2,k\!-\!1,k\}$. 
Consider $z_{k,l,m}^{1,i^{\ast}}$, $z_{k,l,m}^{2,i^{\ast}}$ and $z_{k,l,m}^{3,i^{\ast}}$ as the distributed counterparts of $z_{k}^{1}$, $z_{k}^{2}$ and $z_{k}^{3}$ in equation \eqref{zk123}, respectively. For distributed implementation of the GPI algorithm from the viewpoint of node $i^{\ast}$, define
\begin{subequations}\label{z123dist}
\begin{align}
z_{k,l,m}^{1,i^{\ast}}\!=\,& \langle \mvec{x}_{k,l,m},\mvec{x}_{k-1,l,m} \rangle, \\
z_{k,l,m}^{2,i^{\ast}}\!=\,& \langle \mvec{x}_{k-1,l,m},\mvec{x}_{k-2,l,m} \rangle, \\
z_{k,l,m}^{3,i^{\ast}}\!=\,& \langle \mvec{x}_{k,l,m},\mvec{x}_{k-2,l,m} \rangle,
\end{align}   
\end{subequations}
for any $k\!\in\!\mathbb{N}$. Consider node $i^{\ast}$ in the convergence analysis as the worst-case scenario for the second source of error. Substituting from \eqref{xklmrealdis23DISRI} into \eqref{z123dist} and only keeping the asymptotically dominant terms result in
\begin{subequations}\label{zklmstr}
\begin{align}
z_{k,l,m}^{1,i^{\ast}}\!=&\tfrac{\cos \theta_{l} + \rho_{l} \cos(\gamma_{l} +(2k-1)\theta_{l})}{\sqrt{(1+\rho_{l}\cos(\gamma_{l}+2k\theta_{l}))(1+\rho_{l}\cos(\gamma_{l}+2(k-1)\theta_{l} ))}}+\mathfrak{z}_{k,l,m}^{1},\\
z_{k,l,m}^{2,i^{\ast}}\!=& \tfrac{\cos \theta_{l} + \rho_{l} \cos(\gamma_{l} +(2k-3)\theta_{l} )}{\sqrt{(1+\rho_{l}\cos(\gamma_{l}+2(k-1)\theta_{l} ))(1+\rho_{l}\cos(\gamma_{l}+2(k-2)\theta_{l}))}}+\mathfrak{z}_{k,l,m}^{2}, \\
z_{k,l,m}^{3,i^{\ast}}\!=&\tfrac{\cos \theta_{l} + \rho_{l} \cos(\gamma_{l} +(2k-2)\theta_{l} )}{\sqrt{(1+\rho_{l}\cos(\gamma_{l}+2k\theta_{l} ))(1+\rho_{l}\cos(\gamma_{l}+2(k-2)\theta_{l} ))}}+\mathfrak{z}_{k,l,m}^{3},
\end{align}
\end{subequations}
where $\gamma_{l}\!:=\!2\eta_{l}+\phi_{l}$ and 
\begin{subequations}
\begin{align}
|\mathfrak{z}_{k,l,m}^{1}|=&  \Theta(\zeta^{k-1}_{\mathcal{I},l})+\Theta(\xi^{m^{\ast}}), \\ 
|\mathfrak{z}_{k,l,m}^{2}|=&  \Theta(\zeta^{k-2}_{\mathcal{I},l})+\Theta(\xi^{m^{\ast}}), \\
|\mathfrak{z}_{k,l,m}^{3}|=&  \Theta(\zeta^{k-2}_{\mathcal{I},l})+\Theta(\xi^{m^{\ast}}),
\end{align}    
\end{subequations}
for any $k\!\in\!\mathbb{N}$. Let the real scalars $\check{d}_{k,l,m}^{i}$ and $\hat{d}_{k,l,m}^{i}$ be defined as the distance between successive pairs of one-dimensional and two-dimensional subspaces, respectively, which are computed by the $i\nth$ node in the distributed implementation of the GPI algorithm, for any $i\!\in\!\mathcal{V}$ and $k\!\in\!\mathbb{N}$. 
By considering the results of Lemma~\ref{lemm:dcdh1} and utilizing \eqref{dhdcDISTdist} and \eqref{z123dist}, $\check{d}_{k,l,m}^{i^{\ast}}$ and $\hat{d}_{k,l,m}^{i^{\ast}}$ are obtained as 
\begin{subequations}\label{dcheckdhat}
\begin{align}
\check{d}_{k,l,m}^{i^{\ast}} & = \sqrt{1-|z_{k,l,m}^{1,i^{\ast}}|^{2} },\\
\hat{d}_{k,l,m}^{i^{\ast}} & = \sqrt{1-\tfrac{\big{|}z_{k,l,m}^{1,i^{\ast}}z_{k,l,m}^{2,i^{\ast}}-z_{k,l,m}^{3,i^{\ast}}\big{|}^{2}}{\big{(}1-|z_{k,l,m}^{1,i^{\ast}}|^{2}\big{)}\big{(}1-|z_{k,l,m}^{2,i^{\ast}}|^{2}\big{)}}},
\end{align}   
\end{subequations}
from the perspective of node $i^{\ast}$ for any $k\!\in\!\mathbb{N}$. After substituting \eqref{zklmstr} into \eqref{dcheckdhat} and only keeping the asymptotically dominant terms, it follows that
\begin{subequations}\label{dchcekcdhat2}    
\begin{align}
(\check{d}_{k,l,m}^{i^{\ast}})^2 & \!=\! \tfrac{\sin^{2}\theta_{l} +\rho^{2}_{l}\big{[}\cos(\gamma_{l}+2k\theta_{l})\cos(\gamma_{l}+(2k-2)\theta_{l})-\cos^{2}(\gamma_{l}+(2k-1)\theta_{l})\big{]}}{(1+\rho_{l} \cos(\gamma_{l}+2k\theta_{l}))(1+\rho_{l} \cos(\gamma_{l}+2(k-1)\theta_{l}))} \nonumber \\
&\;\; + \Theta(\zeta_{\mathcal{I},l}^{k-1})+\Theta(\xi^{m^{\ast}}), \\
(\hat{d}_{k,l,m}^{i^{\ast}})^2 & \!=\! \Theta(\zeta_{\mathcal{I},l}^{k-2}) + \Theta(\xi^{m^{\ast}}).
\end{align}
\end{subequations}
Since $0<\xi<1$ (from \eqref{xidef} and considering Assumption~\ref{Assump1}), it follows that $\xi^{m^{\ast}}\!\rightarrow\!0$ as $k\!\rightarrow\!\infty$ (note that $m^{\ast}\!:=\!k$).
Using \eqref{tetmast} and the expression in \eqref{istrdef}, it follows from \eqref{dchcekcdhat2} that
\begin{subequations}\label{This-one}
\begin{align}
& \lim_{k\rightarrow \infty} \check{d}_{k,l,m}^{i}\!=\! \lim_{k\rightarrow \infty} \check{d}_{k,l,m}^{i^{\ast}}\;\;\; \forall i\in \mathcal{V}, \\
& \lim_{k\rightarrow \infty} \hat{d}_{k,l,m}^{i}\!=\! \lim_{k\rightarrow \infty} \hat{d}_{k,l,m}^{i^{\ast}}\;\;\; \forall i\in \mathcal{V}.
\end{align}
\end{subequations} 
By considering the definitions of $\tilde{\mmat{L}}$ and $\tilde{\mmat{L}}_{l^{\ast}}$ in \eqref{eqn:MLM} and \eqref{eqn:Ll11}, respectively, and on noting that $k\!\rightarrow\! \infty$ implies $l^{\ast}\!\rightarrow\! \infty$, one has 
\begin{subequations}\label{EVLEVEl}
\begin{alignat}{2}
& \lim_{k\rightarrow \infty}\!\lambda_{n}(\tilde{\mmat{L}}_{l^{\ast}})\!=\!\lambda_{n}(\tilde{\mmat{L}}), \;&& \lim_{k\rightarrow \infty}\!\lambda_{n-1}(\tilde{\mmat{L}}_{l^{\ast}})\!=\!\lambda_{n-1}(\tilde{\mmat{L}}), \\
& \lim_{k\rightarrow \infty}\!\mvec{v}_{n}(\tilde{\mmat{L}}_{l^{\ast}})\!=\!\mvec{v}_{n}(\tilde{\mmat{L}}), \;&& \lim_{k\rightarrow \infty}\!\mvec{v}_{n-1}(\tilde{\mmat{L}}_{l^{\ast}})\!=\!\mvec{v}_{n-1}(\tilde{\mmat{L}}).
\end{alignat}
\end{subequations}
It then follows from \eqref{noldef}, \eqref{zetaIdef}, \eqref{withldefDIS}, \eqref{zetaIdefDIS} and \eqref{EVLEVEl} that
\begin{subequations}
\begin{alignat}{4}
& \lim_{k\rightarrow \infty}r_{l}\!=\!r, \,&& \lim_{k\rightarrow \infty}\theta_{l}\!=\!\theta, \,&& \lim_{k\rightarrow \infty}\rho_{l}\!=\!\rho, \,&& \lim_{k\rightarrow \infty}\phi_{l}\!=\!\phi, \\
& \lim_{k\rightarrow \infty}c_{l}\!=\!c, \,&& \lim_{k\rightarrow \infty}\eta_{l}\!=\!\eta, \,&& \lim_{k\rightarrow \infty}\gamma_{l}\!=\!\gamma, \,&&
\lim_{k\rightarrow \infty}\zeta_{\mathcal{I},l}\!=\!\zeta_{\mathcal{I}}.
\end{alignat}
\end{subequations}
Given that $0\!<\!\zeta_{\mathcal{I}}\!<\!1$ (by Assumptions~\ref{Assump1} and \ref{Assump2}), it results that $\zeta_{\mathcal{I},l}^{k}\!\rightarrow\!0$ as $k\!\rightarrow\!\infty$. Now, equations \eqref{dkdkI} and \eqref{dchcekcdhat2} as well as the final values in \eqref{This-one} lead to 
\begin{subequations}\label{distIprf}
\begin{align}
&\lim\limits_{k\rightarrow \infty}\check{d}_{k,l,m}^{i} \!=\! \lim\limits_{k\rightarrow \infty}\check{d}_{k},\;\;\; \forall\, i\!\in\! \mathcal{V}, \\
& \lim\limits_{k\rightarrow \infty}\hat{d}_{k,l,m}^{i} \!=\! \lim\limits_{k\rightarrow \infty}\hat{d}_{k},\;\;\; \forall\, i\!\in\! \mathcal{V},
\end{align}
\end{subequations}
on noting that $\zeta_{\mathcal{I},l}^{k-1}\!\rightarrow\!0$, $\zeta_{\mathcal{I},l}^{k-2}\!\rightarrow\!0$ and $\xi^{m^{\ast}}\!\rightarrow\!0$ as $k\!\rightarrow\!\infty$. 
Use \eqref{dk+1dist}, and let $d_{k+1,l,m}^{i}$ be defined as
\begin{equation}\label{dchk1dist}
d_{k+1,l,m}^{i}=\min\{\check{d}_{k,l,m}^{i},\hat{d}_{k,l,m}^{i}\},    
\end{equation}
for any $i\!\in\!\mathcal{V}$ and $k\!\in\!\mathbb{N}$. It then results from \eqref{dkdkI2}, \eqref{distIprf} and \eqref{dchk1dist} for scenario $\mathcal{I}$ that
\begin{equation}\label{This-one2}
\lim\limits_{k\rightarrow \infty}d_{k+1,l,m}^{i}= \lim\limits_{k\rightarrow \infty}\hat{d}_{k} =0, \;\;\; \forall\, i\!\in\! \mathcal{V},    
\end{equation}
which guarantees the termination of the distributed GPI algorithm from the viewpoint of every node after finite iterations in scenario $\mathcal{I}$, noting that the positive threshold $\epsilon$ is used in the termination condition of the algorithm. Moreover, this indicates that the GPI algorithm successfully estimates the magnitude of the complex conjugate eigenvalues $\lambda_{n}(\tilde{\mmat{L}})$ and $\lambda_{n-1}(\tilde{\mmat{L}})$ as the pair of dominant eigenvalues of $\tilde{\mmat{L}}$ in a distributed fashion from the perspective of all nodes in scenario $\mathcal{I}$. 
Use \eqref{Thislambda1} and \eqref{Thislambda2}, and let $\check{\lambda}_{k,l,m}^{i}$ and $\hat{\lambda}_{k,l,m}^{i}$ be defined as the magnitudes of the dominant eigenvalue of $\tilde{\mmat{L}}_{l^{\ast}}$ in scenarios $\mathcal{R}$ and $\mathcal{I}$, respectively. It then follows from \eqref{This-one2} that 
\begin{equation}\label{This-one25}
\lim\limits_{k\rightarrow \infty}\hat{\lambda}_{k,l,m}^{i}=|\lambda_{n}(\tilde{\mmat{L}})|,    
\end{equation}
for any $i\!\in\!\mathcal{V}$. Moreover, the pair of complex conjugate dominant eigenvalues $\lambda_{n}(\tilde{\mmat{L}})$ and $\lambda_{n-1}(\tilde{\mmat{L}})$ correspond to the two-dimensional subspace $\mathcal{W}^{\ast}:=\mathrm{span}\{\mvec{v}_{n-1}(\tilde{\mmat{L}}),\mvec{v}_{n}(\tilde{\mmat{L}})\}$ in scenario $\mathcal{I}$, indicating that the subspace sequence 
$\{\mathcal{W}_{k,l,m}\}_{k\in \mathbb{N}}$, $\mathcal{W}_{k,l,m}=\mathrm{span}\{\mvec{x}_{k-1,l,m},\mvec{x}_{k,l,m}\}$, converges to $\mathcal{W}^{\ast}$ while the subspace sequence $\{\mathcal{V}_{k,l,m}\}_{k\in \mathbb{N}}$, $\mathcal{V}_{k,l,m}=\mathrm{span}\{\mvec{x}_{k,l,m}\}$, is not convergent as $k\rightarrow \infty$. 
Use \eqref{lambdak1dist} and define $\tilde{\lambda}_{k+1,l,m}^{i}$ as 
\begin{equation}\label{This-one3}
\tilde{\lambda}_{k+1,l,m}^{i} \!=\!
\begin{cases}
\tfrac{1}{\delta}\big{[}1-\log\small{(}\check{\lambda}_{k,l,m}^{i}\small{)}\big{]}, \!\!& \mathrm{if}\; d_{k+1,l,m}^{i}\!=\!\check{d}_{k,l,m}^{i}, \\
\tfrac{1}{\delta}\big{[}1-\log\small{(}\hat{\lambda}_{k,l,m}^{i}\small{)}\big{]},   \!\!& \mathrm{if}\; d_{k+1,l,m}^{i}\!=\!\hat{d}_{k,l,m}^{i},
\end{cases}    
\end{equation}
for any $i\!\in\!\mathcal{V}$ and $k\!\in\!\mathbb{N}$.
The asymptotic estimation of the GAC of the network by all nodes follows from \eqref{This-one2}, \eqref{This-one25} and \eqref{This-one3}, i.e., 
\begin{equation}\label{FineqDIS}
\lim\limits_{k\rightarrow \infty}\tilde{\lambda}_{k,l,m}^{i} \!=\!  \tilde{\lambda}(\mmat{L}), \; \forall i\!\in\! \mathcal{V}.   
\end{equation}
This proves the validity of \eqref{Prdist} and completes the convergence proof in scenario $\mathcal{I}$.

Consider now scenario $\mathcal{R}$, in which the dominant eigenvalue of $\tilde{\mmat{L}}$ is a real scalar. Using the proposed distributed GPI algorithm and following a procedure similar to the one used in scenario $\mathcal{I}$, let the real scalars $r_{l}$ and $c_{0,l}$ be defined as
\begin{subequations}
\begin{align}
r_{l}=&\, \lambda_{n}(\tilde{\mmat{L}}_{l^{\ast}}), \\
c_{0,l}=&\,|c_{n,0,l}|,
\end{align}    
\end{subequations}
for the initial state vector $\mvec{x}_{0}\!\in\!\mathbb{C}^{n}$. As a result, the initial state vector $\mvec{x}_{0}$ in \eqref{x0dist} is rewritten as 
\begin{equation}
\mvec{x}_{0} = c_{0,l}\,\mvec{v}_{n}(\tilde{\mmat{L}}_{l^{\ast}}) +\sum_{j=1}^{n-1}c_{j,0,l}\,\mvec{v}_{j}(\tilde{\mmat{L}}_{l^{\ast}}),
\end{equation}
where $c_{0,l}\neq 0$ due to Assumption~\ref{Assump3}. By successive implementation of the GPI update procedure in scenario $\mathcal{R}$ after the inclusion of the first source of error
only, the intermediate state vector $\bar{\mvec{x}}_{k,l}\!\in\!\mathbb{C}^{n}$ and the state vector $\mvec{x}_{k,l}\!\in\!\mathbb{C}^{n}$ at the $k\nth$ iteration are generated in a recursive manner as follows
\begin{subequations}
\begin{align}
\bar{\mvec{x}}_{k} =&\, c_{k-1,l}\,r_{l}\,\mvec{v}_{n}(\tilde{\mmat{L}}_{l^{\ast}})\nonumber \\
&+ \tfrac{1}{r_{l}^{k-1}}\sum_{j=1}^{n-1} c_{j,k-1,l}\,\lambda_{j}^{k}(\tilde{\mmat{L}}_{l^{\ast}})\,\mvec{v}_{j}(\tilde{\mmat{L}}_{l^{\ast}}), \\
\mvec{x}_{k,l} =&\, c_{k,l}\,\mvec{v}_{n}(\tilde{\mmat{L}}_{l^{\ast}}) + \tfrac{1}{r_{l}^{k}} \sum_{j=1}^{n-1} c_{j,k,l} \,\lambda_{j}^{k}(\tilde{\mmat{L}}_{l^{\ast}})\,\mvec{v}_{j}(\tilde{\mmat{L}}_{l^{\ast}}), \label{xkmaindefrRDIS}
\end{align}    
\end{subequations}
where
\begin{subequations}\label{TotalRl}
\begin{align}
&c_{k,l}= \tfrac{1}{\sqrt{1+h_{\mathcal{R},k,l}}},\\ 
&c_{j,k,l}= \tfrac{c_{j,0,l}}{c_{0,l}\sqrt{1+ h_{\mathcal{R},k,l}}},\\
&h_{\mathcal{R},k,l}= 2\sum_{j=1}^{n-1}\Re\!\!\Big{[} \tfrac{c_{j,k-1,l}}{c_{k-1,l}}\tfrac{\lambda_{j}^{k}(\tilde{\mmat{L}}_{l^{\ast}})}{r_{l}^{k}} \mvec{v}_{n}^{\hr}(\tilde{\mmat{L}}_{l^{\ast}}) \mvec{v}_{j}(\tilde{\mmat{L}}_{l^{\ast}})\Big{]}  \label{hikdefRDIS}\\
& \!+\!\sum_{j=1}^{n-1}\sum_{p=1}^{n-1}\!\tfrac{c_{j,k-1,l}^{\hr}\,c_{p,k-1,l}}{(c_{k-1,l})^{2}} \tfrac{\lambda_{j}^{k,\hr}(\tilde{\mmat{L}}_{l^{\ast}}) \lambda_{p}^{k}(\tilde{\mmat{L}}_{l^{\ast}})}{r_{l}^{2k}}\mvec{v}_{j}^{\hr}(\tilde{\mmat{L}}_{l^{\ast}}) \mvec{v}_{p}(\tilde{\mmat{L}}_{l^{\ast}}), \nonumber
\end{align}    
\end{subequations}
for any $j\!\in\!\mathbb{N}_{n-1}$ and $k\!\in\!\mathbb{N}$. Using \eqref{xkmaindefrRDIS} and \eqref{TotalRl}, the vector $\mvec{x}_{k,l}$ is obtained as 
\begin{equation}\label{skthisoneR0l}
\mvec{x}_{k,l} = \tfrac{1}{\sqrt{1+h_{\mathcal{R},k,l}}}
\Big{[}\mvec{v}_{n}(\tilde{\mmat{L}}_{l^{\ast}})\!+\!\sum_{j=1}^{n-1}\tfrac{c_{j,0,l}}{c_{0,l}}\tfrac{\lambda_{j}^{k}(\tilde{\mmat{L}}_{l^{\ast}})}{r_{l}^{k}}\mvec{v}_{j}(\tilde{\mmat{L}}_{l^{\ast}})\Big{]},    
\end{equation}
for any $k\!\in\!\mathbb{N}$. 
By defining $\zeta_{\mathcal{R},l}$ as 
\begin{equation}\label{zetaIdefRDIS}
\zeta_{\mathcal{R},l}=\Big{|}\tfrac{\lambda_{n-1}(\tilde{\mmat{L}}_{l^{\ast}})}{\lambda_{n}(\tilde{\mmat{L}}_{l^{\ast}})}\Big{|}=\tfrac{|\lambda_{n-1}(\tilde{\mmat{L}}_{l^{\ast}})|}{r_{l}},    
\end{equation}
it follows from Assumptions~\ref{Assump1} and \ref{Assump2} that $0\!<\!\zeta_{\mathcal{R},l}\!<\!1$. 
It then follows from \eqref{hikdefRDIS} and \eqref{zetaIdefRDIS} that $h_{\mathcal{R},k,l}$ is a real scalar satisfying $h_{\mathcal{R},k,l}=\Theta(\zeta_{\mathcal{R},l}^{k})$ for any $k\!\in\!\mathbb{N}$, given that $\tfrac{|\lambda_{j}(\tilde{\mmat{L}}_{l^{\ast}})|}{r_{l}}\!\leq\!\zeta_{\mathcal{R},l}\!<\!1$ for all $j\!\in\! \mathbb{N}_{n-1}$. 
%
As a result, $\mvec{x}_{k,l}$ in \eqref{skthisoneR0l} can be rewritten as
\begin{equation}\label{skthisoneRDIS}
\mvec{x}_{k,l} = \tfrac{\mvec{v}_{n}(\tilde{\mmat{L}}_{l^{\ast}})+ a_{k,l}\,\mvec{t}_{k,l}}{\sqrt{1+h_{\mathcal{R},k,l}}},   
\end{equation}
where $a_{k,l}=\Theta(\zeta_{\mathcal{R},l}^{k})$ and $\mvec{t}_{k,l}\!\in\! \mathbb{C}^{n}$ is defined as a unit vector along $\sum_{j=1}^{n-1} c_{j,0,l}\,\lambda_{j}^{k}(\tilde{\mmat{L}}_{l^{\ast}})\,\mvec{v}_{j}(\tilde{\mmat{L}}_{l^{\ast}})$ for any $k\!\in\!\mathbb{N}$. 
Since $l^{\ast}\!:=\!k$ and $0\!<\!\zeta_{\mathcal{R},l}\!<\!1$, if follows that $\zeta^{k}_{\mathcal{R},l}\rightarrow 0$ as $k\rightarrow \infty$. By linear approximation of the function in the right-hand side of \eqref{skthisoneRDIS} around $a_{k,l}\!=\!0$ and $h_{\mathcal{R},k,l}\!=\!0$, one arrives at
\begin{equation}\label{x1234aRDIS}
\mvec{x}_{k,l}=  \mvec{v}_{n}(\tilde{\mmat{L}}_{l^{\ast}}) +b_{k,l}\,\hat{\mvec{t}}_{k,l},
\end{equation}
where $b_{k,l}=\Theta(\zeta^{k}_{\mathcal{R},l})$ and the unit vector $\hat{\mvec{t}}_{k,l}\!\in\! \mathbb{C}^{n}$ has the same direction as that of $2a_{k,l}\mvec{t}_{k,l}-h_{\mathcal{R},k,l}\mvec{v}_{n}(\tilde{\mmat{L}}_{l^{\ast}})$. Following a procedure similar to the above, the unit state vectors $\mvec{x}_{k-1,l}$ and $\mvec{x}_{k-2,l}$ are also obtained as 
\begin{subequations}\label{x1234RDIS}
\begin{align}
\mvec{x}_{k-1,l} =& \, \mvec{v}_{n}(\tilde{\mmat{L}}_{l^{\ast}}) + b_{k-1,l}\, \hat{\mvec{t}}_{k-1,l}, \label{x1234bRDIS} \\
\mvec{x}_{k-2,l} =& \,\mvec{v}_{n}(\tilde{\mmat{L}}_{l^{\ast}}) + b_{k-2,l}\, \hat{\mvec{t}}_{k-2,l}, \label{x1234cRDIS}
\end{align}   
\end{subequations}
for two unit vectors $\hat{\mvec{t}}_{k-1,l}, \hat{\mvec{t}}_{k-2,l}\!\in\!\mathbb{C}^{n}$ and two scalars $b_{k-1,l}\!=\!\Theta(\zeta^{k-1}_{\mathcal{R},l})$ and $b_{k-2,l}\!=\!\Theta(\zeta^{k-2}_{\mathcal{R},l})$. 
To incorporate the effect of the second source of error on the performance of the distributed GPI algorithm in addition to the first one, the state vector $\mvec{x}_{k,l,m}=[x_{k,l,m}^{i}]\!\in\!\mathbb{C}^{n}$ is defined at the $k\nth$ iteration of the algorithm. Following a procedure similar to the one used for equations \eqref{xklmiast} and \eqref{xklmiast2}, $\mvec{x}_{k,l,m}$ is obtained as
\begin{equation}\label{xkintermDISR}
\mvec{x}_{k,l,m} \!=\! \mvec{x}_{k,l}+ b_{k,l,m}\, \hat{\mvec{t}}_{k,l,m},   
\end{equation}
where $|b_{k,l,m}|\!=\!\Theta(\xi^{m^{\ast}})$ and $\hat{\mvec{t}}_{k,l,m}$ represents a unit vector in $\mathbb{C}^{n}$ for any $k\!\in\!\mathbb{N}$. 
After substituting $\mvec{x}_{k,l}$ from \eqref{x1234aRDIS} into \eqref{xkintermDISR} and following a similar procedure for the vectors $\mvec{x}_{k-1,l,m}$ and $\mvec{x}_{k-2,l,m}$, one arrives at
\begin{subequations}\label{xklmrealdis23DISR}
\begin{align}
\mvec{x}_{k,l,m}\!=\!&\,\mvec{v}_{n}(\tilde{\mmat{L}}_{l^{\ast}}) \!+\! b_{k,l}\hat{\mvec{t}}_{k,l}\!+\!b_{k,l,m}\hat{\mvec{t}}_{k,l,m}, \\ 
\mvec{x}_{k-1,l,m}\!=\!&\mvec{v}_{n}(\tilde{\mmat{L}}_{l^{\ast}}) \!+\! b_{k-1,l} \hat{\mvec{t}}_{k-1,l} \!+\!b_{k-1,l,m}\hat{\mvec{t}}_{k-1,l,m},\\ 
\mvec{x}_{k-2,l,m}\!=\!&\,\mvec{v}_{n}(\tilde{\mmat{L}}_{l^{\ast}}) \!+\! b_{k-2,l} \hat{\mvec{t}}_{k-2,l}\!+\!b_{k-2,l,m}\hat{\mvec{t}}_{k-2,l,m},
\end{align}    
\end{subequations}
where $b_{j,l}\!=\!\Theta(\zeta_{\mathcal{R},l}^{j})$, $|b_{j,l,m}|\!=\!\Theta(\xi^{m^{\ast}})$, and  $\hat{\mvec{t}}_{j,l},\hat{\mvec{t}}_{j,l,m}\!\in\! \mathbb{C}^{n}$ represent a set of unit vectors, for any $j\!\in\!\{k-2,k-1,k\}$ and $k\!\in\!\mathbb{N}$. 
Considering \eqref{istrdef}, let $z_{k,l,m}^{1,i^{\ast}}$, $z_{k,l,m}^{2,i^{\ast}}$ and $z_{k,l,m}^{3,i^{\ast}}$ be defined as the distributed counterparts of $z_{k}^{1}$, $z_{k}^{2}$ and $z_{k}^{3}$ in \eqref{zk123}, respectively, from the viewpoint of node $i^{\ast}$. Substituting from \eqref{xklmrealdis23DISR} into \eqref{z123dist} and keeping the asymptotically dominant terms only, the values of these inner products are obtained as
\begin{equation}\label{innerpdistRDIS}
z_{k,l,m}^{j,i^{\ast}}\!= \!1-\mathfrak{z}_{k,l,m}^{j},
\end{equation}
for any $j\!\in\!\mathbb{N}_{3}$, where
\begin{subequations}
\begin{align}
|\mathfrak{z}_{k,l,m}^{1}|\!=& \Theta(\zeta^{k-1}_{\mathcal{R},l})+\Theta(\xi^{m^{\ast}}), \\
|\mathfrak{z}_{k,l,m}^{2}|\!=& \Theta(\zeta^{k-2}_{\mathcal{R},l})+\Theta(\xi^{m^{\ast}}), \\
|\mathfrak{z}_{k,l,m}^{3}|\!=&\Theta(\zeta^{k-2}_{\mathcal{R},l})+\Theta(\xi^{m^{\ast}}),
\end{align}
\end{subequations}
for prespecified integers $l^{\ast}$ and $m^{\ast}$ and any $k\!\in\!\mathbb{N}$, such that all the inner products in \eqref{innerpdistRDIS} belong to the interval $[-1,1]$. After substituting \eqref{innerpdistRDIS} into \eqref{dcheckdhat}, $\check{d}_{k,l,m}^{i^{\ast}}$ and $\hat{d}_{k,l,m}^{i^{\ast}}$ are obtained as
\begin{subequations}\label{ThieoneT}
\begin{align}
\check{d}_{k,l,m}^{i^{\ast}} =&\,
\sqrt{1-\big{|}1- \mathfrak{z}_{k,l,m}^{1}\big{|}^2}, \\
\hat{d}_{k,l,m}^{i^{\ast}} =&\, \sqrt{1-\tfrac{\big{|} (1-\mathfrak{z}_{k,l,m}^{1})(1-\mathfrak{z}_{k,l,m}^{2})-(1-\mathfrak{z}_{k,l,m}^{3})\big{|}^2}{\big{(} 1-|1-\mathfrak{z}_{k,l,m}^{1}|^2\big{)} \big{(} 1-|1-\mathfrak{z}_{k,l,m}^{2}|^2\big{)}}}.  \label{Thieone33} 
\end{align}    
\end{subequations}
By keeping only the asymptotically dominant terms, using the definition of function $\Theta(\cdot)$ and on noting that $m^{\ast}:=k$ at the $k\nth$ iteration, it follows from \eqref{ThieoneT} that
\begin{subequations}\label{Thieone3}
\begin{align}
(\check{d}_{k,l,m}^{i^{\ast}})^{2} =&\,
\Theta(\zeta_{\mathcal{R},l}^{k-1}) \!+\!\Theta(\xi^{m^{\ast}}), \\
(\hat{d}_{k,l,m}^{i^{\ast}})^{2} =&\, 1-\tfrac{\Theta(\zeta_{\mathcal{R},l}^{2k-4})+\Theta(\xi^{2m^{\ast}})}{\Theta(\zeta_{\mathcal{R},l}^{2k-3})+\Theta(\xi^{2m^{\ast}})}=1-\Theta(1).  \label{Thieone32} 
\end{align}    
\end{subequations}
On noting that $l^{\ast}\!:=\!k$, it follows that  $\lim_{k\rightarrow \infty} \zeta_{\mathcal{R},l}\!=\!\zeta_{\mathcal{R}}$. Given that $0\!<\!\zeta_{\mathcal{R}}\!<\!1$ and $0\!<\!\xi\!<\!1$ (according to Assumptions~\ref{Assump1} and \ref{Assump2}), it follows that $\zeta_{\mathcal{R},l}^{k-1}\!\rightarrow\!0$ and $\xi^{m^{\ast}}\!\rightarrow\!0$ as $k\!\rightarrow\!\infty$. 
Due to \eqref{This-one} and the definition of function $\Theta(\cdot)$, it results from \eqref{Thisone5} and \eqref{Thieone3} that there exists a positive constant $\kappa$ such that
\begin{subequations}\label{Thisone7R}
\begin{align}
& \lim\limits_{k\rightarrow \infty} \check{d}_{k,l,m}^{i}=\lim\limits_{k\rightarrow \infty} \check{d}_{k}=0,\;\;\; \forall \,i\in \mathcal{V}, \\
& \lim\limits_{k\rightarrow \infty} \hat{d}_{k,l,m}^{i}=\lim\limits_{k\rightarrow \infty} \hat{d}_{k}>\kappa,\;\;\; \forall \,i\in \mathcal{V},
\end{align}
\end{subequations}
in scenario $\mathcal{R}$. 
By defining $d_{k+1,l,m}^{i}$ based on \eqref{dchk1dist}, one concludes from \eqref{Thisone7R} that for scenario $\mathcal{R}$
\begin{equation}\label{This-one2R}
\lim\limits_{k\rightarrow \infty}d_{k+1,l,m}^{i}= \lim\limits_{k\rightarrow \infty}\check{d}_{k} =0, \;\;\; \forall\, i\!\in\! \mathcal{V},    
\end{equation}
which guarantees the termination of the distributed GPI algorithm after a finite number of iterations in this scenario from the viewpoint of all nodes, given the positive threshold $\epsilon$ used in the termination condition of the algorithm. Additionally, the distributed GPI algorithm asymptotically estimates the magnitude of the real eigenvalue $\lambda_{n}(\tilde{\mmat{L}})$ as the dominant eigenvalue of $\tilde{\mmat{L}}$ in a distributed manner from the viewpoint of all nodes in scenario $\mathcal{R}$ such that 
\begin{equation}\label{This-one25b}
\lim\limits_{k\rightarrow \infty}\check{\lambda}_{k,l,m}^{i}=|\lambda_{n}(\tilde{\mmat{L}})|,    
\end{equation}
for any $i\in\mathcal{V}$. Since the dominant eigenvalue $\lambda_{n}(\tilde{\mmat{L}})$ corresponds to the one-dimensional subspace $\mathcal{V}^{\ast}:=\mathrm{span}\{\mvec{v}_{n}(\tilde{\mmat{L}})\}$ in this scenario, the subspace sequence $\{\mathcal{V}_{k,l,m}\}_{k\in \mathbb{N}}$, $\mathcal{V}_{k,l,m}=\mathrm{span}\{\mvec{x}_{k,l,m}\}$, approaches $\mathcal{V}^{\ast}$, while the subspace sequence $\{\mathcal{W}_{k,l,m}\}_{k\in \mathbb{N}}$, $\mathcal{W}_{k,l,m}=\mathrm{span}\{\mvec{x}_{k-1,l,m},\mvec{x}_{k,l,m}\}$, is not convergent as $k\rightarrow \infty$. 
Define $\tilde{\lambda}_{k+1,l,m}^{i}$ based on \eqref{This-one3} for any $i\!\in\!\mathcal{V}$ and $k\!\in\!\mathbb{N}$. Using \eqref{This-one2R} and \eqref{This-one25b}, the asymptotic estimation of the GAC of the network by all the nodes is concluded, as described in \eqref{FineqDIS}. This proves the validity of \eqref{Prdist} and completes the convergence proof in scenario $\mathcal{R}$. 
\end{proof}

\begin{remark}
In addition to the iteration index $k$, used to characterize the main loop of the distributed GPI procedure in Algorithm~\ref{AlgGPIdist}, iteration indices $l$ and $m$ are utilized for two additional inner loops nested within the main loop of Algorithm~\ref{AlgGPIdist}, each repeated $l^{\ast}$ and $m^{\ast}$ times, respectively.  
Given that the convergence proof of the distributed GPI algorithm requires $l^{\ast}$ and $m^{\ast}$ to unboundedly increase as $k\rightarrow \infty$, $l^{\ast}:=k$ and $m^{\ast}:=k$ are used to evaluate them at the $k\nth$ iteration for any $k\!\in\!\mathbb{N}$. 
Since this approach can lead to unnecessarily excessive computational complexity, termination conditions can be defined for each one of the two inner loops based on a desired level of precision, instead of fixing the number of nested iterations to $k$ in the $k\nth$ iteration of the main loop.  
To achieve this, $\epsilon_{L}$ and $\epsilon_{M}$ are introduced as two sufficiently small positive values, and the upper bounds $l_{max}$ and $m_{max}$ on the number of iterations for the two nested loops in the $k\nth$ iteration can be defined as the smallest integers $l$ and $m$, respectively, for which the following inequalities hold
\begin{subequations}
\begin{align}
&\max_{i\in\mathcal{V}}\left|y^i(l)-y^i(l\!-\!1)\right|<\epsilon_L, \\ 
&\max_{i\in\mathcal{V},s\in\mathbb{N}_4} |z^{s,i}_{k}(m)-z^{s,i}_{k}(m\!-\!1)|<\epsilon_M.
\end{align}
\end{subequations}
Then, line~\ref{Thisline} of Algorithm~\ref{AlgGPIdist} can be replaced with
\begin{equation}
l^{\ast}\!\gets\!\min\{k,l_{max}\},\;\;m^{\ast}\!\gets\!\min\{k,m_{max}\}. 
\end{equation}
\end{remark}

\begin{remark}
The advantage of the GPI algorithm in computing the connectivity of asymmetric networks compared to the algorithms in \cite{AsadiTechRep_17,AsadiTSMC_20} becomes clear by using the \textrm{CONGEST} model as a well-known distributed computing framework to implement the GPI algorithm in a distributed manner~\cite{Peleg_00,Hillel_19}. 
In this computing model, all nodes exchange messages with their neighbors in a synchronous manner while the size of the message that can be exchanged over any communication link of the network in one iteration is of order $O(\log_{2}(n))$ bits for a network of size $n$. 
Due to this limitation in the CONGEST model, to implement the distributed algorithms with a message length of order larger than $O(\log_{2}(n))$ bits, a long message can be divided into smaller packets and sent over a sequence of extra iterations~\cite{Hillel_18,Romanow_95}. Although this solution satisfies the constraint on the maximum message length of each communication link in every iteration, it comes at the cost of increasing the number of communication rounds of the entire algorithm by adding a sufficient number of iterations to the algorithm, resulting in higher message complexity. 
Since the proposed distributed GPI algorithm has a maximum message length of order $O(1)$ bits, it already satisfies this constraint, and consequently, does not require any additional iterations when implemented using the CONGEST model. On the other hand, the distributed algorithms with the maximum message length of order $O(n)$ bits can only be implemented at the cost of adding $O(\tfrac{n}{\log_{2}(n)})$ extra number of iterations to each single iteration of the algorithm for the purpose of exchanging messages between the nodes.
This superiority of the proposed distributed GPI algorithm over other algorithms in terms of its lower message complexity becomes more noticeable in large networks (i.e., for considerable values of $n$). 
\end{remark}

\section{Simulation Results}\label{Sec:VIII}

\begin{example}\label{Exp:1}
Consider an asymmetric network composed of six nodes represented by a strongly connected weighted digraph $\mathcal{G}=(\mathcal{V},\mathcal{E},\mmat{W})$,  demonstrated in Fig.~\ref{Fig:Ex12}, with the weight matrix 
\begin{equation}
\mmat{W}=\begin{bmatrix}
     0    &  0   &  0.78 & 0.71 & 0.93 & 0.73 \\
     0    &  0   &  0    & 0    & 0.90 & 0.88 \\
     0.98 &  0   &  0    & 0.76 & 0.55 & 0    \\
     0    &  0.10&  0    & 0    & 0    & 0.75 \\
     0    &  0   &  0.61 & 0.77 & 0    & 0    \\
     0    &  0   &  0    & 0    & 0.61 & 0
\end{bmatrix}.
\end{equation}
\vspace{-10pt}
\begin{figure}[htbp]
\centering
\includegraphics[width=0.35\textwidth]{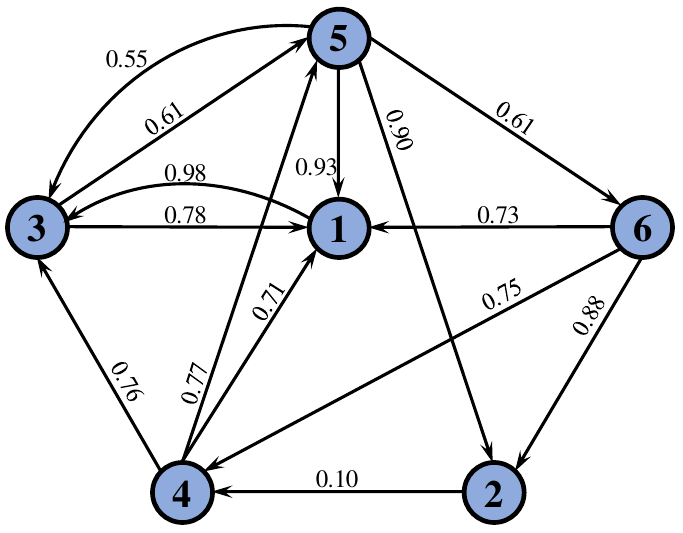}
\caption{The weighted digraph $\mathcal{G}$ in Example~\ref{Exp:1}.}
\label{Fig:Ex12}
\end{figure}

\noindent In this example, the GAC of the network can be obtained directly by finding the spectrum of the Laplacian matrix $\mmat{L}$. More precisely, $\tilde{\lambda}(\mmat{L})\!=\!1.192$, which corresponds to the pair of complex conjugate eigenvalues $1.192\pm \mathrm{j} 0.630$ of $\mmat{L}$. The performance of Algorithms~\ref{AlgGPI1} and \ref{AlgGPIdist} in centralized and distributed estimation of the network's GAC is demonstrated in Fig.~\ref{Fig:GPI-I} by choosing $\delta\!=\!0.235$, $\epsilon\!=\!5\times 10^{-4}$, $l^{\ast}\!:=\!k$, $m^{\ast}\!:=\!k$ and using $\mvec{x}_{0}=[0.0976\;0.2323\;0.2316\;0.8137\;0.1618\;0.4411]^{\tr}$ as the initial state vector of the network. 

Since the GAC of the network in this example is associated with a pair of complex conjugate eigenvalues of $\mmat{L}$, the GPI algorithm is required to identify scenario $\mathcal{I}$, where the network's GAC is associated with a two-dimensional subspace $\mathcal{W}^{\ast}=\mathrm{span}\{\mvec{v}_{5}(\tilde{\mmat{L}}), \mvec{v}_{6}(\tilde{\mmat{L}})\}$. The estimation of the network's GAC via the centralized GPI algorithm is demonstrated in Fig.~\ref{Fig:I11}, as a sequence $\{\tilde{\lambda}_{k}\}_{k\in \mathbb{N}}$ converging to $\tilde{\lambda}(\mmat{L})$ under Algorithm~\ref{AlgGPI1}, which terminates after 46 iterations. The evolution of $\check{d}_{k}$ and $\hat{d}_{k}$, representing the distance between the pairs of successive one-dimensional and two-dimensional subspaces, respectively, at the $k\nth$ iteration of the GPI algorithm, is depicted in Fig.~\ref{Fig:I12}. This figure demonstrates that the sequence $\{\hat{d}_{k}\}_{k\in \mathbb{N}}$ converges to zero while the sequence $\{\check{d}_{k}\}_{k\in \mathbb{N}}$ is not convergent in scenario $\mathcal{I}$. This confirms in this example that the GPI algorithm correctly associates the dominant eigenvalue of $\tilde{\mmat{L}}$ with the desired two-dimensional subspace $\mathcal{W}^{\ast}$ in scenario $\mathcal{I}$. It also follows from Fig.~\ref{Fig:I13} that the sequence $\{\hat{\lambda}_{k}\}_{k\in \mathbb{N}}$ converges to the magnitude of the dominant eigenvalue of $\tilde{\mmat{L}}$, i.e. $|\lambda_{6}(\tilde{\mmat{L}})|=2.055$, while the sequence $\{\check{\lambda}_{k}\}_{k\in \mathbb{N}}$ is not convergent. In addition, Fig.~\ref{Fig:I14} shows that scenario $\mathcal{I}$ is correctly identified by the centralized GPI algorithm after 11 iterations.

\begin{figure*}
\hspace{-15pt}  
  \begin{tabular}[c]{cccc}
    \begin{subfigure}[b]{0.24\textwidth}
      \includegraphics[width=\textwidth]{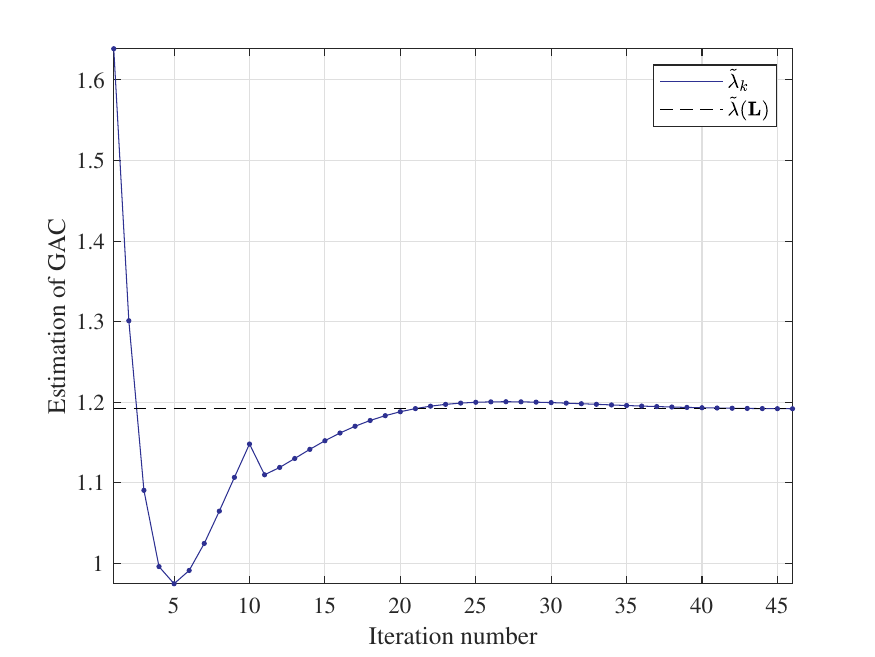}
      \caption{}
      \label{Fig:I11}
    \end{subfigure}&
    \begin{subfigure}[b]{0.24\textwidth}
      \includegraphics[width=\textwidth]{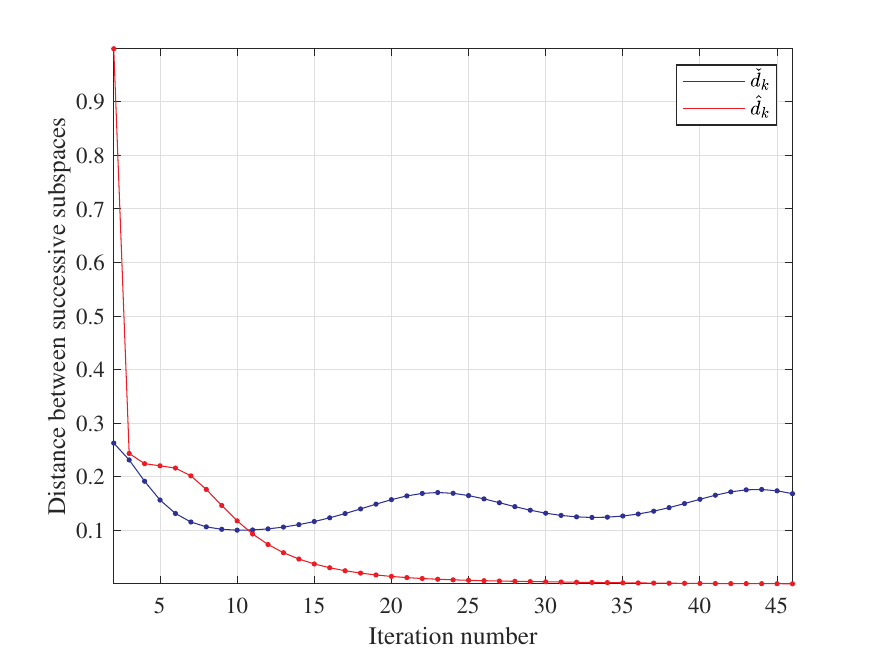}
      \caption{}
      \label{Fig:I12}
    \end{subfigure}&
    \begin{subfigure}[b]{0.24\textwidth}
      \includegraphics[width=\textwidth]{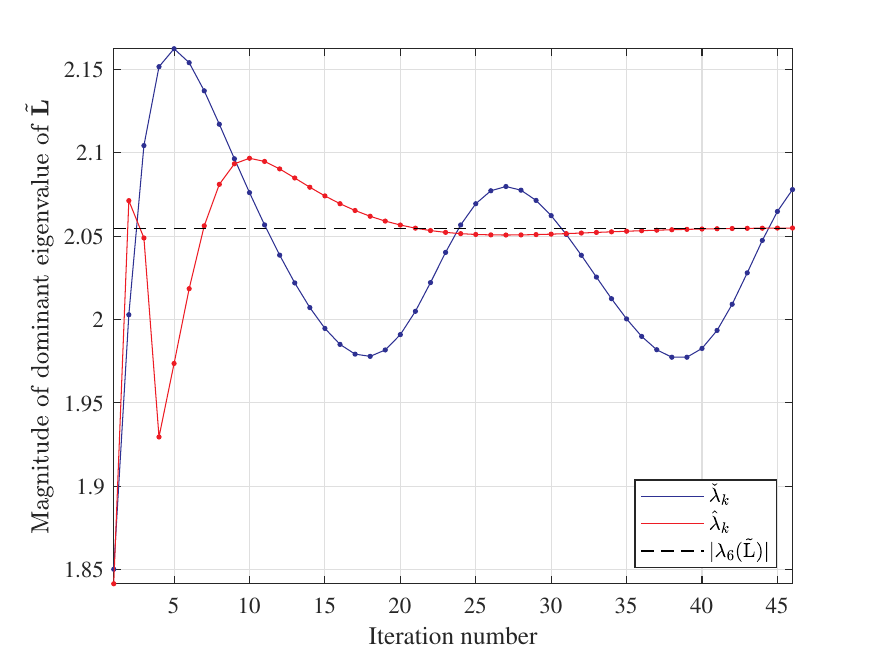}
      \caption{}
      \label{Fig:I13}
    \end{subfigure}&
    \begin{subfigure}[b]{0.24\textwidth}
      \includegraphics[width=\textwidth]{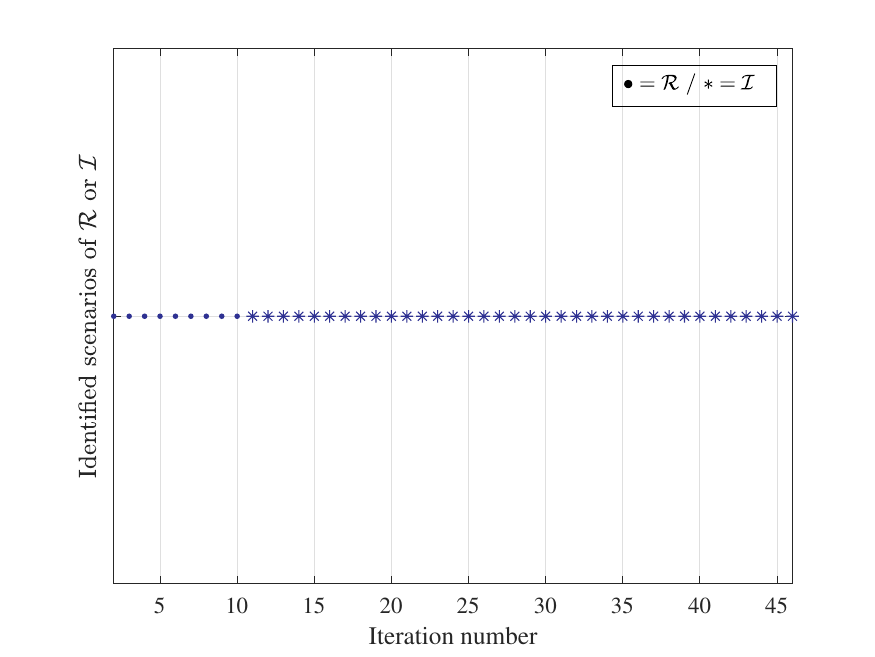}
      \caption{}
      \label{Fig:I14}
    \end{subfigure} \\
    \begin{subfigure}[b]{0.24\textwidth}
      \includegraphics[width=\textwidth]{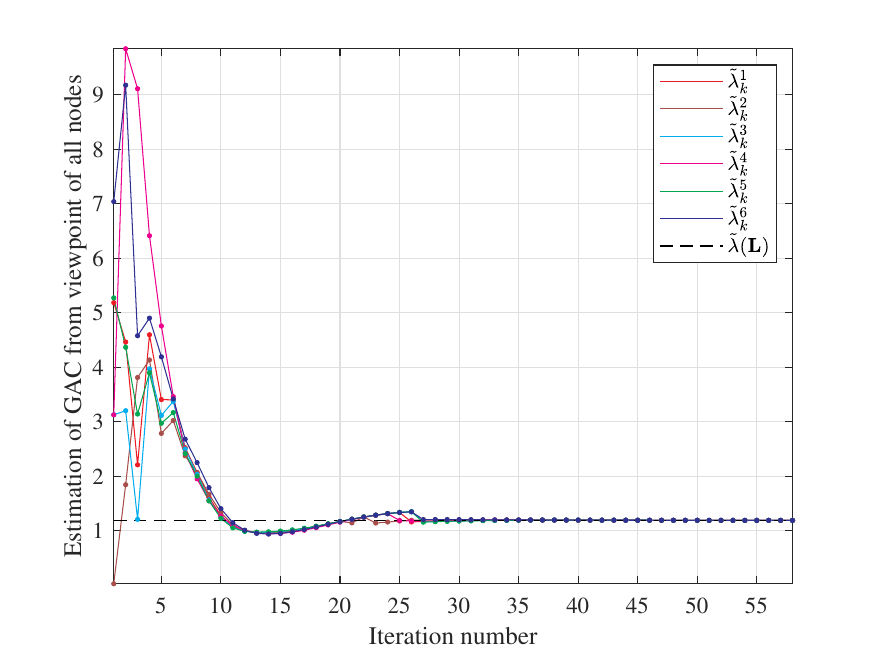}
      \caption{}
      \label{Fig:I21}
    \end{subfigure}&
    \begin{subfigure}[b]{0.24\textwidth}
      \includegraphics[width=\textwidth]{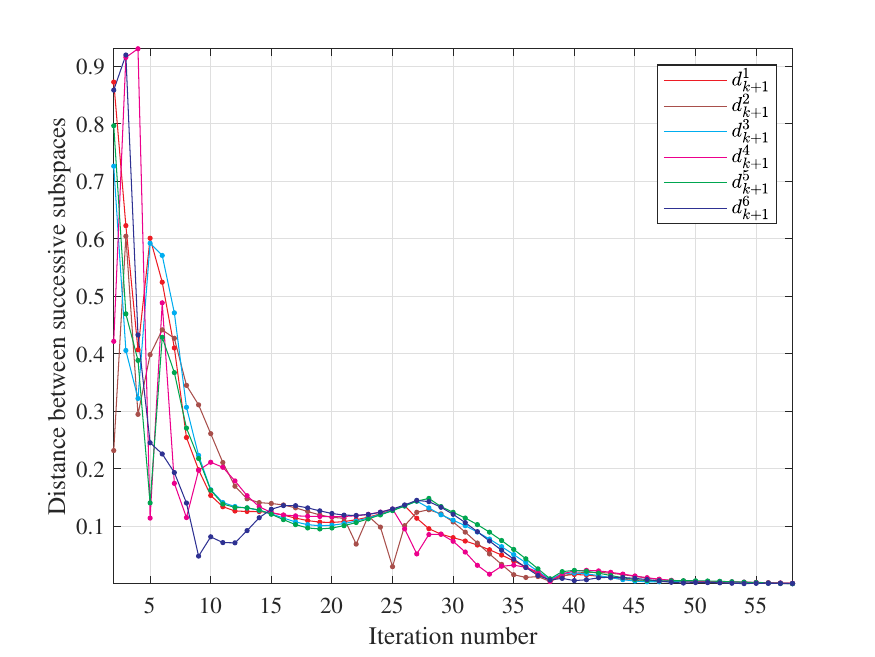}
      \caption{}
      \label{Fig:I22}
    \end{subfigure}&
    \begin{subfigure}[b]{0.24\textwidth}
      \includegraphics[width=\textwidth]{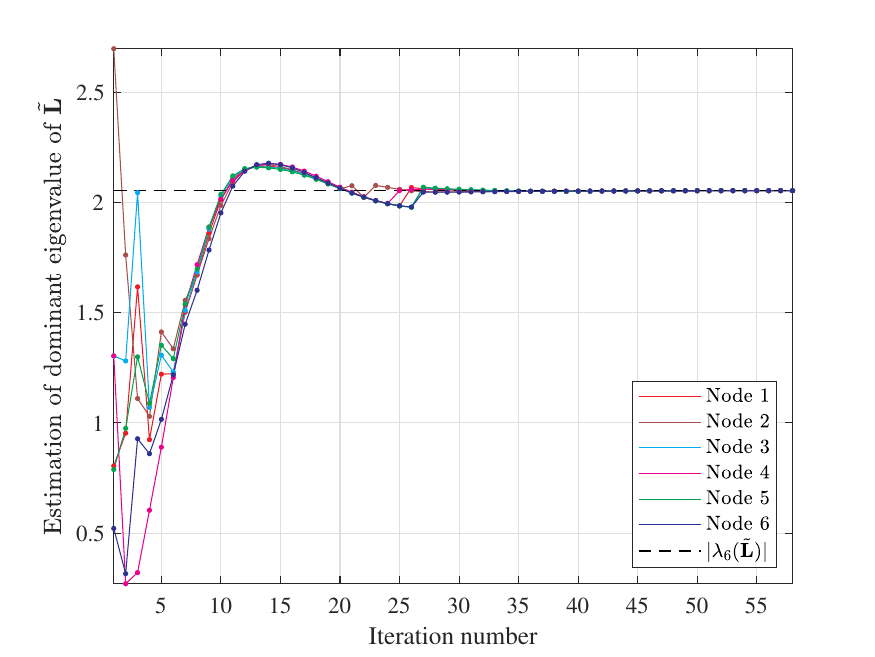}
      \caption{}
      \label{Fig:I23}
    \end{subfigure}&
    \begin{subfigure}[b]{0.24\textwidth}
      \includegraphics[width=\textwidth]{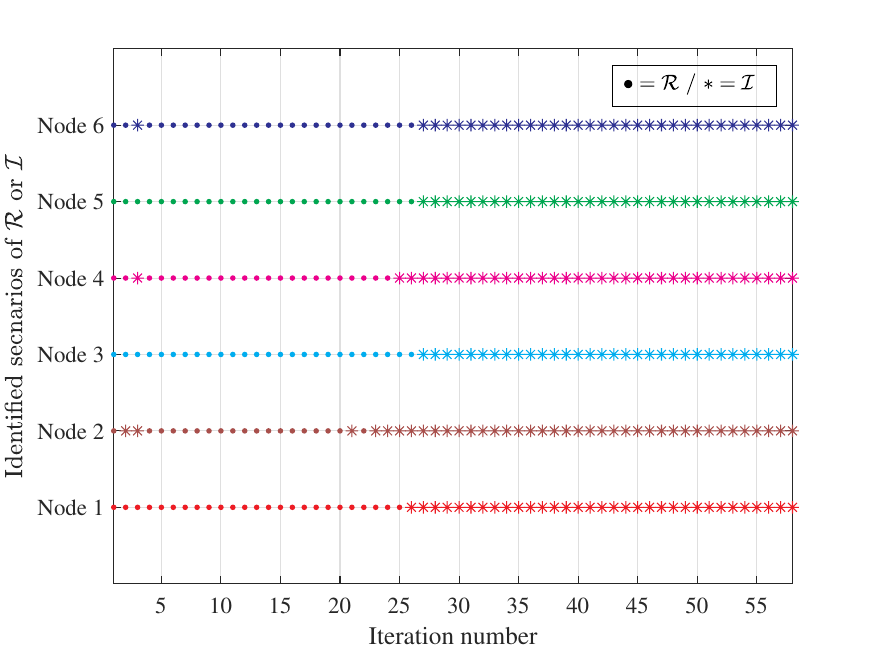}
      \caption{}
      \label{Fig:I24}
    \end{subfigure}
  \end{tabular}
  \caption{Performance of the GPI algorithm in estimating the network's GAC in scenario $\mathcal{I}$: centralized implementation using Algorithm~\ref{AlgGPI1} in \ref{Fig:I11}-\ref{Fig:I14}, and distributed implementation using Algorithm~\ref{AlgGPIdist} in \ref{Fig:I21}-\ref{Fig:I24}.}\label{Fig:GPI-I}
\end{figure*}
\begin{figure}[htbp]
\centering
\begin{subfigure}[b]{0.24\textwidth}
\includegraphics[width=\textwidth]{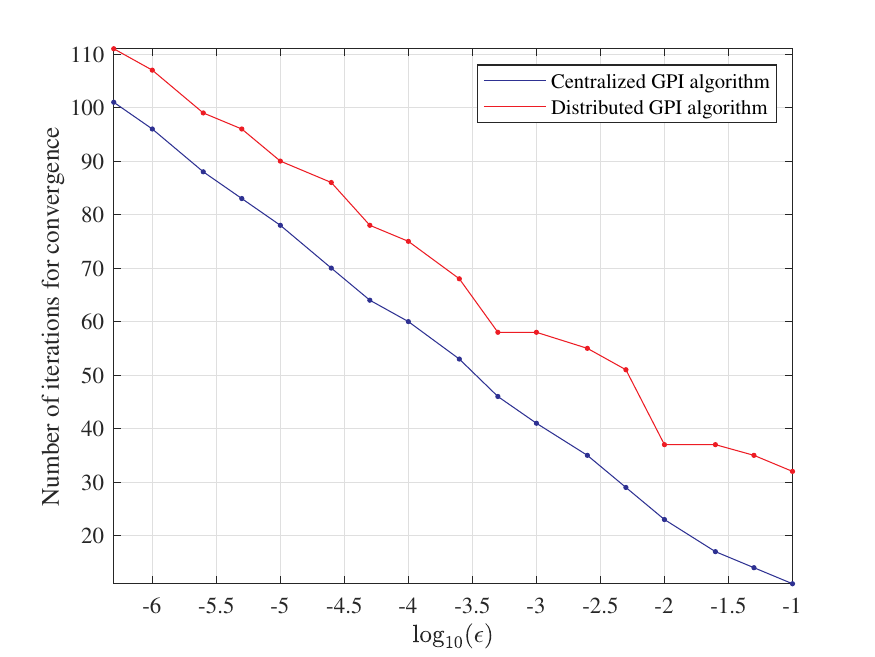}
\caption{}
\label{ISpeed}
\end{subfigure}\hfill
\begin{subfigure}[b]{0.24\textwidth}
\includegraphics[width=\textwidth]{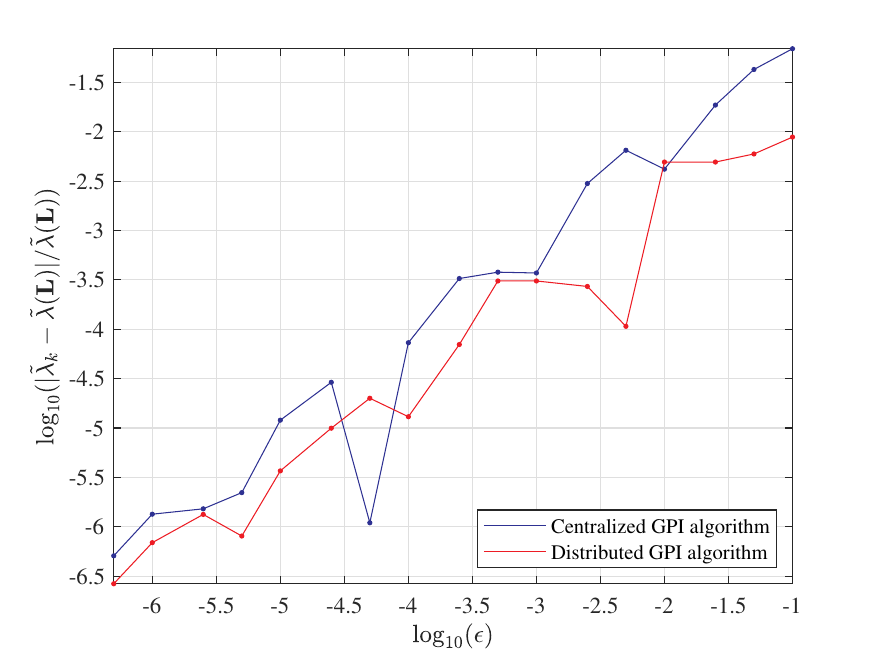}
\caption{}
\label{IAccur}
\end{subfigure}
\caption{Impact of parameter $\epsilon$ on the convergence speed (\ref{ISpeed}) and estimation accuracy (\ref{IAccur}) of the GPI algorithm in Example~\ref{Exp:1}.}
\label{Ieps2}
\end{figure}
The performance of the distributed GPI algorithm in Example~\ref{Exp:1} is depicted in Figs.~\ref{Fig:I21}-\ref{Fig:I24} for scenario $\mathcal{I}$. The estimation of the network's GAC in a distributed manner from the viewpoint of every node is depicted in Fig.~\ref{Fig:I21}. Using Algorithm~\ref{AlgGPIdist}, the sequence $\{\tilde{\lambda}_{k}^{i}\}_{k\in \mathbb{N}}$ converges to $\tilde{\lambda}(\mmat{L})$ after 58 iterations, for any $i\!\in\!\mathbb{N}_{6}$. The evolution of sequence $\{d_{k+1}^{i}\}_{k\in \mathbb{N}}$, for any $i\in \mathbb{N}_{6}$ (obtained based on the distance between subsequent one-dimensional and two-dimensional subspaces), is plotted in Fig.~\ref{Fig:I22}, which shows all six curves converge to zero. 
The estimated magnitude of the dominant eigenvalue of $\tilde{\mmat{L}}$ from the viewpoint of each node is depicted in Fig.~\ref{Fig:I23}, which demonstrates all of them converge to $|\lambda_{6}(\tilde{\mmat{L}})|=2.055$. Moreover, scenario $\mathcal{I}$ is correctly identified by all nodes in a distributed manner after at most 27 iterations of Algorithm~\ref{AlgGPIdist}, as shown in Fig.~\ref{Fig:I24}. 

Since the stopping condition of the GPI algorithm is characterized by the threshold value $\epsilon$, the effect of this parameter on the convergence speed and estimation accuracy of the GPI algorithm in scenario $\mathcal{I}$ is investigated in Fig.~\ref{Ieps2} for both centralized and distributed approaches. 
In this figure, the number of iterations required for the main loop of the GPI algorithm to converge and the relative error $|\tilde{\lambda}_{k}-\tilde{\lambda}(\mmat{L})|/\tilde{\lambda}(\mmat{L})$ (or $\max_{i\in \mathcal{V}}|\tilde{\lambda}_{k}^{i}-\tilde{\lambda}(\mmat{L})|/\tilde{\lambda}(\mmat{L})$ for the  distributed implementation) are respectively used as the measures of the convergence speed and estimation accuracy of the GPI algorithm. 
More specifically, Fig.~\ref{ISpeed} shows that as $\log_{10}(\epsilon)$ increases, the required number of iterations for the convergence of the GPI algorithm in both centralized and distributed approaches decreases almost linearly. 
In addition, Fig.~\ref{IAccur} demonstrates the impact of parameter $\epsilon$ on the estimation accuracy of the network's GAC, where the relative error increases almost linearly as $\epsilon$ increases, for both centralized and distributed implementations of the GPI algorithm.


\end{example}


\begin{example}\label{Exp:2}
Let $\mathcal{G}\!=\!(\mathcal{V},\mathcal{E},\mmat{W})$ denote the strongly connected weighted digraph depicted in Fig.~\ref{Fig:Ex22}, representing an asymmetric network composed of six nodes. Let the weight matrix $\mmat{W}$ be given by
\begin{equation}
\mmat{W}=\begin{bmatrix}
     0    &  0   &  0 & 0.61 & 0.75 & 0    \\
     0.60 &  0   &  0 & 0.97 & 0    & 0.71 \\
     0    &  0.86&  0 & 0.77 & 0    & 0    \\
     0    &  0   &  0 & 0    & 0.74 & 0.72 \\
     0.85 &  1   &  0 & 0    & 0    & 1    \\
     0    &  0   &0.76& 0    & 0.58 & 0
\end{bmatrix}.
\end{equation}
\begin{figure}[htbp]
\centering
\includegraphics[width=0.35\textwidth]{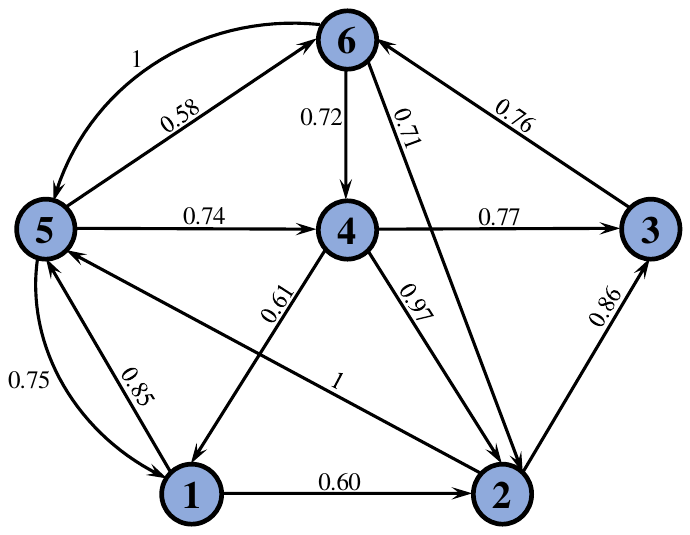}
\caption{The weighted digraph $\mathcal{G}$ in Example~\ref{Exp:2}.}
\label{Fig:Ex22}
\end{figure}

\noindent The GAC of this network is $\tilde{\lambda}(\mmat{L})=1.255$, corresponding to a real eigenvalue of $\mmat{L}$. This indicates that the GPI algorithm is required to correctly identify scenario $\mathcal{R}$, in which the network's GAC is associated with a one-dimensional subspace $\mathcal{V}^{\ast}\!=\!\mathrm{span}\{\mvec{v}_{6}(\tilde{\mmat{L}})\}$. By considering the parameters $\delta\!=\!0.269$, $\epsilon\!=\!5\times 10^{-4}$, $l^{\ast}\!:=\!k$, $m^{\ast}\!:=\!k$ and the initial state vector as $\mvec{x}_{0}=[0.1423\;0.4528\;0.6571\;0.0866\;0.5208\;0.2534]^{\tr}$, the performance of Algorithms~\ref{AlgGPI1} and \ref{AlgGPIdist} in computing the GAC of the network in scenario $\mathcal{R}$ is depicted in Fig.~\ref{Fig:GPI-R}. 

The estimation of the network's GAC in a centralized setting is provided in Fig.~\ref{Fig:R11}, which shows the sequence $\{\tilde{\lambda}_{k}\}_{k\in \mathbb{N}}$ converges to $\tilde{\lambda}(\mmat{L})$, and Algorithm~\ref{AlgGPI1} terminates after 49 iterations. The evolution of $\check{d}_{k}$ and $\hat{d}_{k}$, representing the distance between the pairs of successive one-dimensional and two-dimensional
subspaces, respectively, at the $k\nth$ iteration, is depicted in Fig.~\ref{Fig:R12}, where the sequence $\{\check{d}_{k}\}_{k\in \mathbb{N}}$ converges to zero, while the sequence $\{\hat{d}_{k}\}_{k\in \mathbb{N}}$ is not convergent.  
It also follows from Fig.~\ref{Fig:R13} that the sequence $\{\check{\lambda}_{k}\}_{k\in \mathbb{N}}$ converges to the magnitude of the dominant eigenvalue of $\tilde{\mmat{L}}$, i.e. $|\lambda_{6}(\tilde{\mmat{L}})|=1.939$, while the sequence $\{\hat{\lambda}_{k}\}_{k\in \mathbb{N}}$ is not convergent. 
As a result, the GPI algorithm correctly associates the dominant eigenvalue of $\tilde{\mmat{L}}$ with the desired one-dimensional subspace $\mathcal{V}^{\ast}$ in scenario $\mathcal{R}$. Moreover, scenario $\mathcal{R}$ is correctly identified by the GPI algorithm after 9 iterations, according to Fig.~\ref{Fig:R14}.

\begin{figure*}
\hspace{-15pt}  
  \begin{tabular}[c]{cccc}
    \begin{subfigure}[b]{0.24\textwidth}
      \includegraphics[width=\textwidth]{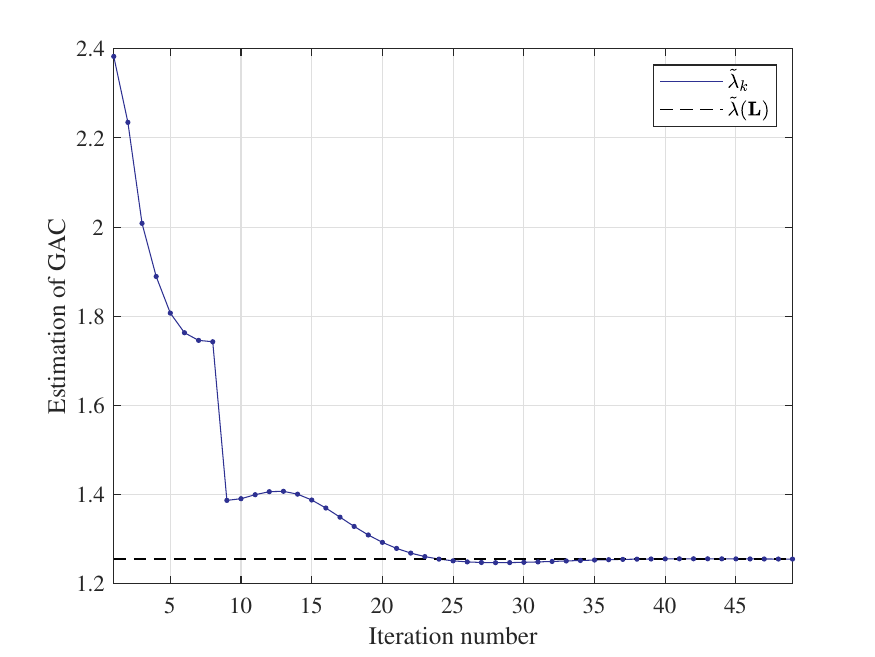}
      \caption{}
      \label{Fig:R11}
    \end{subfigure}&
    \begin{subfigure}[b]{0.24\textwidth}
      \includegraphics[width=\textwidth]{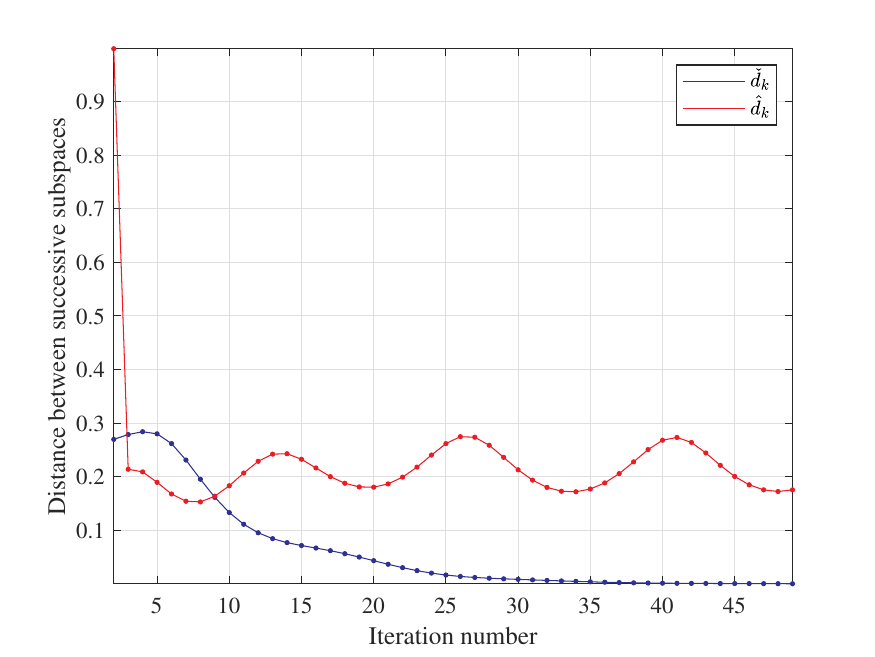}
      \caption{}
      \label{Fig:R12}
    \end{subfigure}&
    \begin{subfigure}[b]{0.24\textwidth}
      \includegraphics[width=\textwidth]{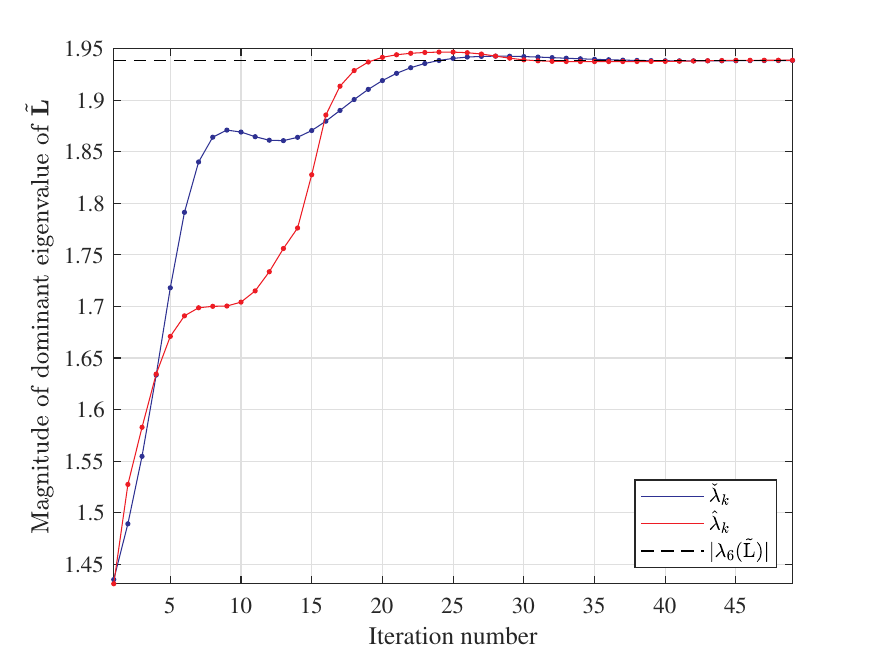}
      \caption{}
      \label{Fig:R13}
    \end{subfigure}&
    \begin{subfigure}[b]{0.24\textwidth}
      \includegraphics[width=\textwidth]{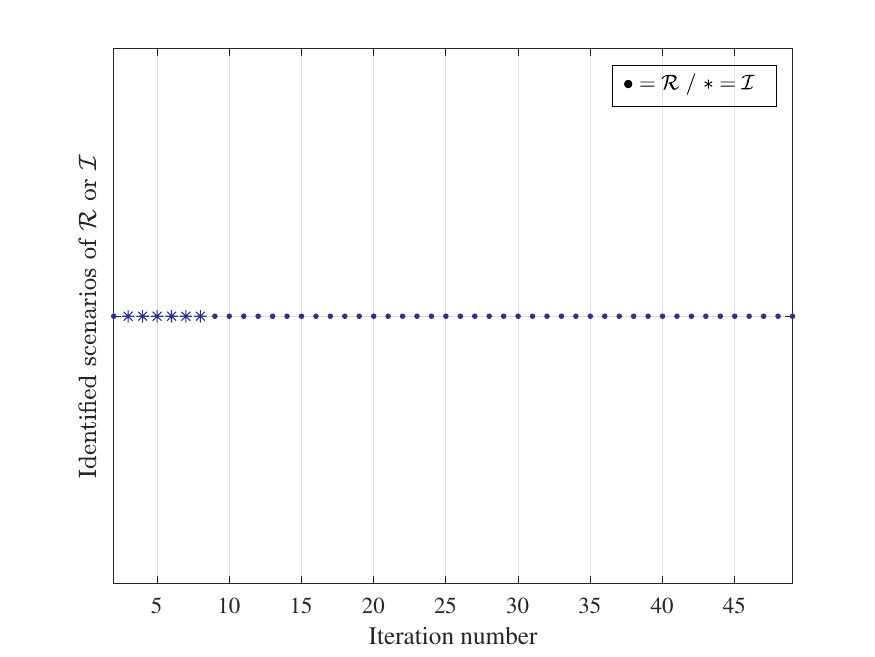}
      \caption{}
      \label{Fig:R14}
    \end{subfigure} \\
    \begin{subfigure}[b]{0.24\textwidth}
      \includegraphics[width=\textwidth]{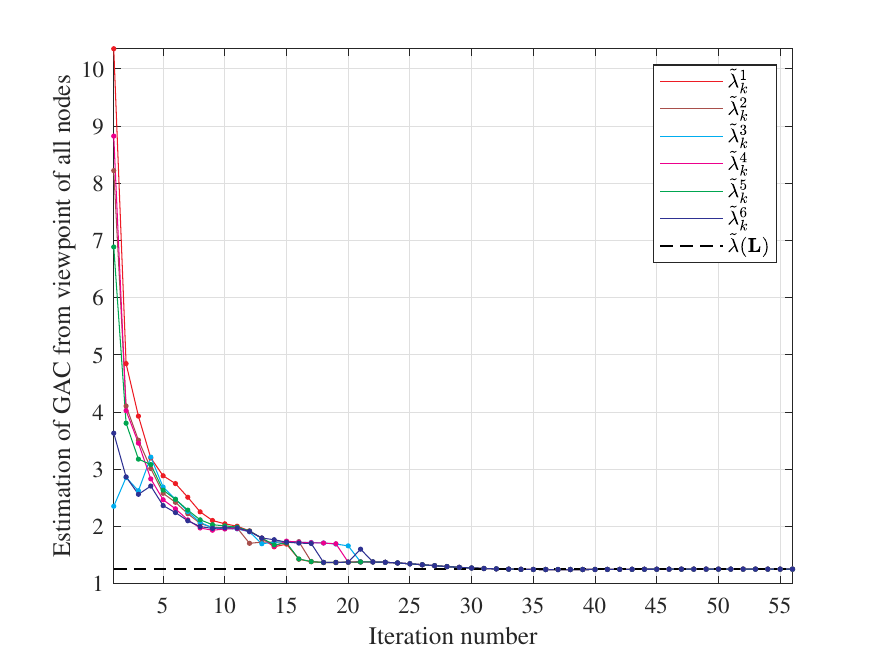}
      \caption{}
      \label{Fig:R21}
    \end{subfigure}&
    \begin{subfigure}[b]{0.24\textwidth}
      \includegraphics[width=\textwidth]{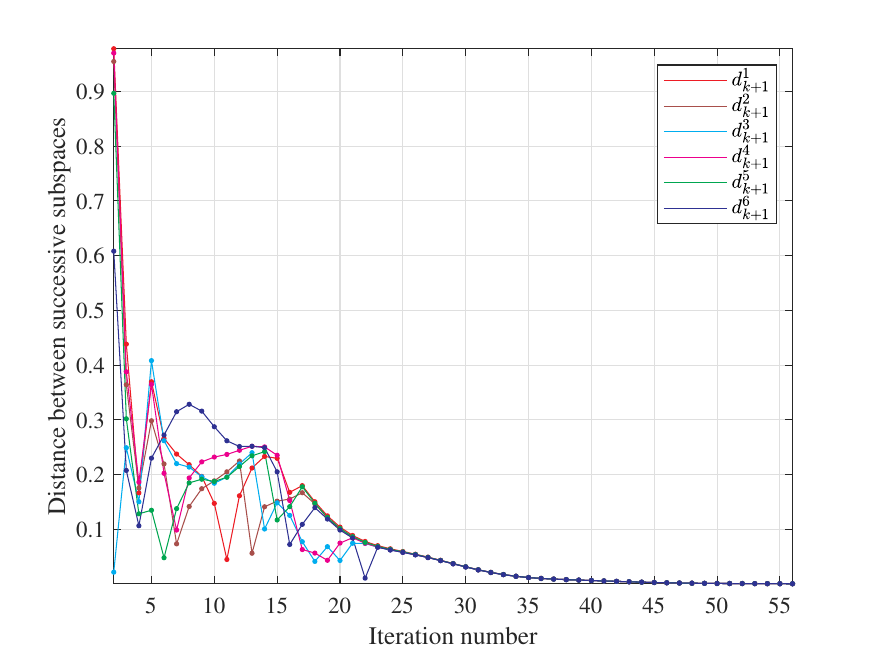}
      \caption{}
      \label{Fig:R22}
    \end{subfigure}&
    \begin{subfigure}[b]{0.24\textwidth}
      \includegraphics[width=\textwidth]{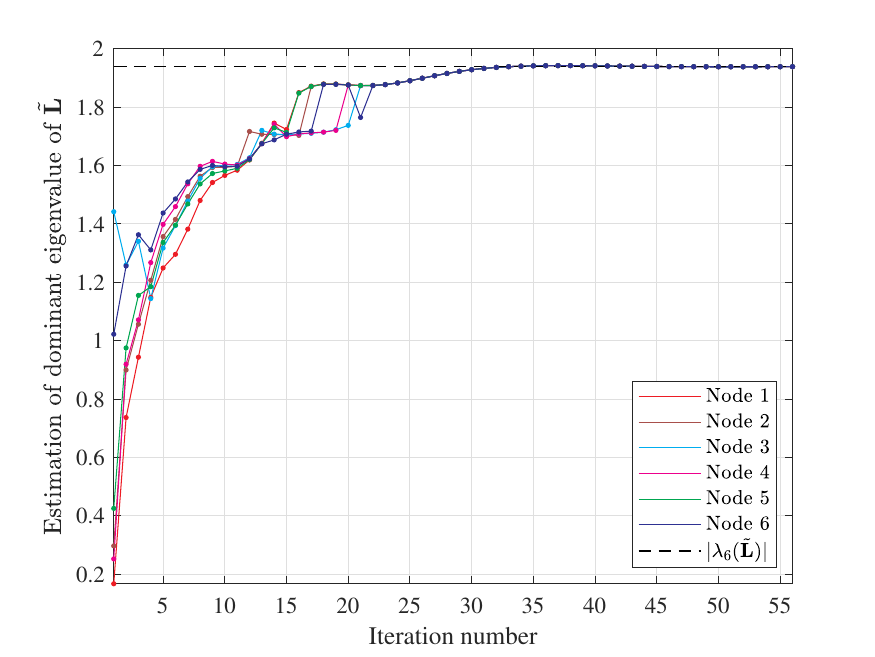}
      \caption{}
      \label{Fig:R23}
    \end{subfigure}&
    \begin{subfigure}[b]{0.24\textwidth}
      \includegraphics[width=\textwidth]{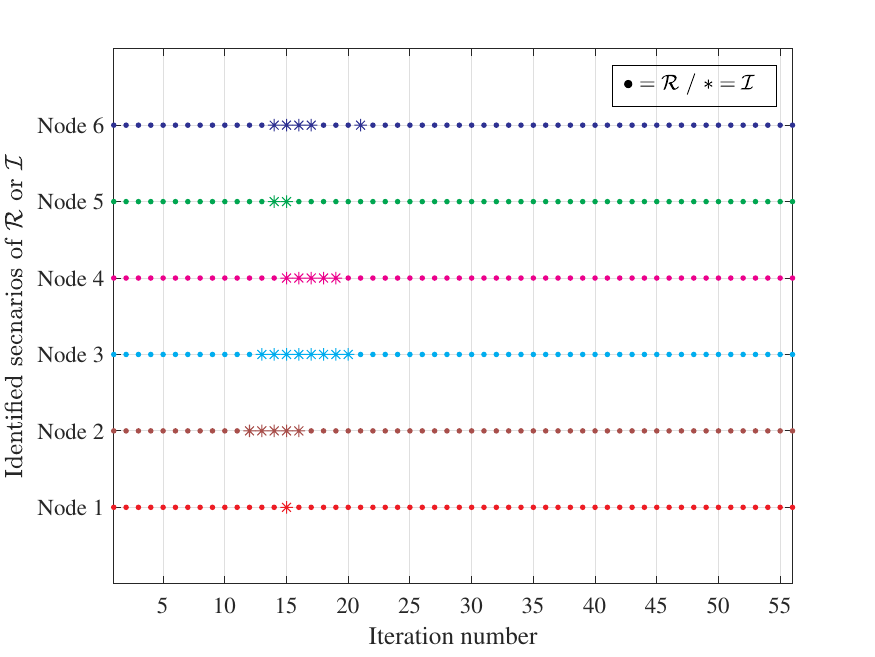}
      \caption{}
      \label{Fig:R24}
    \end{subfigure}
  \end{tabular}
  \caption{Performance of the GPI algorithm in estimating the network's GAC in scenario $\mathcal{R}$: centralized implementation using Algorithm~\ref{AlgGPI1} in \ref{Fig:R11}-\ref{Fig:R14}, and distributed implementation using Algorithm~\ref{AlgGPIdist} in \ref{Fig:R21}-\ref{Fig:R24}.}\label{Fig:GPI-R}
\end{figure*}
\begin{figure}[htbp]
\centering
\begin{subfigure}[b]{0.24\textwidth}
\includegraphics[width=\textwidth]{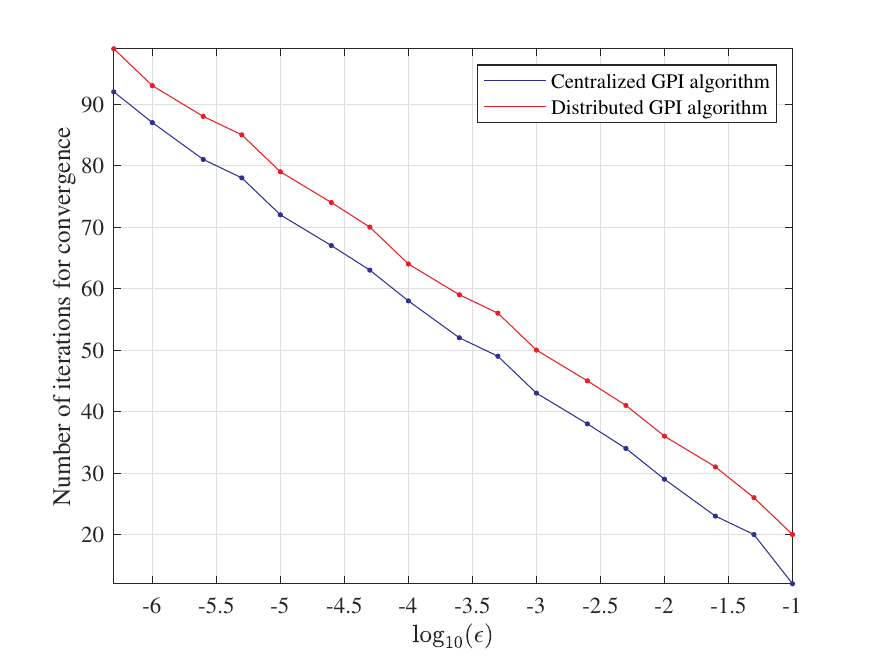}
\caption{}
\label{RSpeed}
\end{subfigure}\hfill
\begin{subfigure}[b]{0.24\textwidth}
\includegraphics[width=\textwidth]{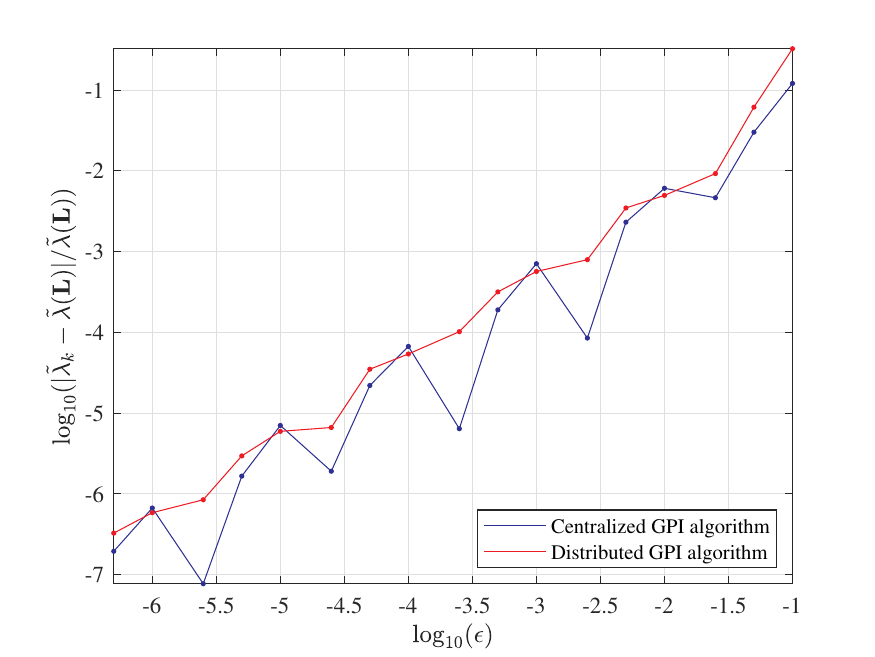}
\caption{}
\label{RAccur}
\end{subfigure}
\caption{Impact of parameter $\epsilon$ on the convergence speed (\ref{RSpeed}) and estimation accuracy (\ref{RAccur}) of the GPI algorithm in Example~\ref{Exp:2}.}
\label{Reps2}
\end{figure}
The performance of the distributed GPI algorithm in Example~\ref{Exp:2} is depicted in Figs.~\ref{Fig:R21}-\ref{Fig:R24}.  
The GAC of the network is estimated in a distributed manner from the viewpoint of every node in Fig.~\ref{Fig:R21}, where the sequence $\{\tilde{\lambda}_{k}^{i}\}_{k\in \mathbb{N}}$ converges to $\tilde{\lambda}(\mmat{L})$ after 56 iterations, for any $i\!\in\!\mathbb{N}_{6}$. The evolution of the sequence $\{d_{k+1}^{i}\}_{k\in \mathbb{N}}$ is provided in Fig.~\ref{Fig:R22}, which shows that all six curves converge to zero.  
The estimated magnitude of the dominant eigenvalue of $\tilde{\mmat{L}}$ from the viewpoint of each node is depicted in Fig.~\ref{Fig:R23}, showing the convergence of every curve to $|\lambda_{6}(\tilde{\mmat{L}})|=1.939$. Moreover, scenario $\mathcal{R}$ is correctly identified by all nodes in the distributed approach after at most 22 iterations of Algorithm~\ref{AlgGPIdist}, according to Fig.~\ref{Fig:R24}.

The effect of the size of the threshold parameter $\epsilon$ on the convergence speed and estimation accuracy of the GPI algorithm in scenario $\mathcal{R}$ is depicted in Fig.~\ref{Reps2}. It follows from Fig.~\ref{RSpeed} that the required number of iterations for the convergence of the GPI algorithm in both centralized and distributed implementations decreases almost linearly, as $\log_{10}(\epsilon)$ increases in scenario $\mathcal{R}$. Moreover, it follows from Fig.~\ref{RAccur} that the accuracy of the GPI algorithm in estimating the GAC of the network decreases almost linearly, as the value of $\epsilon$ increases in scenario $\mathcal{R}$ for both centralized and distributed implementations. 

\end{example}
\begin{example}\label{Exp:3}
In this example, the performance of the proposed distributed GPI algorithm is compared with that of the subspace consensus approach in \cite{AsadiTechRep_17} using Monte Carlo simulation. By employing the CONGEST computing model, these two distributed procedures are implemented on randomly generated strongly connected asymmetric networks of different sizes. 
For each network size, the weight matrix $\mmat{W}$ is generated as a matrix of uniformly distributed random numbers between 0 and 1. 
Algorithm~\ref{AlgGPIdist} and the distributed procedure in \cite{AsadiTechRep_17} are then applied to the same asymmetric network using $\delta\!=\!\Delta^{-1}-0.01$, $\epsilon\!=\!0.01$ and the same randomly-generated initial state vector. The same procedure is repeated for 20 randomly-generated strongly connected digraphs, and the number of communication rounds required for the convergence of each algorithm is recorded. The average values obtained from simulations is deemed the desired number of communication rounds for each algorithm (with the same network size under different topologies). It is shown in Fig.~\ref{Reps3} that as the network size increases, the number of average communication rounds remains more or less constant for the distributed GPI algorithm, while it increases almost linearly for the  subspace consensus method, confirming the superiority of the proposed approach as far as message complexity is concerned.

\begin{figure}[htbp]
\centering
\begin{subfigure}[b]{0.24\textwidth}
\includegraphics[width=\textwidth]{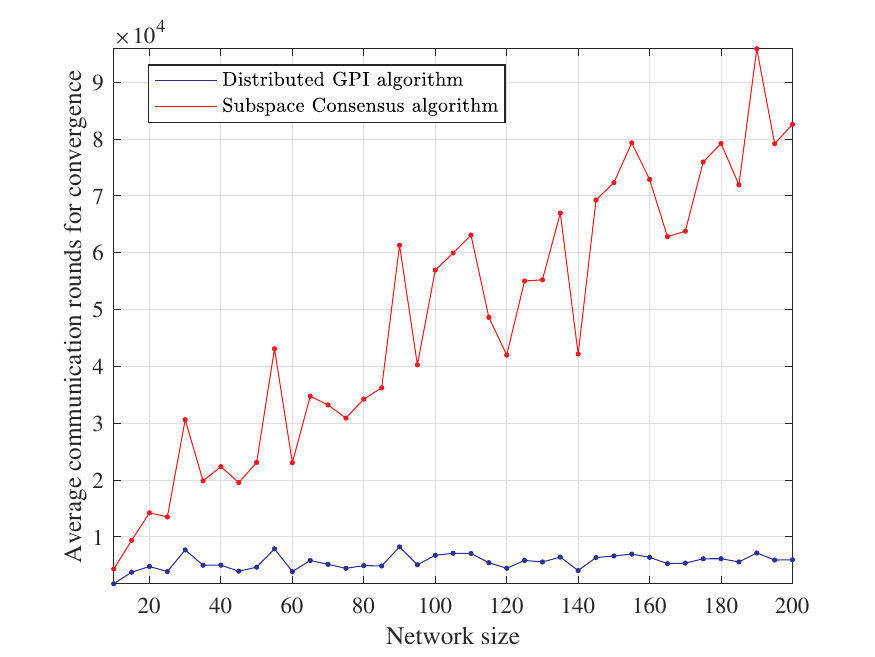}
\caption{}
\label{4.1}
\end{subfigure}\hfill
\begin{subfigure}[b]{0.24\textwidth}
\includegraphics[width=\textwidth]{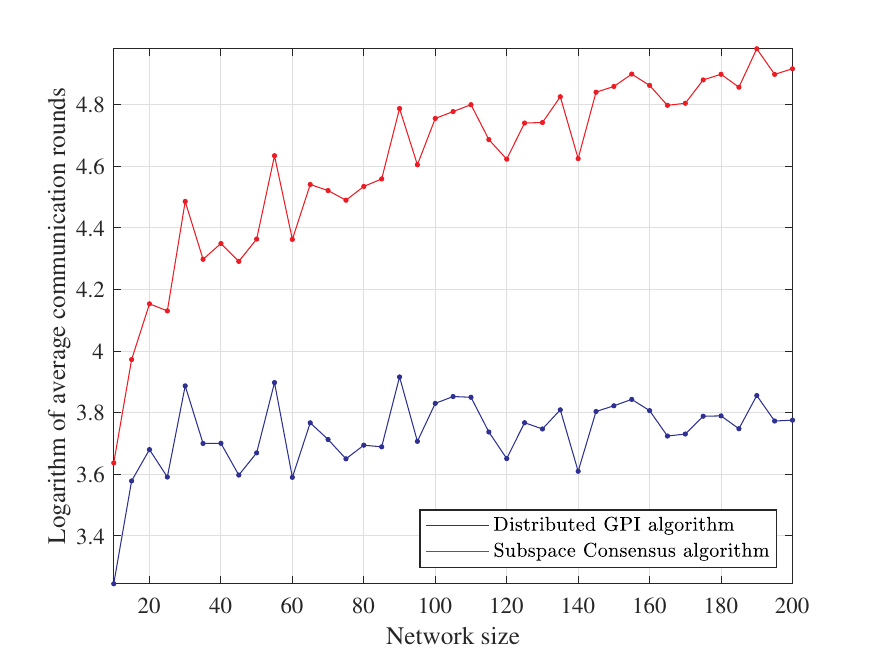}
\caption{}
\label{4.2}
\end{subfigure}
\caption{Impact of network size on the average number of communication rounds (\ref{4.1}) and its decimal  logarithm (\ref{4.2}) required for convergence using CONGEST model in Example~\ref{Exp:3}.}
\label{Reps3}
\end{figure}

\end{example}

\section{Conclusions}\label{Sec:IX}

A generalized power iteration (GPI) algorithm is introduced in this work as a novel approach to compute the generalized algebraic connectivity (GAC) of asymmetric networks in both centralized and distributed manners. The GPI algorithm generalizes the use of power iteration approach to asymmetric networks by generating sequences of one-dimensional and two-dimensional subspaces in both centralized and distributed fashions. 
As one of its major advantages, the distributed implementation of the GPI algorithm is truly scalable with a fixed message length per node irrespective of the network's size. 
It is shown that under some standard assumptions one of the two subspace sequences converges to the subspace characterizing the GAC of the network from the viewpoint of every node. After detailed convergence analyses of the GPI algorithm in both centralized and distributed implementations, its efficacy in computing the connectivity measure of asymmetric networks is verified by simulations.


\bibliographystyle{IEEEtran}
\bibliography{References}

\end{document}